\documentclass[aip,preprint,amsmath]{revtex4-1}

\draft % marks overfull lines with a black rule on the right

\usepackage{graphicx}
\usepackage{subcaption}
\usepackage{color}

\draft % marks overfull lines with a black rule on the right

\begin{document}

%\preprint{}

\title{Grid patterns, spatial inter-scan variations and scanning reading repeatability in radiochromic film dosimetry} 

\author{I. M{\'e}ndez}
\email[]{nmendez@onko-i.si}
\author{\v{Z}. \v{S}ljivi{\'c}}
\author{R. Hudej}
\author{A. Jenko}
\author{B. Casar}
\affiliation{Department of Medical Physics, Institute of Oncology Ljubljana, Zalo\v{s}ka cesta 2, Ljubljana 1000, Slovenia}

\begin{abstract}

\textbf{Purpose:}
When comparing different scans of the same radiochromic film, several patterns can be observed. These patterns are caused by different sources of uncertainty, which affect the repeatability of the scanner. The purpose of this work was to study these uncertainties.

\noindent\textbf{Methods:}
The variance of the scanner noise, as a function of the pixel position, was studied for different resolutions. The inter-scan variability of the scanner response was analyzed taking into account spatial discrepancies. Finally, the distance between the position of the same point in different scans was examined.

\noindent\textbf{Results:}
The variance of noise follows periodical patterns in both axes, causing the grid patterns. These patterns were identified for resolutions of 50, 72 and 96 dpi, but not for 150 dpi. Specially recognizable is the sinusoidal shape with a period of 8.5 ${\rm mm}$ that is produced with 72 dpi. Inter-scan variations of the response caused systematic relative dose deviations larger than 1\% in 5\% of the red channel images, 9\% of the green and 51\% of the blue. No systematic deviation larger than 1\% was found after applying response corrections. The initial positioning and the speed of the scanner lamp vary between scans. 

\noindent\textbf{Conclusions:}
Three new sources of uncertainty, which influence radiochromic film dosimetry with flatbed scanners, have been identified and analyzed in this work: grid patterns, spatial inter-scan variations and scanning reading repeatability. A novel correction method is proposed, which mitigates spatial inter-scan variations caused by deviations in the autocalibration of the individual Charge Coupled Device detectors.

\end{abstract}


\maketitle %\maketitle must follow title, authors, abstract and \pacs

\section{Introduction}

The system composed of radiochromic films and a flatbed scanner is the dosimeter of choice for many applications in radiology and radiation therapy\cite{Devic:2011}. This dosimetry system is affected by several sources of uncertainty. Some of them involve only the film: for example, the thickness variations of the active layer\cite{hartmann:2010}, the change in film darkening as a function of post-irradiation time\cite{andres:2010}, the influence of humidity and temperature\cite{rink:2008, Girard:2012}, the UV-induced polymerization\cite{aapm:55}, etc. Some other uncertainties are a consequence of the interaction of the characteristics of both the film and scanner: for example, the lateral artifact\cite{Schoenfeld:2014, vanbattum:2015}, Newton rings\cite{dreindl:2014}, the dependency with the orientation of the film on the scanner bed\cite{butson:2009}, the cross talk effect\cite{vanbattum:2015}, the dependency on film-to-light source distance\cite{lewis:2015, palmer:2015}, etc. Finally, other uncertainties are intrinsic to the scanner: for example, noise\cite{bouchard:2009, vanHoof:2012}, the inter-scan variability of the scanner response\cite{lewis:2015}, warming-up of the lamp\cite{Paelinck:2007, ferreira:2009}, differences between color channels\cite{AMicke:2011, mayer:2012, mendez:2014, perez:2014}, etc. 

Despite all those perturbations, GAFChromic films (Ashland Inc., Wayne, NJ) have been repeatedly found to be capable of delivering accurate dose measurements\cite{lewis:2012, palmer:2013, perez:2014, mendez:2015}. Still, to further improve the accuracy of the dosimetry system, thorough knowledge of its uncertainties is necessary.

GAFChromic EBT3 films were used in this study, in combination with the Epson Expression 10000XL scanner (Seiko Epson Corporation, Nagano, Japan). In the literature, the Epson Expression 10000XL scanner has been selected numerous times\cite{vanbattum:2015, Schoenfeld:2014, lewis:2015, lewis:2015b, Martisikova:2008, andres:2010} for radiochromic film dosimetry. In this work, the repeatability of this scanner has been examined. As a result, three new artifacts have been identified and analyzed: grid patterns, spatial inter-scan variations and scanning reading repeatability.   

\section{Methods and materials}

GAFChromic EBT3 films from lot 06061401 were employed. They were irradiated with a Novalis Tx accelerator (Varian, Palo Alto, CA, USA). The darkening of the films was measured with an Epson Expression 10000XL scanner. The scanner was powered on 30 min before readings and five scans were taken to warm up its lamp. The films were placed on the center of the scanner with an opaque frame. To avoid the Callier effect \cite{lewis:2015b, palmer:2015}, a glass sheet, with a thickness of 3 mm, was placed on top of the films. They were scanned in portrait orientation ({\it i.e.}, the short side of the film parallel to the scanner lamp) and transmission mode. Images were acquired using the Epson Scan v3.49a software, in 48-bit RGB (16 bit per channel) format, while processing tools were turned off. Images were saved as TIFF files. Data analysis was performed with the R programming language \cite{R:software}. 

\subsection{Preliminary test}

A film was placed at a depth of 11 cm in an IBA MULTICube phantom (IBA Dosimetry GmbH, Schwarzenbruck, Germany). Source-to-film distance was 100 cm. The film was irradiated with a step pattern composed of six stripes with doses of 0.25, 1, 4, 8, 2 and, again, 0.25 Gy. It was scanned ten consecutive times 24 h after irradiation, with a resolution of 72 dpi. The mean of the ten scans was calculated. For each scan and color channel, the difference image between the scan and the mean scan was also computed.

\begin{figure*}
\begin{minipage}[b]{0.32\linewidth}
\centering
\includegraphics[width=\linewidth]{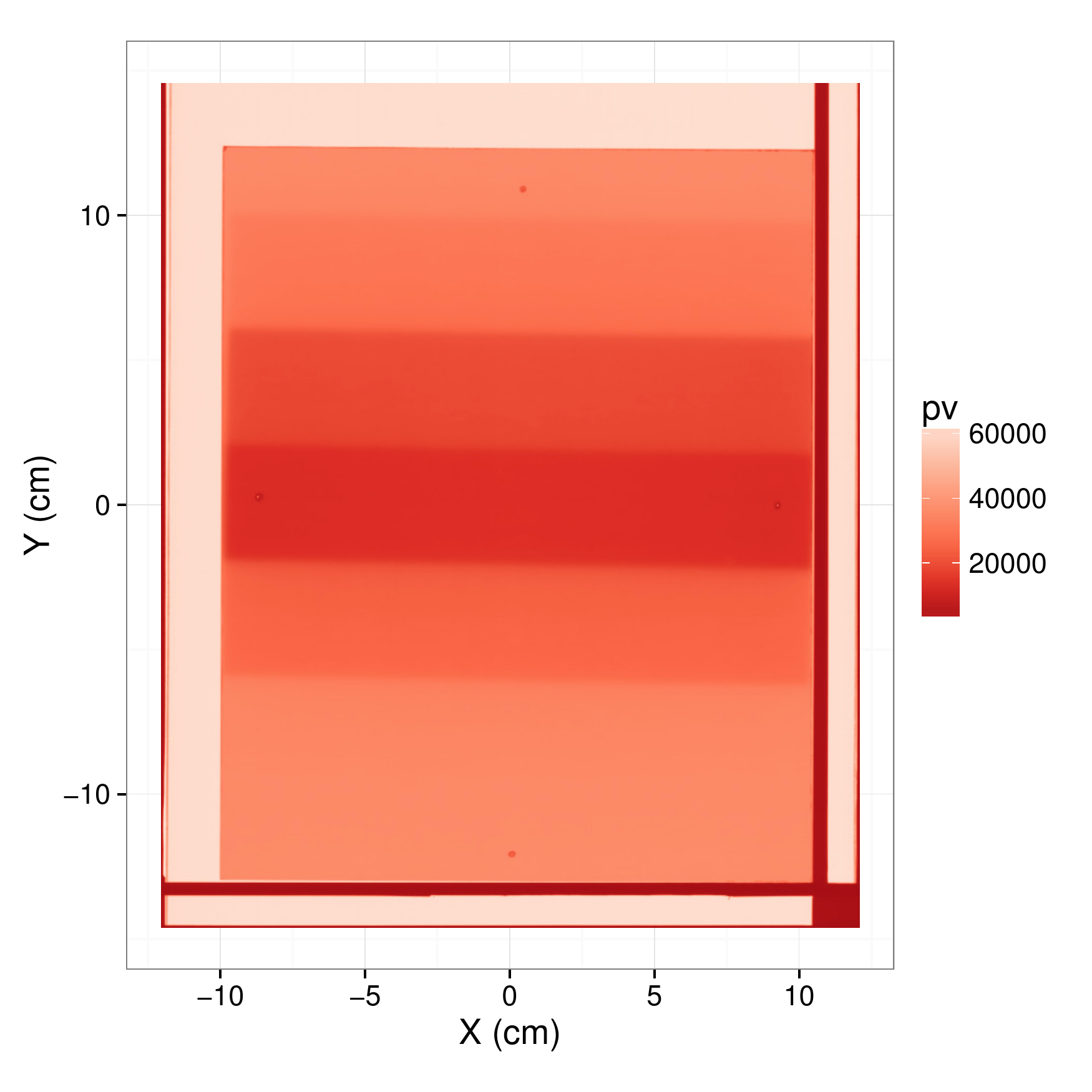}\\
(a)
\end{minipage}
\hfill
\begin{minipage}[b]{0.32\linewidth}
\centering
\includegraphics[width=\linewidth]{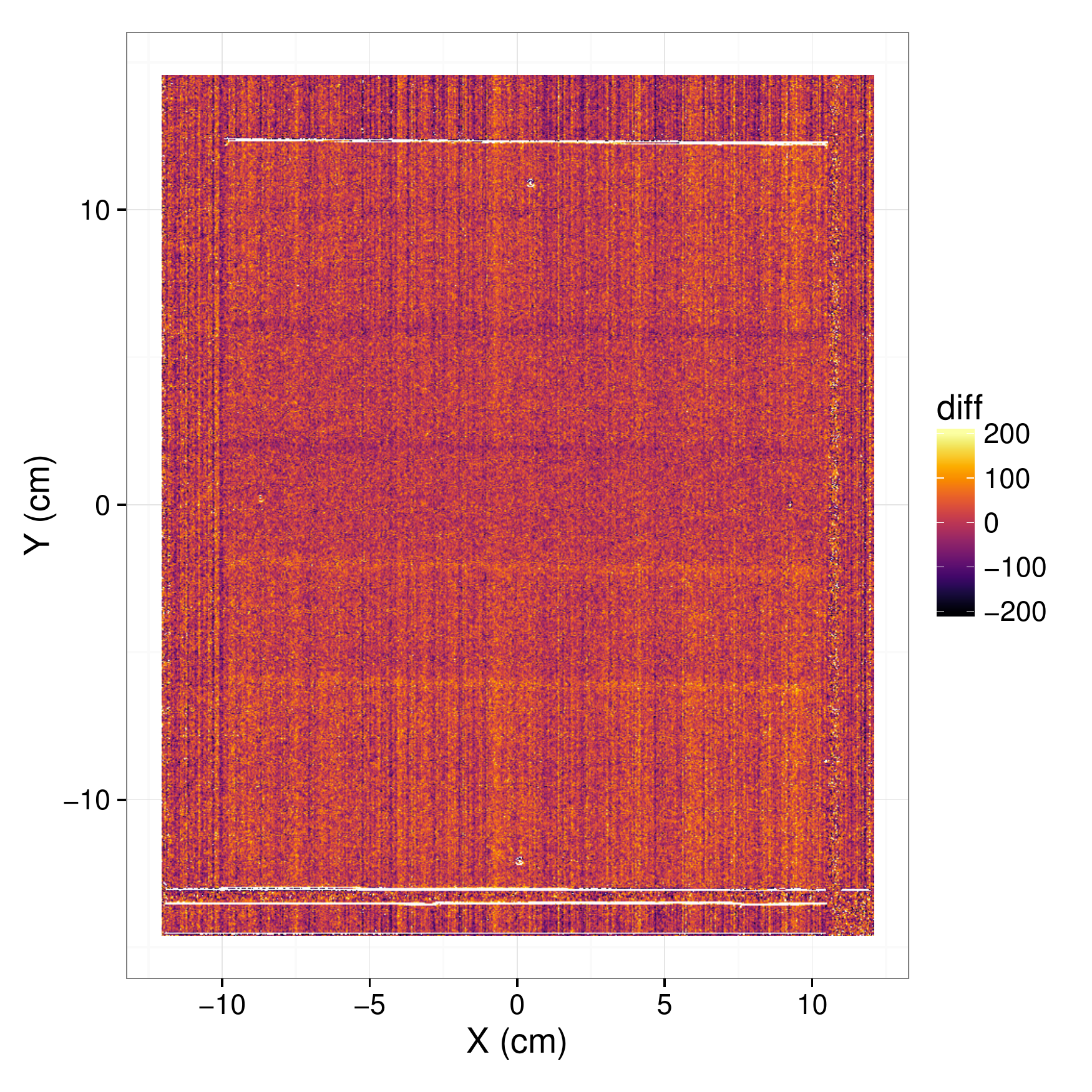}\\
(b)
\end{minipage}
\hfill
\begin{minipage}[b]{0.32\linewidth}
\centering
\includegraphics[width=\linewidth]{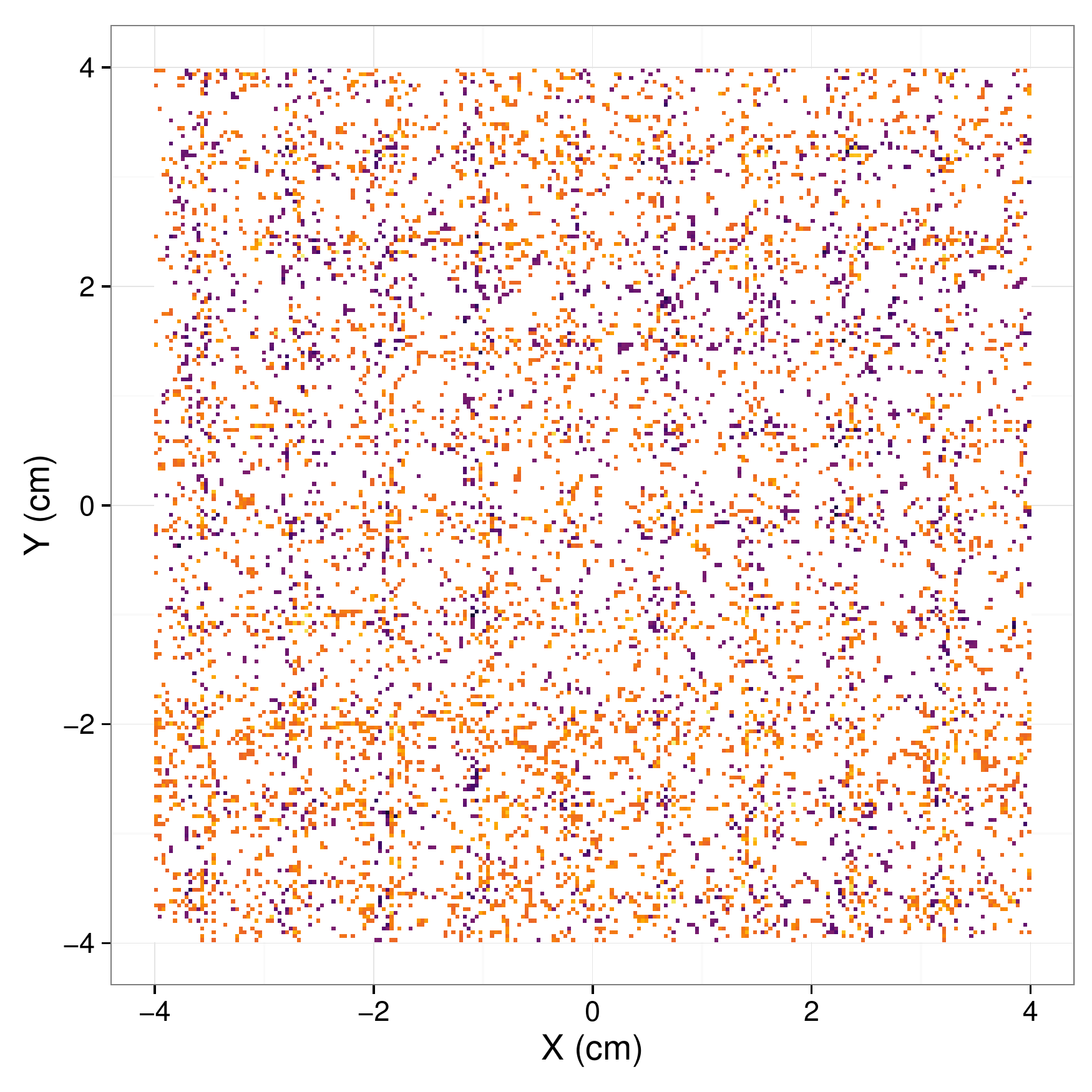}\\
(c)
\end{minipage}
\caption{\label{fig:preliminary} Pixel values (pv) in the red channel for a) one of the scans of the step pattern, b) the difference between this scan and the mean scan, excluding absolute differences larger than 200 pv, and c) a zoom of the difference image, excluding absolute differences larger than 200 pv or smaller than 60 pv.}
\end{figure*}

Figure~\ref{fig:preliminary} shows pixel values in the red channel of one scan image as well as the difference between this scan and the mean scan. Several patterns can be observed in the difference image: for example, the edges of the steps generate thick lines approximately parallel to the scanner lamp (axis X). There are many thin linear patterns perpendicular to the scanner lamp (axis Y). In addition, there is a grid pattern, which can be better perceived in Figure~\ref{fig:preliminary}c. These artifacts were present in most of the difference images. They were also found using the Epson Scan v3.41 software, as well as with another Epson Expression 10000XL scanner. The following tests were developed to analyze them.

\subsection{Grid pattern}

Four different setups were studied. In the first one, an unexposed film was scanned. In the second one, without the presence of the film, the light transmitted through the flattening glass sheet was measured, with the image referred to as  white background. In the third one, the bed of the scanner, except for the calibration area, was covered with a black opaque plastic in order to avoid the transmission of light to the detectors; this was called the black background. In the last one, three previously irradiated film stripes were scanned; their dimensions were 20.3 ${\rm cm}$ $\times$ 4 ${\rm cm}$ and had received homogeneous doses of 100, 200, and 400 cGy, respectively. Each setup was scanned with resolutions of 50, 72, 96 and 150 dpi. While the Epson Expression 10000XL scanner has an optical resolution of 2400 dpi, these resolutions were regarded as the most commonly used for film dosimetry. For each resolution, 20 scans were taken. 

For each of the four setups, resolution and color channel, the mean scan image was calculated. The difference between each scan and the corresponding mean image was obtained. Pixel value differences were grouped by column (X axis) and row (Y axis), while the mean absolute deviations (MADs) of the differences were computed. The MAD is a measure of statistical dispersion which is more robust to outliers than the standard deviation. If the sample is normally distributed, as was generally the case, the MAD is an estimator proportional to the standard deviation of the population. Hence, the objective of this test was to obtain the dispersion of the measures of the scanner ({\it i.e.}, the noise) as a function of the pixel position.

For the three irradiated stripes, relative dose uncertainties resulting from repeated scans were calculated. Dose uncertainties for each pixel were obtained as the product of the standard deviation of the pixel value, which can be determined from the MAD, times the derivative of the dose with respect to the pixel value.

\subsection{Spatial inter-scan variability}

\subsubsection{Measurements}

\begin{figure}
\includegraphics[width=0.47\linewidth]{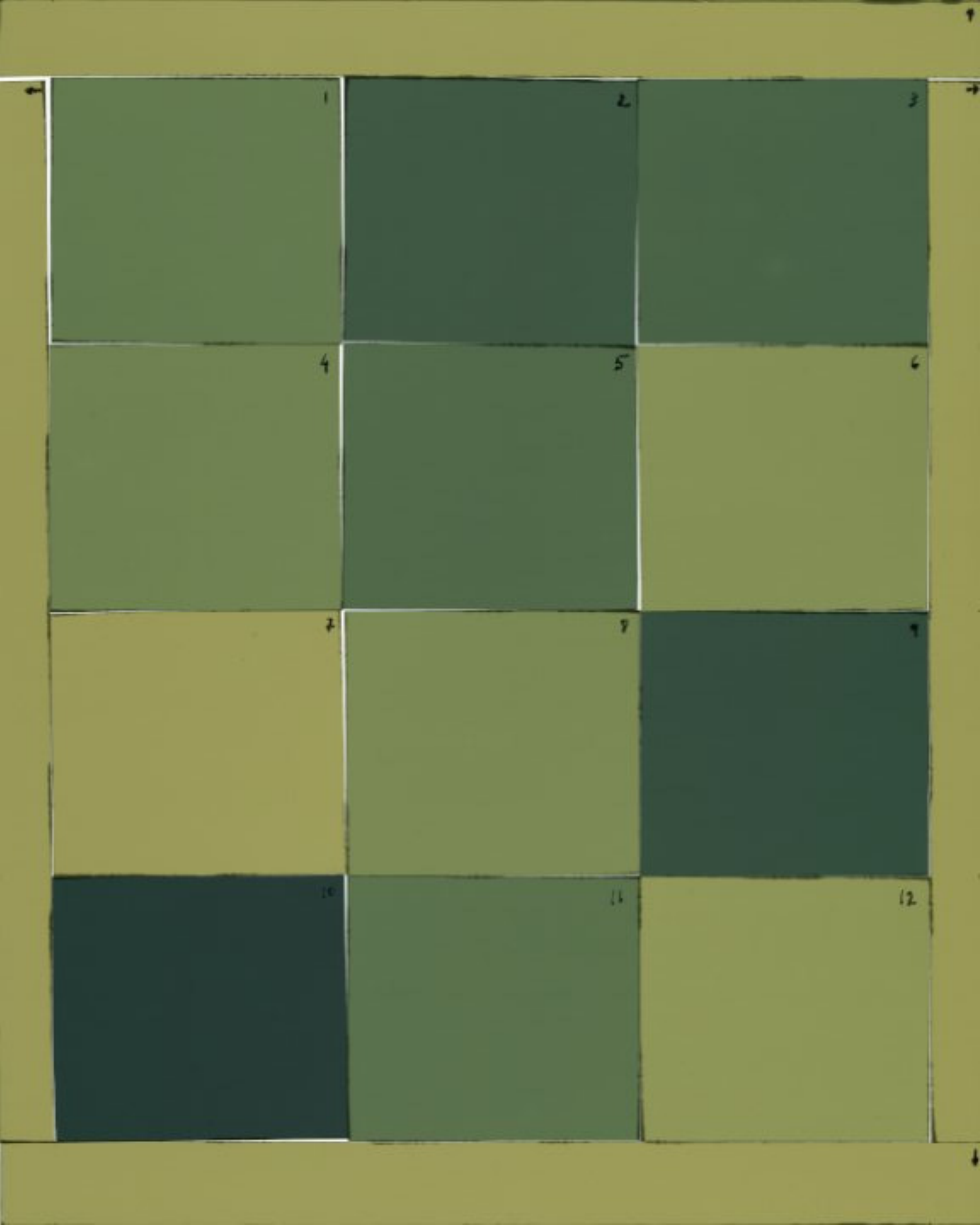}
\caption{\label{fig:intrascan} Spatial inter-scan variability: setup of film fragments.}
\end{figure}

\begin{figure}
\includegraphics[width=0.47\linewidth]{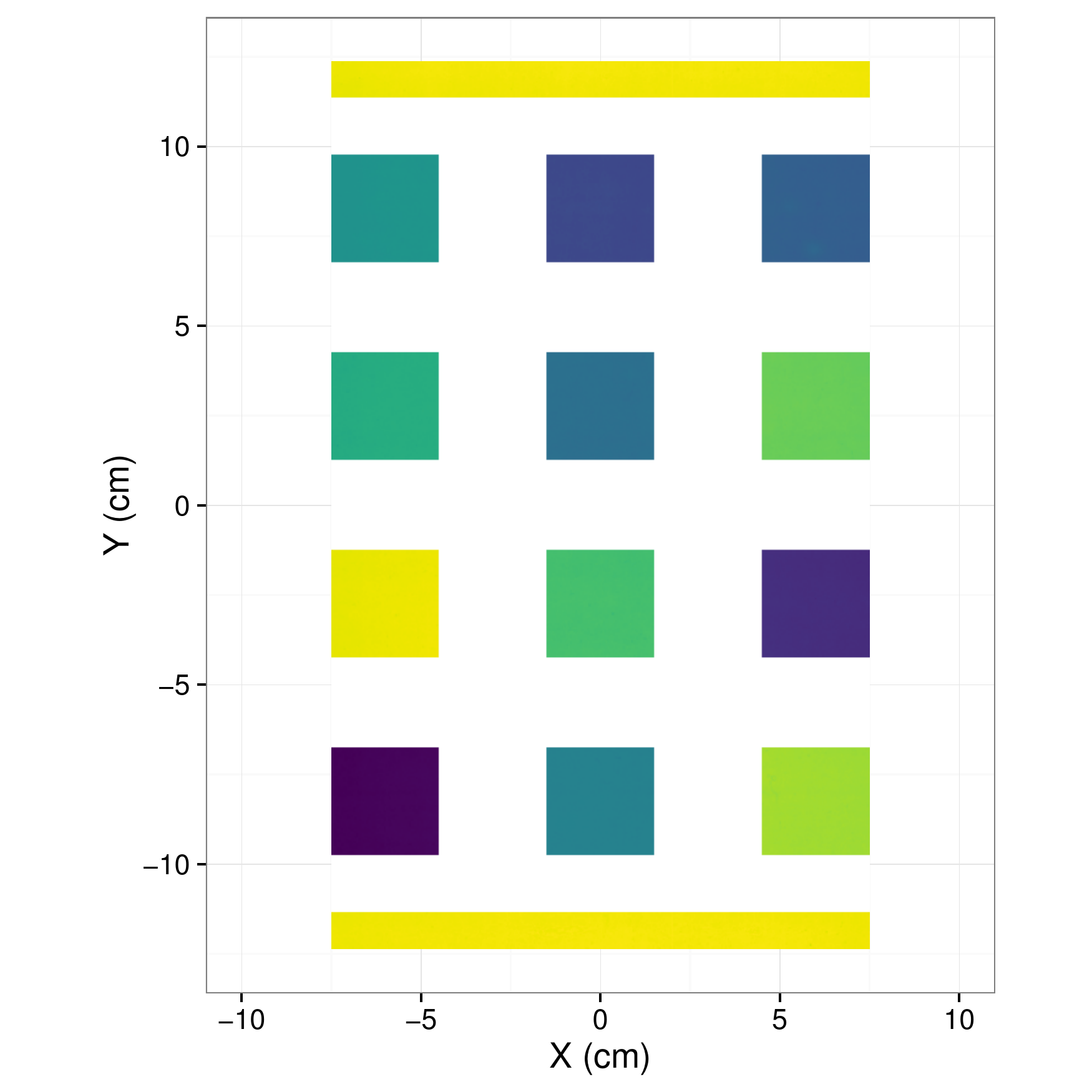}
\caption{\label{fig:intrascan_rois} Spatial inter-scan variability: ROIs analyzed.}
\end{figure}

Inter-scan variations of the scanner response produce discrepancies in the dose-response relationship between the calibration and subsequent scans, which, if not corrected, can cause important errors in film dosimetry\cite{lewis:2015}. One of the objectives of this test was to investigate spatial variations of the inter-scan repeatability. In order to do so, a film was divided in 16 fragments. Twelve of them, the central fragments, were 6.0 ${\rm cm}$ $\times$ 5.5 ${\rm cm}$. The superior and inferior margins were 20.3 ${\rm cm}$ $\times$ 1.7 ${\rm cm}$. The other two fragments, the lateral margins, were 1.2 ${\rm cm}$ $\times$ 22.0 ${\rm cm}$. 

Each central fragment was separately placed at a depth of 7 cm in a 14$\times$30$\times$30 ${\rm cm^3}$ Plastic Water\texttrademark phantom (Computerized Imaging Reference Systems Inc. Norfolk, VA, USA) at 100 ${\rm cm}$ SSD (source-to-surface distance). They were irradiated with a a 10 ${\rm cm}$ $\times$ 10 ${\rm cm}$ field, at 150, 400, 300, 100, 250, 50, 0, 75, 500, 750, 200 and 25 cGy (fragments 1-12). Doses were randomly distributed to prevent misleading patterns arising.
  
The film was reassembled, as shown in Figure~\ref{fig:intrascan}, and scanned with resolutions of 50, 72, 96 and 150 dpi, four months after irradiation. Each resolution was scanned 20 consecutive times. For each resolution and color channel, the mean scan image was calculated. 

Regions of interest (ROIs), with dimensions of 3 ${\rm cm}$ $\times$ 3 ${\rm cm}$ centered on each of the central fragments were selected, while two ROIs were centered on the superior and inferior margins with dimensions of 15 ${\rm cm}$ $\times$ 1 ${\rm cm}$. They are shown in Figure~\ref{fig:intrascan_rois}. Only pixels contained in the ROIs were analyzed to avoid the edges of the fragments. 

\subsubsection{Corrections}

Another objective of this test was to find the most accurate model to correct the inter-scan variability, taking into account spatial differences.

Even though, in clinical dosimetry, the reference dose-response relationship should be the sensitometric curve obtained during the calibration, in analysis of the inter-scan variations we can select any image or combination of images as reference. In this study, the reference image was considered to be the mean scan. Applying the correction to a scan image should reduce the differences between it and the reference. Several corrections were examined and two of them were finally chosen: the mean correction and the column correction. The superior and inferior margin ROIs, which were unexposed, were used as the reference material (Ref ROI) to derive the corrections.

The mean correction was calculated as follows:  

\begin{equation}
\label{inter-scan corr}
M(i, j) = v (i, j)  \left\langle \frac{M(i_{Ref}, j_{Ref})}{v(i_{Ref}, j_{Ref})} \right\rangle
\end{equation}

where $(i, j)$ symbolizes the pixel position in the image ($i$ is the row and $j$ the column), $M$ is the value of the pixel in the mean scan, $v$ is the value in the scan being corrected, and $(i_{Ref}, j_{Ref})$ is a pixel in the Ref ROI. Therefore, the mean correction is the average of the factors applied to each of the pixels in the Ref ROI to obtain the values of the mean image. The mean correction is constant for every pixel of the scan, it is spatially invariant.
   
The column correction can be described as follows:  

\begin{equation}
\label{inter-scan corr}
M(i, j) = v (i, j)  \left\langle \frac{M(i_{Ref}, j)}{v(i_{Ref}, j)} \right\rangle
\end{equation}

Thus, the column correction only averages the factors of the pixels in the Ref ROI which are in the column of the pixel being corrected. In this way, the deviations of the individual charge-coupled device (CCD) detectors are rectified. The column correction depends on the scan and on the position of the pixel in the scan, it is a spatial correction.

\subsection{Scanning reading repeatability}

\begin{figure}
\includegraphics[width=0.47\linewidth]{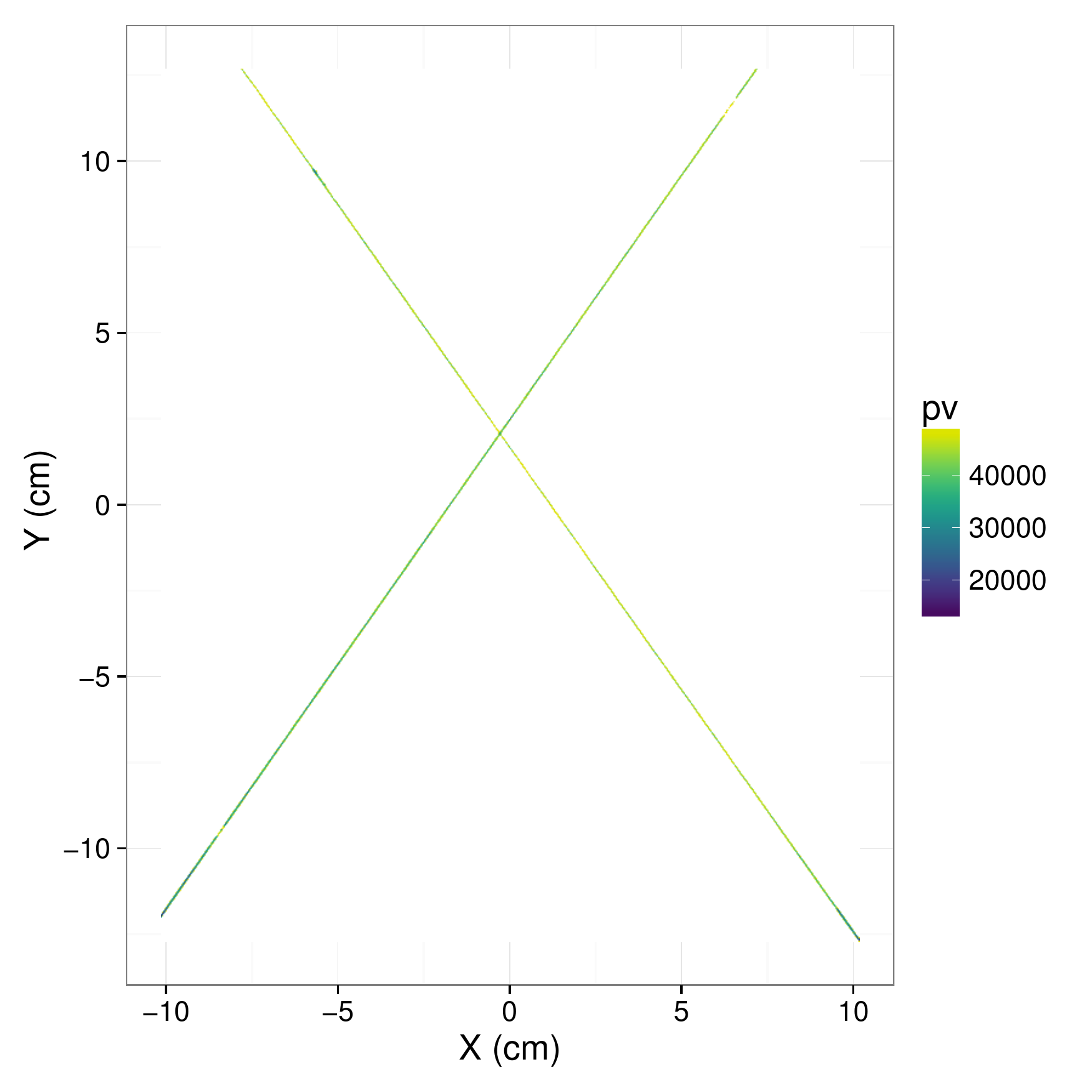}
\caption{\label{fig:speed_x} Scanning reading repeatability
: the cross shape.}
\end{figure}

A cross shape was drawn on a transparency sheet. It was placed on the center of the scanner under the flattening glass. Fifty scans were taken for each resolution, employing resolutions of 50, 72, 96 and 150 dpi. To select only the pixels of the cross shape, pixel values higher than 50000 were removed. All three color channels were combined in a single image. One of the scans can be seen in Figure~\ref{fig:speed_x}. 

The arms of the cross, which can be several pixels thick, were transformed into lines. To do this, the weighted mean column position of the pixels was calculated for each arm separately and each row of the scan. Thus, each row of the scan was associated with the most likely positions of the cross shape, namely, two positions, one for  each arm, with the exception of the point where both arms cross.

Although the inverse of the pixel value was employed as weight to compute the most likely positions, different weights were tested with negligible influence in the results. In an analogous fashion, row positions of the cross shape were associated with the columns of the scan. Additionally, the mean or reference cross for each resolution was computed by combining all the pixels of each scan, and calculating the weighted mean positions of the arms.

For each row and each column of the scan (or, equivalently, for each X and Y position), the distance, in each axis, between the most likely positions of the reference cross shape and of the cross shape of each scan was calculated.   

\section{Results}

\subsection{Grid pattern}

\begin{figure*}
\begin{minipage}[b]{0.24\linewidth}
\centering
\includegraphics[width=\linewidth]{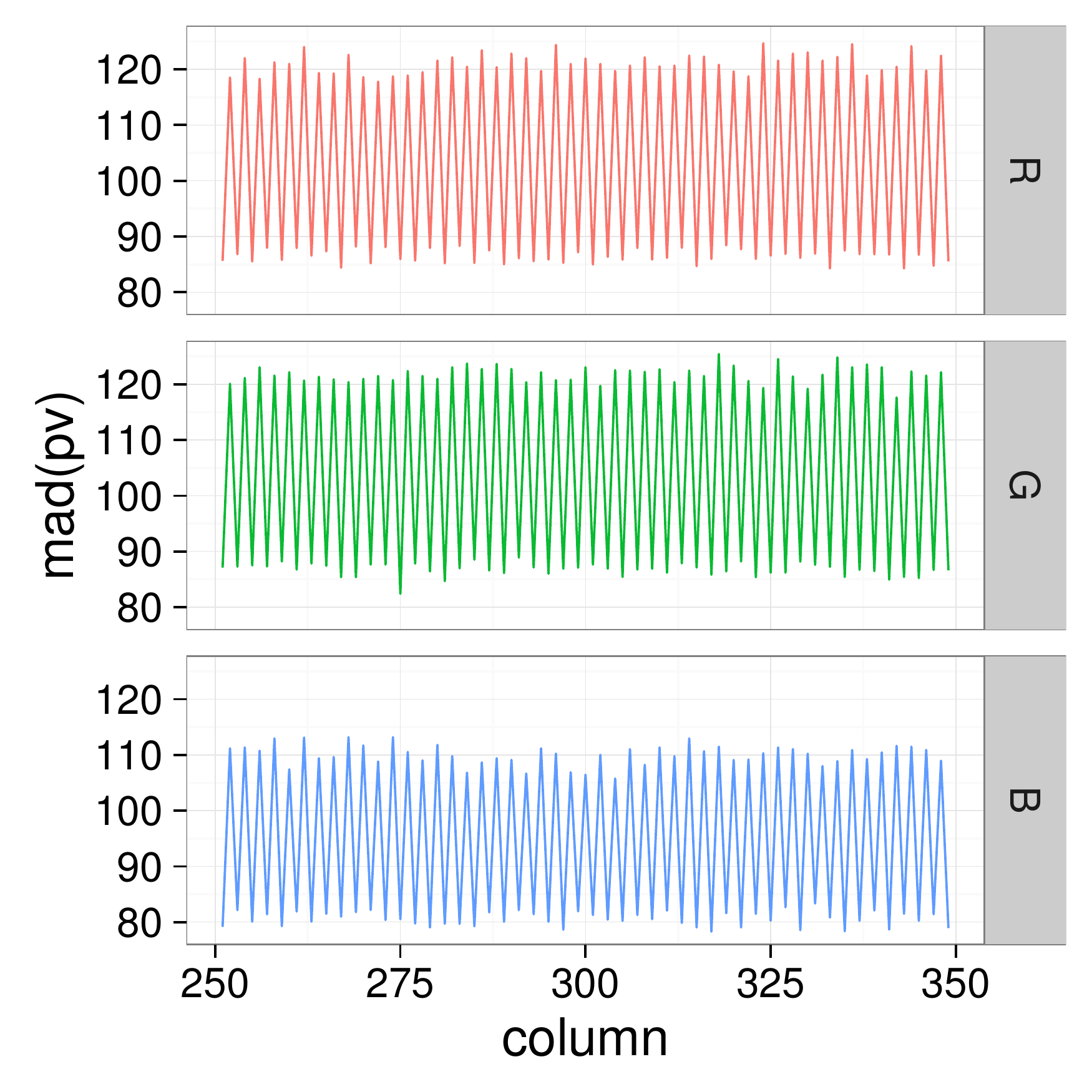}\\
(a1)
\end{minipage}
\hfill
\begin{minipage}[b]{0.24\linewidth}
\centering
\includegraphics[width=\linewidth]{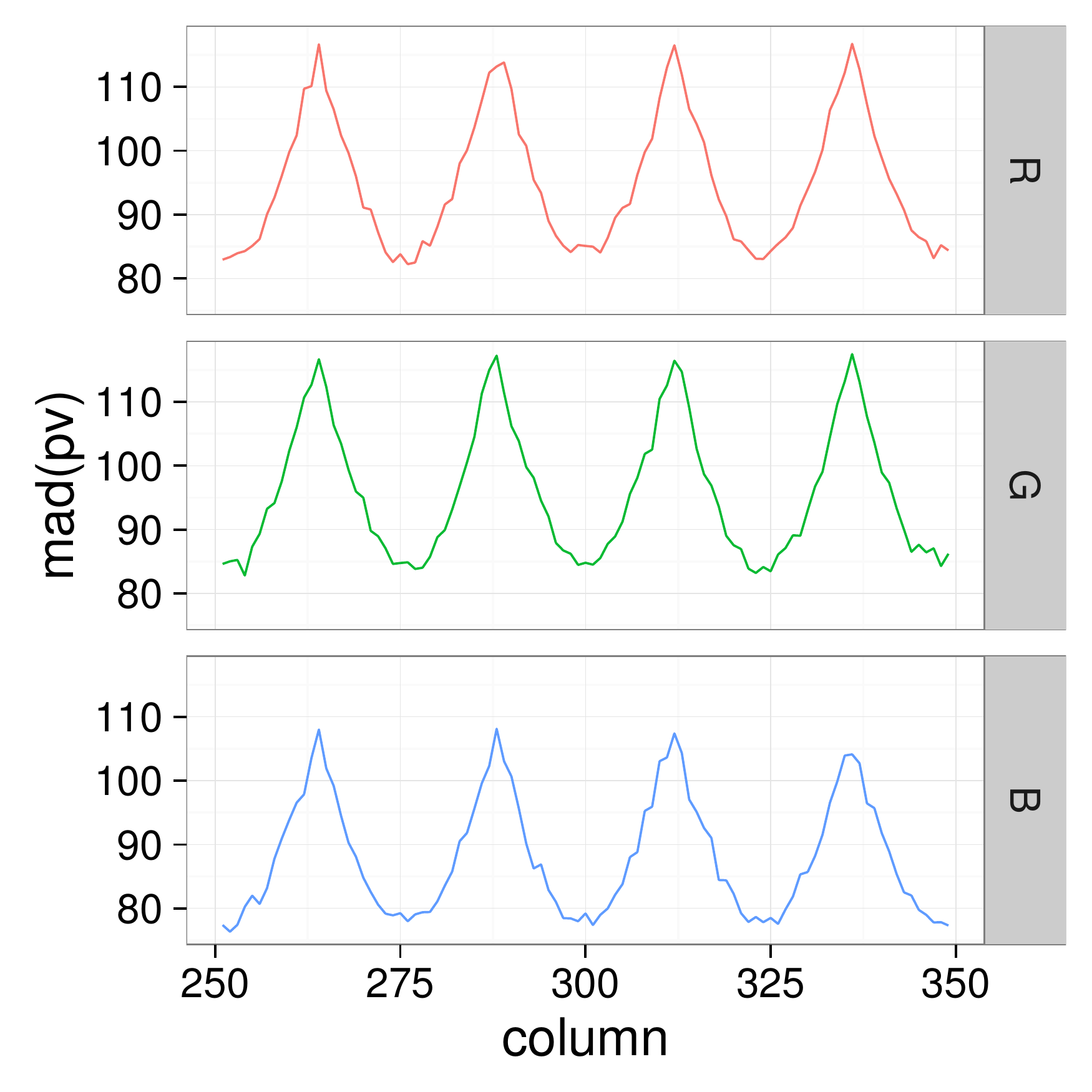}\\
(b1)
\end{minipage}
\hfill
\begin{minipage}[b]{0.24\linewidth}
\centering
\includegraphics[width=\linewidth]{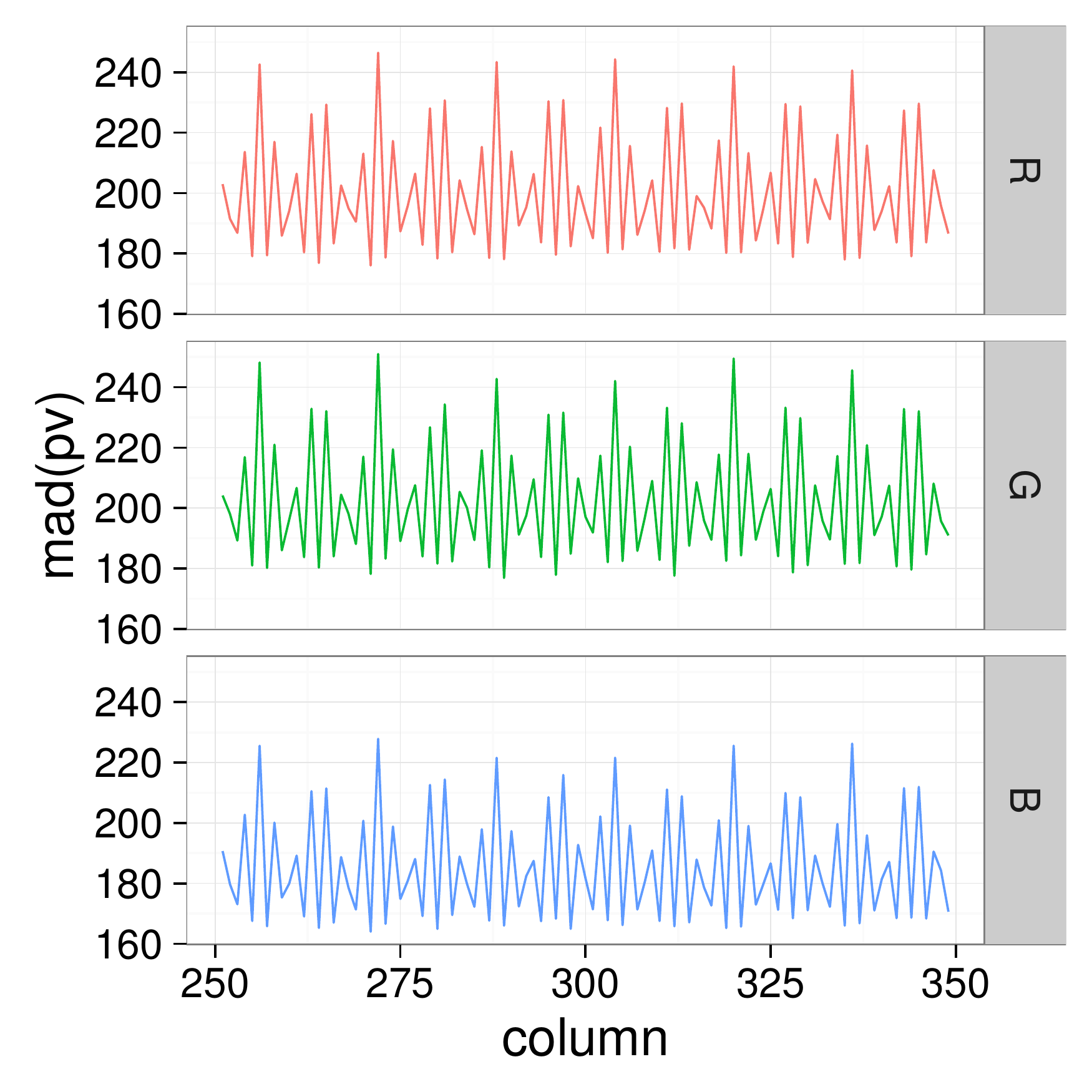}\\
(c1)
\end{minipage}
\hfill
\begin{minipage}[b]{0.24\linewidth}
\centering
\includegraphics[width=\linewidth]{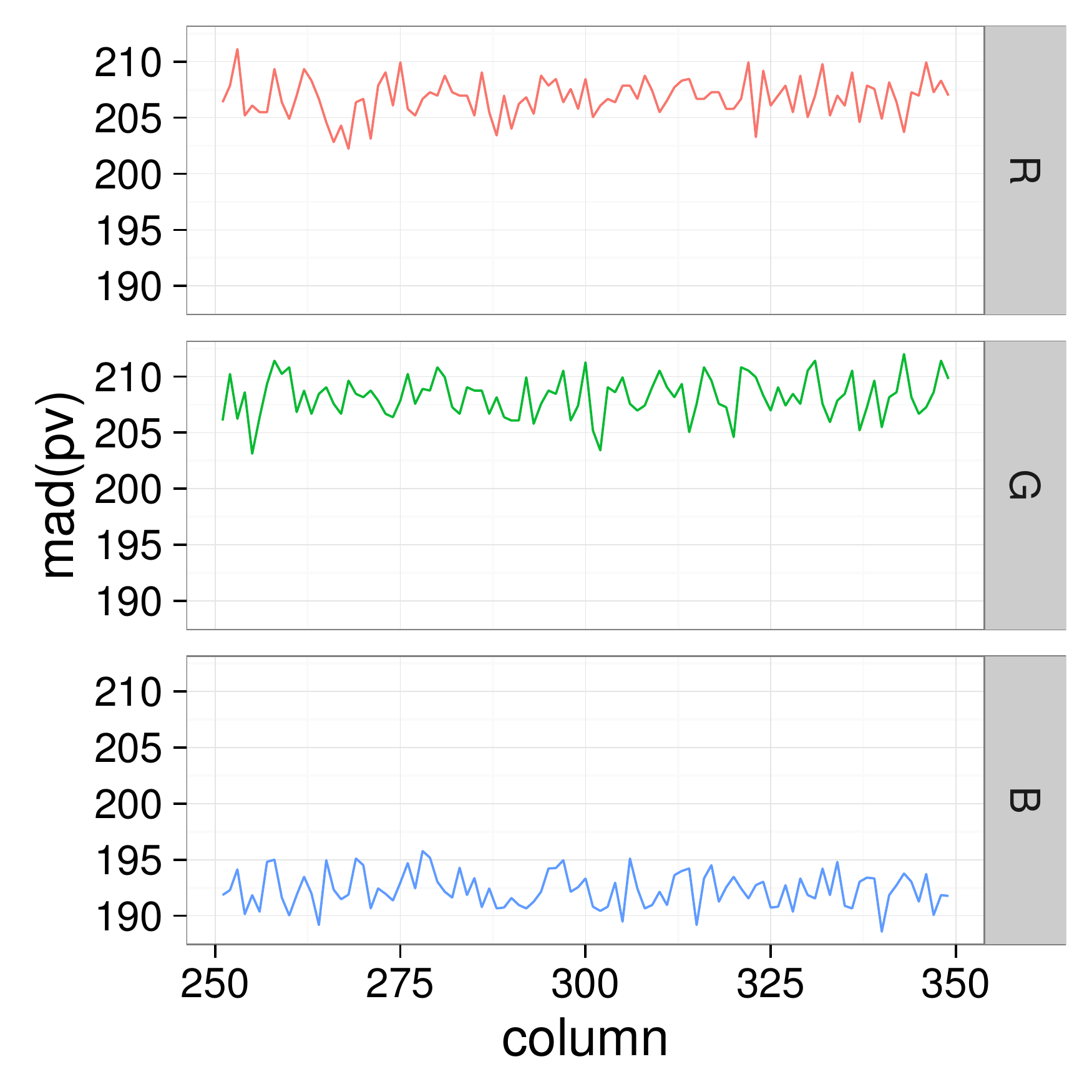}\\
(d1)
\end{minipage}
\begin{minipage}[b]{0.24\linewidth}
\vspace{3ex}
\centering
\includegraphics[width=\linewidth]{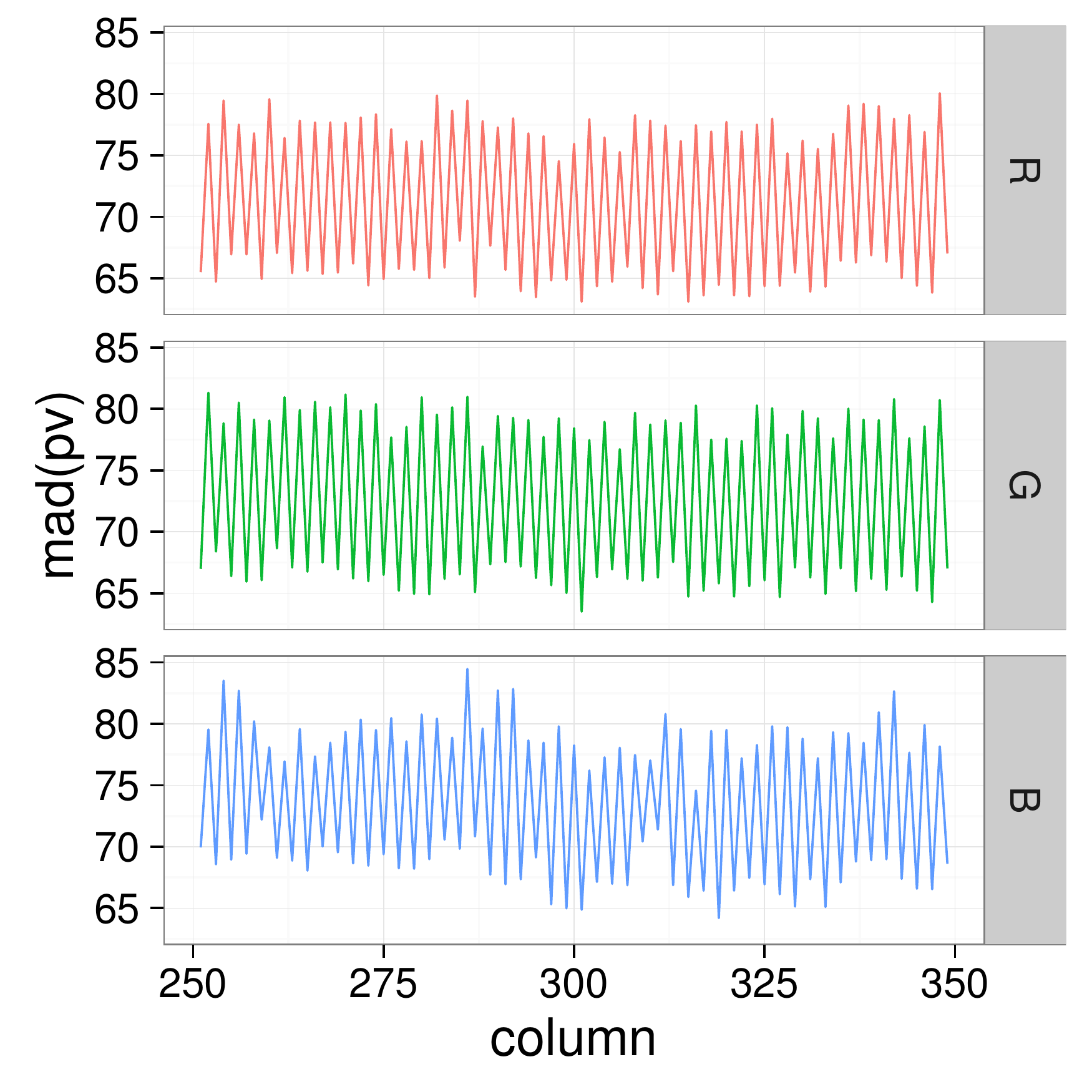}\\
(a2)
\end{minipage}
\hfill
\begin{minipage}[b]{0.24\linewidth}
\centering
\includegraphics[width=\linewidth]{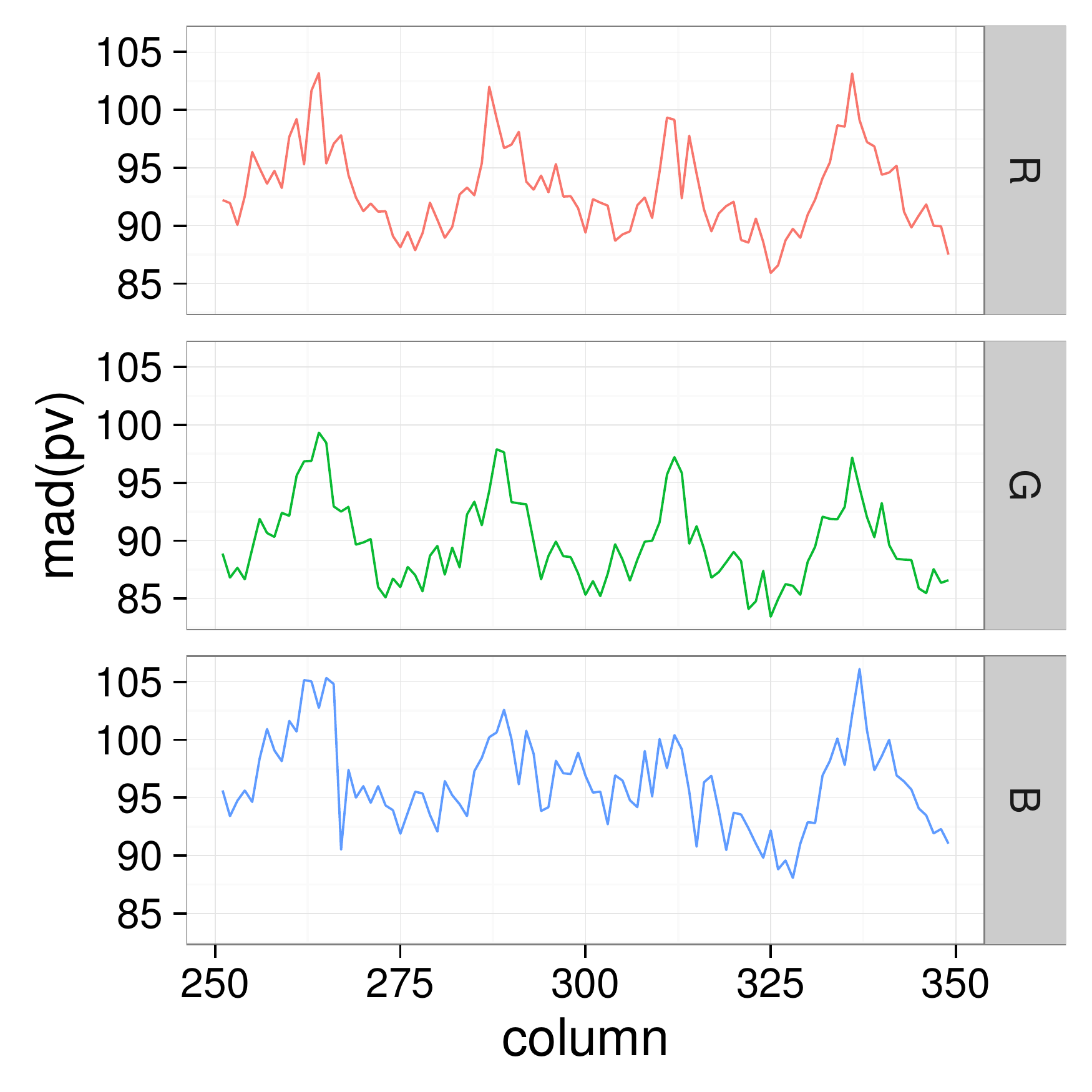}\\
(b2)
\end{minipage}
\hfill
\begin{minipage}[b]{0.24\linewidth}
\centering
\includegraphics[width=\linewidth]{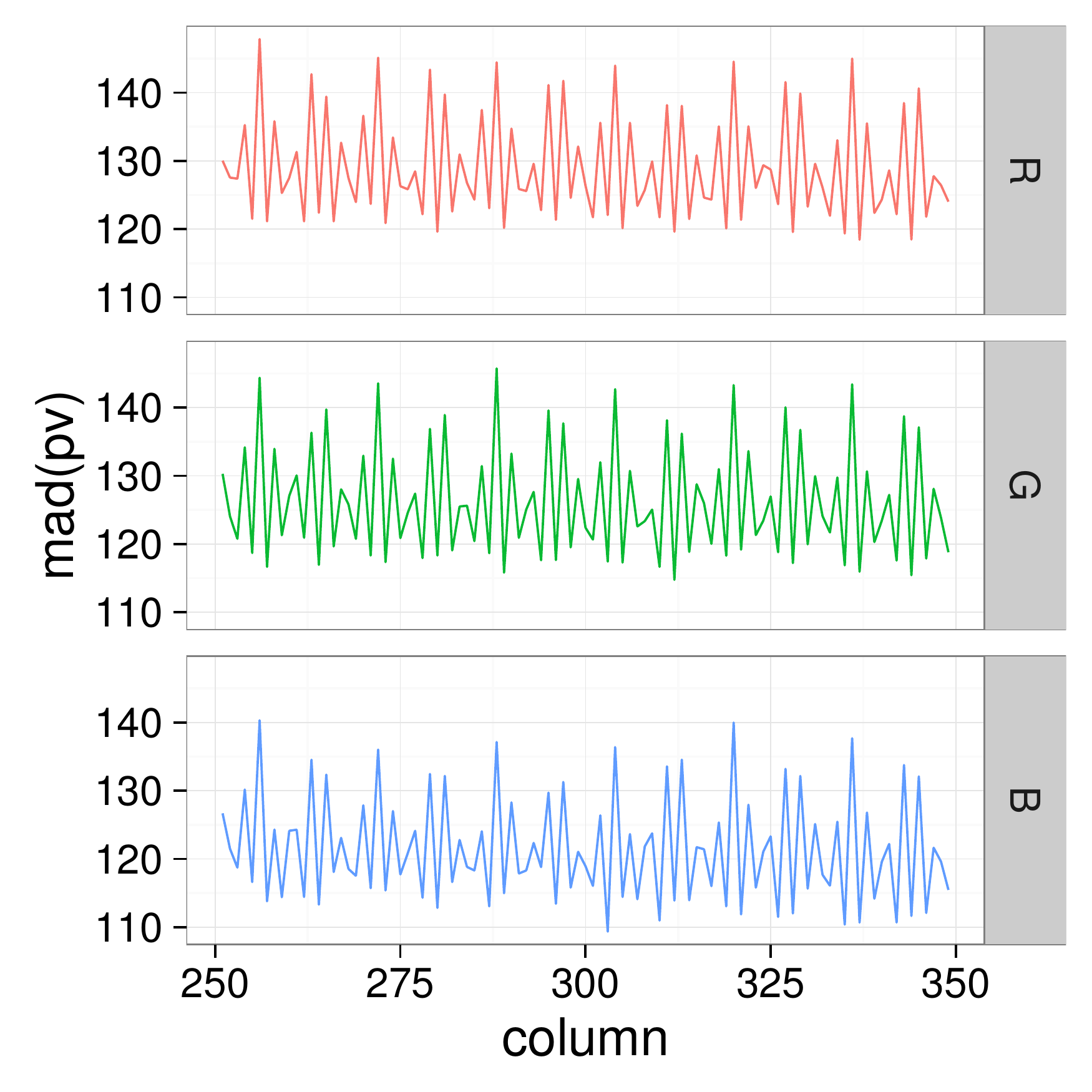}\\
(c2)
\end{minipage}
\hfill
\begin{minipage}[b]{0.24\linewidth}
\centering
\includegraphics[width=\linewidth]{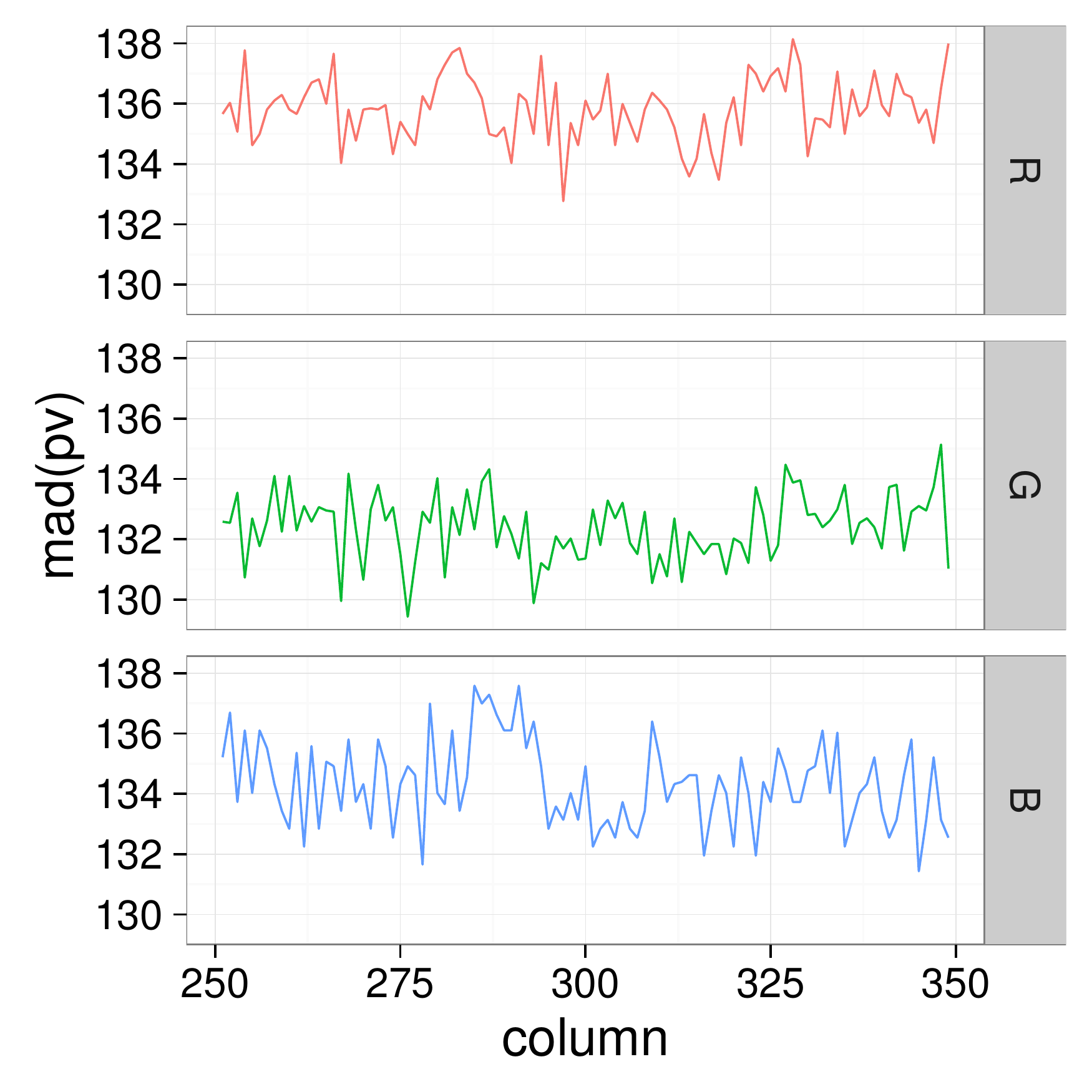}\\
(d2)
\end{minipage}
\begin{minipage}[b]{0.24\linewidth}
\vspace{3ex}
\centering
\includegraphics[width=\linewidth]{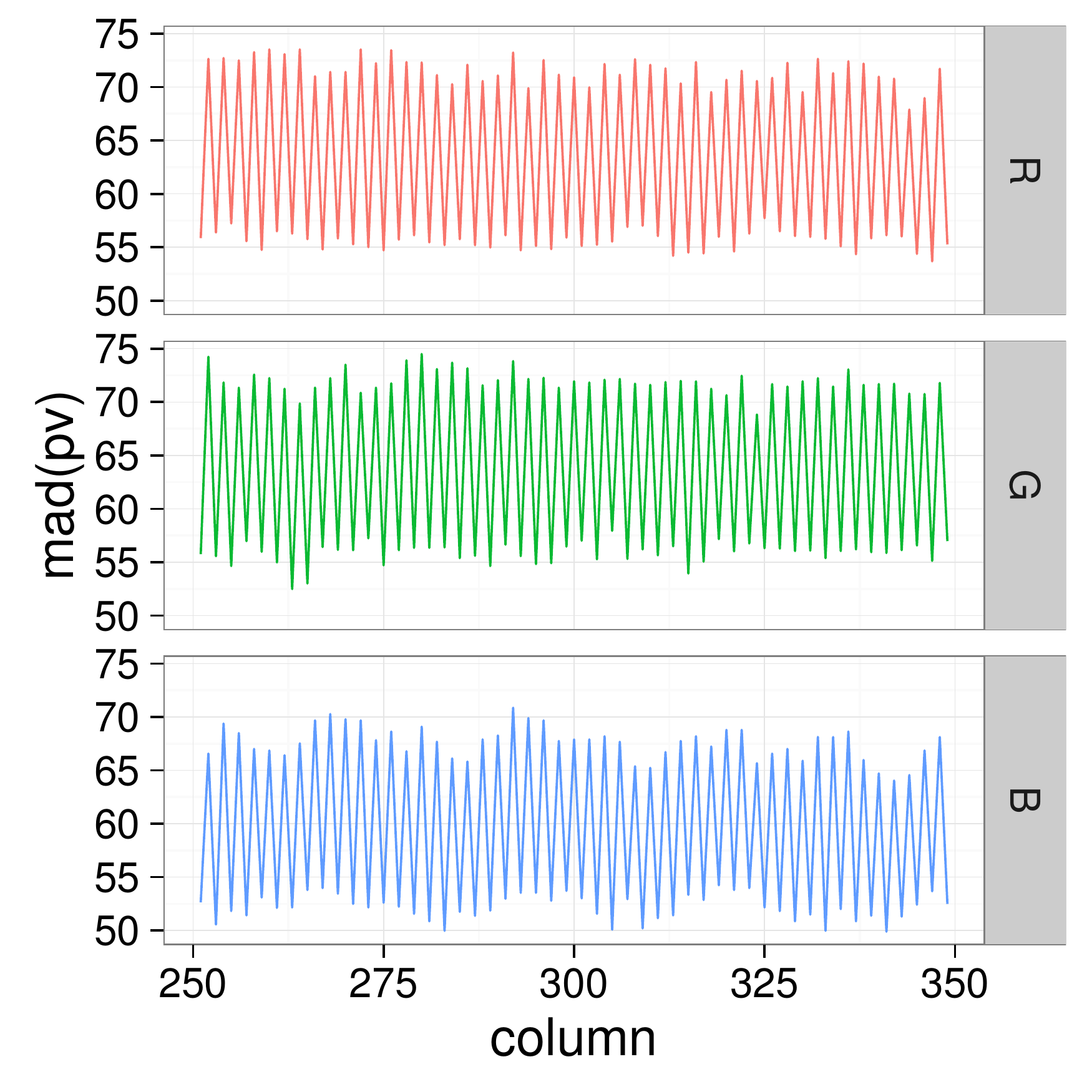}\\
(a3)
\end{minipage}
\hfill
\begin{minipage}[b]{0.24\linewidth}
\centering
\includegraphics[width=\linewidth]{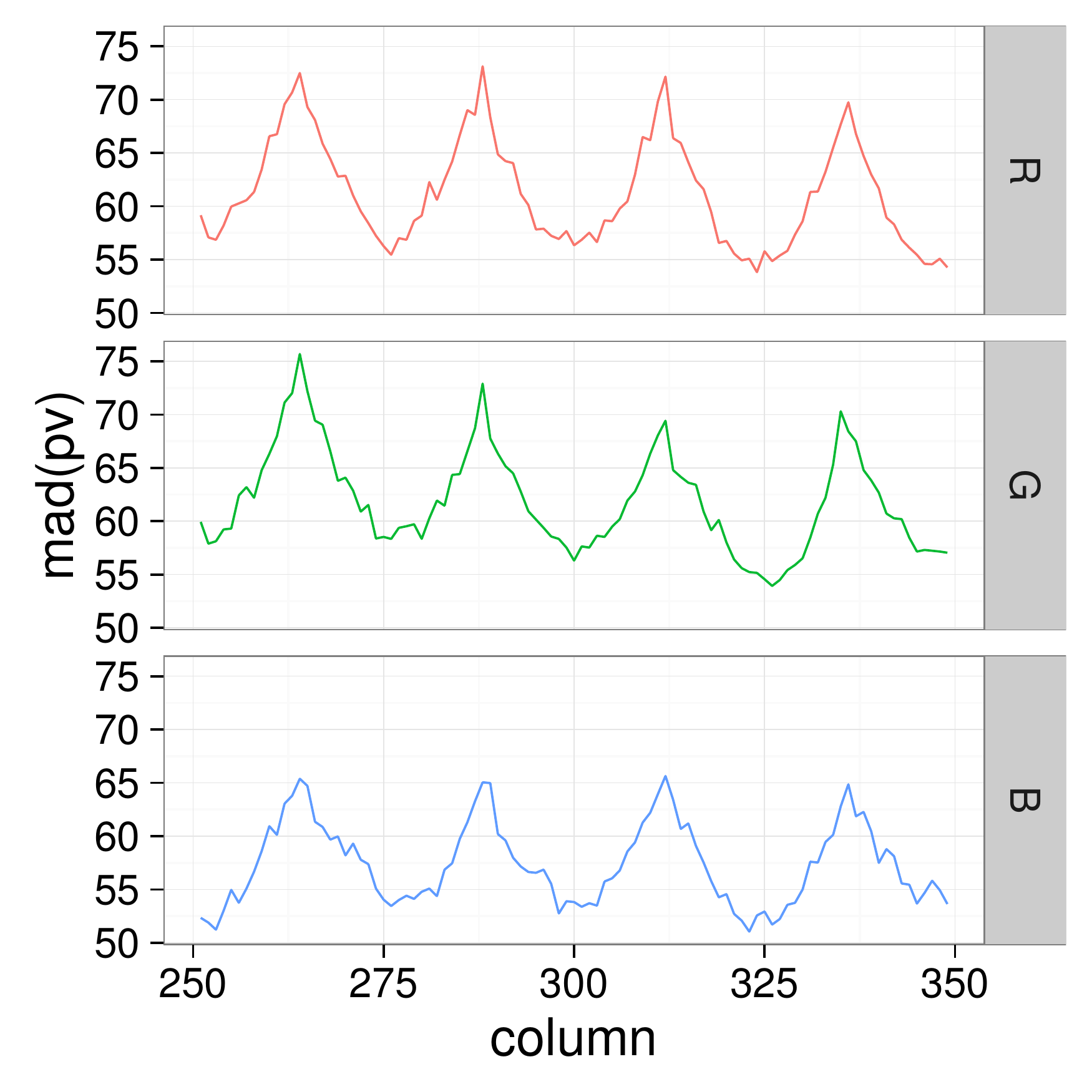}\\
(b3)
\end{minipage}
\hfill
\begin{minipage}[b]{0.24\linewidth}
\centering
\includegraphics[width=\linewidth]{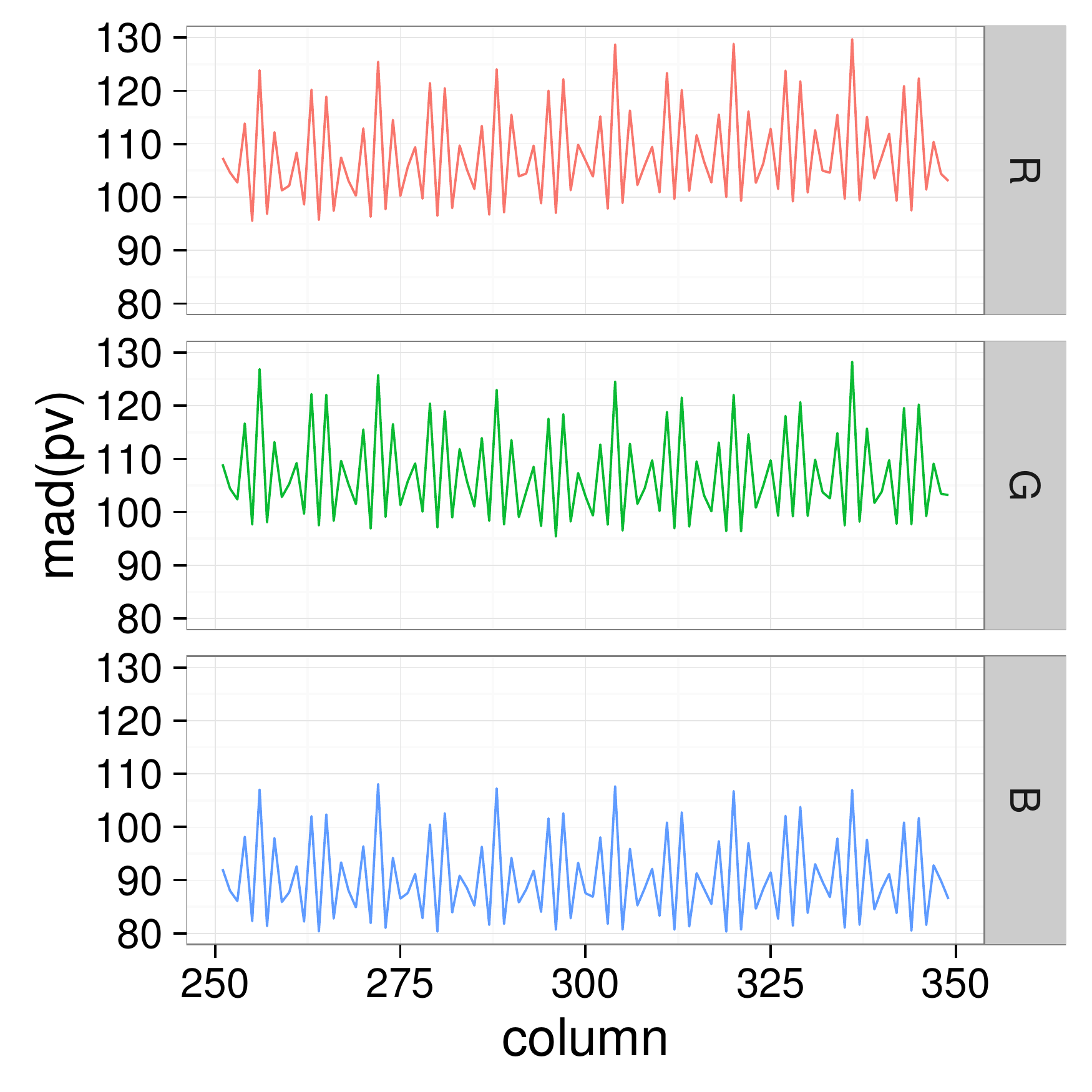}\\
(c3)
\end{minipage}
\hfill
\begin{minipage}[b]{0.24\linewidth}
\centering
\includegraphics[width=\linewidth]{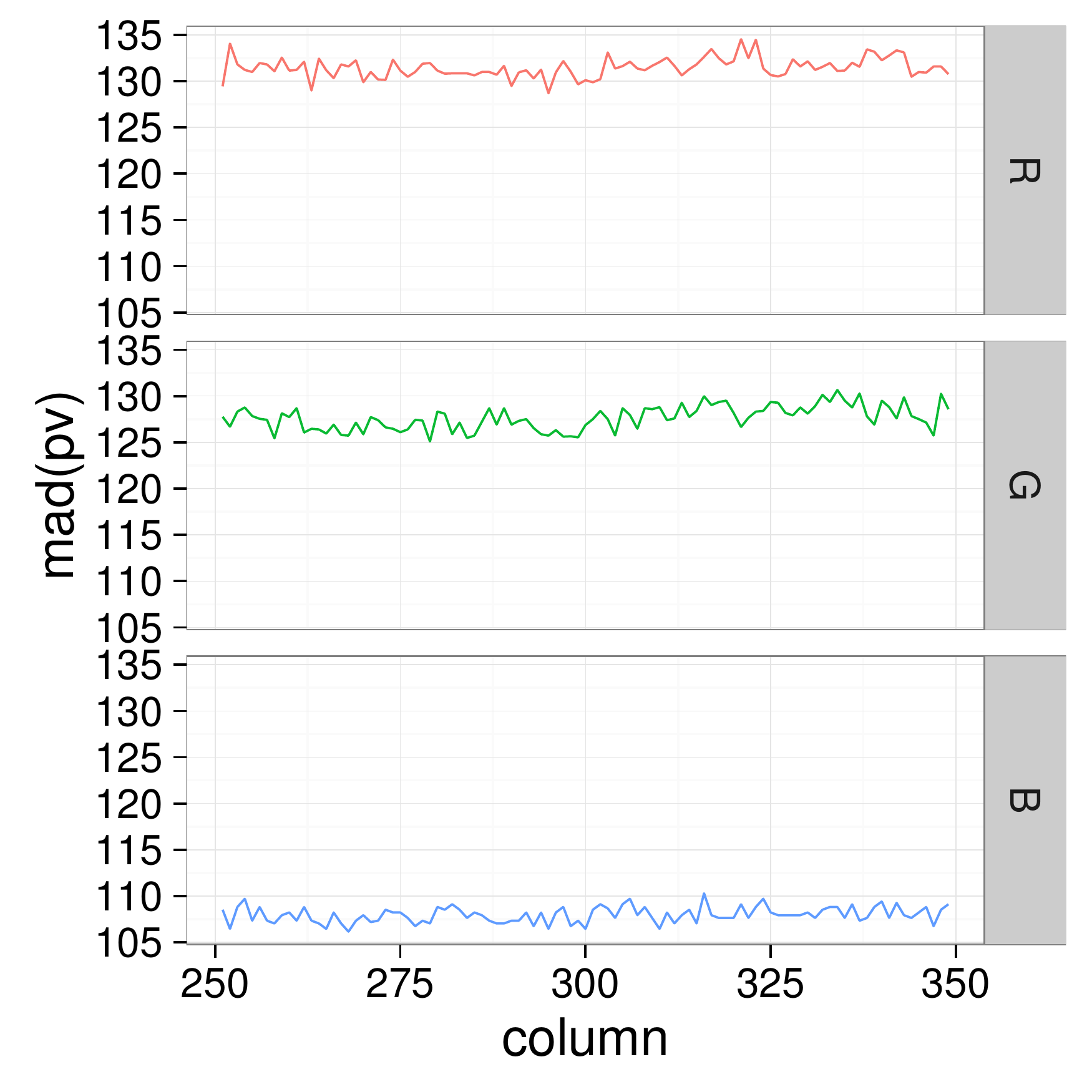}\\
(d3)
\end{minipage}

\caption{\label{fig:res_grid} Mean absolute deviations (MADs) of the differences in pixel value with respect to the mean image as a function of the column for each setup (black background: 1, white background: 2, unexposed film: 3), resolution (50: a, 72: b, 96: c, 150: d) and color channel (red: R, green: G, blue: B).}
\end{figure*}

Figure~\ref{fig:res_grid} plots the MAD of the differences in pixel value with respect to the mean image as a function of the column, resolution and color channel for the unexposed film, white and black backgrounds. For the sake of clarity, only 100 columns are included. Nevertheless, the same patterns with the same periodicity appear in the rows and in the rest of the columns.

To discard that the patterns found in the black background were caused by scattered light, measurements were repeated covering the scanner with opaque plastics, as well as preventing the transmission of light to the detectors with different opaque materials. Similar results were obtained in every case.

\begin{figure*}
\begin{minipage}[b]{0.24\linewidth}
\centering
\includegraphics[width=\linewidth]{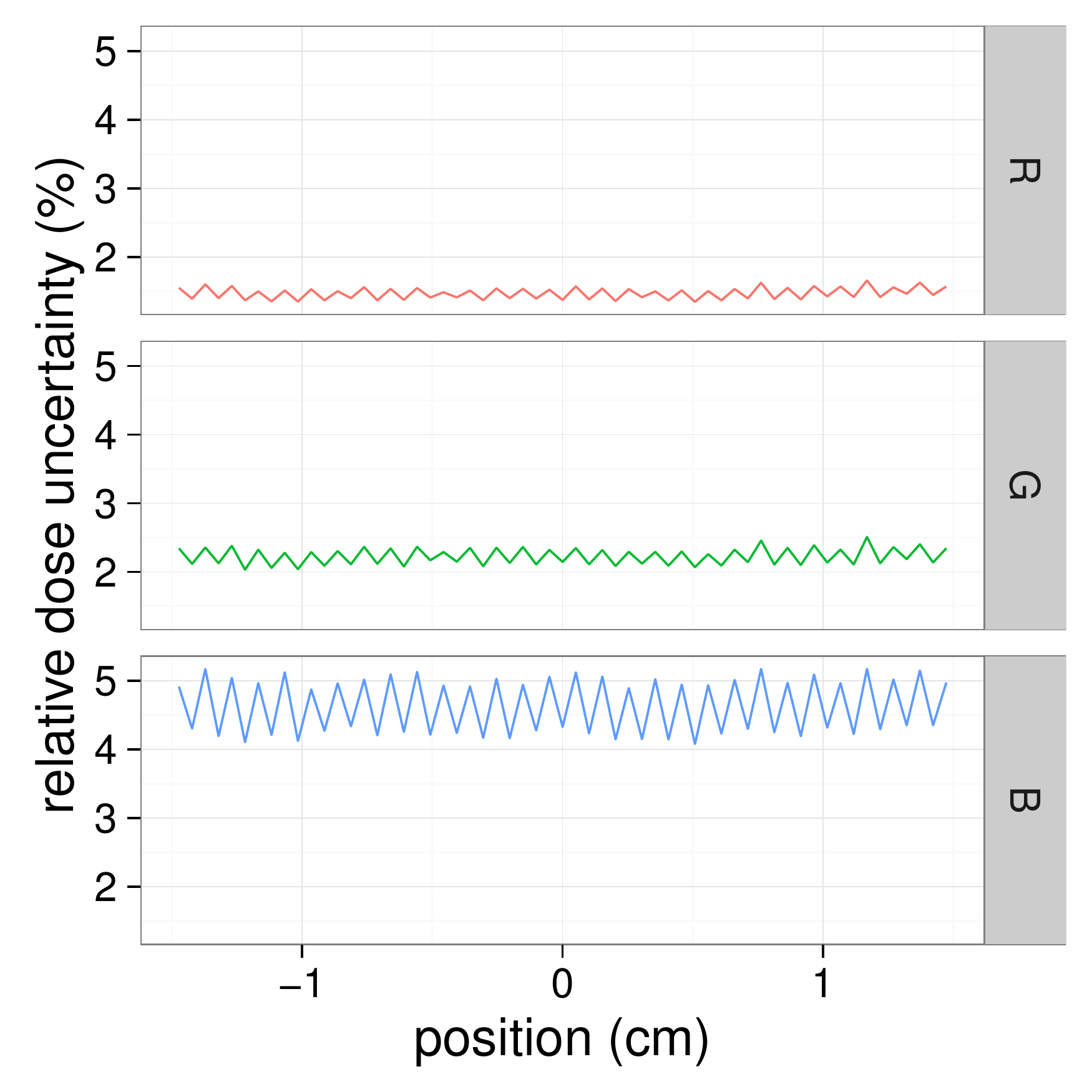}\\
(a1)
\end{minipage}
\hfill
\begin{minipage}[b]{0.24\linewidth}
\centering
\includegraphics[width=\linewidth]{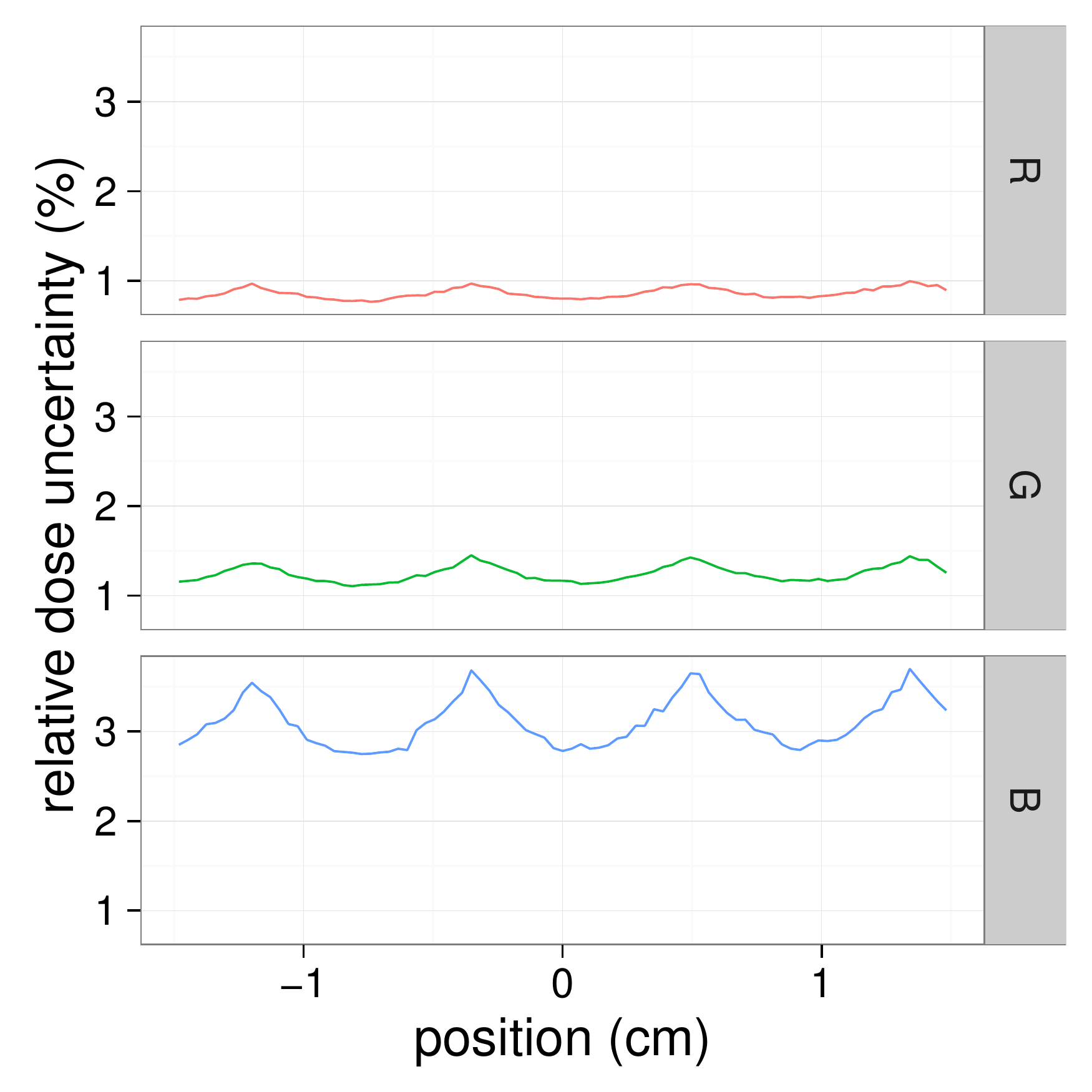}\\
(b1)
\end{minipage}
\hfill
\begin{minipage}[b]{0.24\linewidth}
\centering
\includegraphics[width=\linewidth]{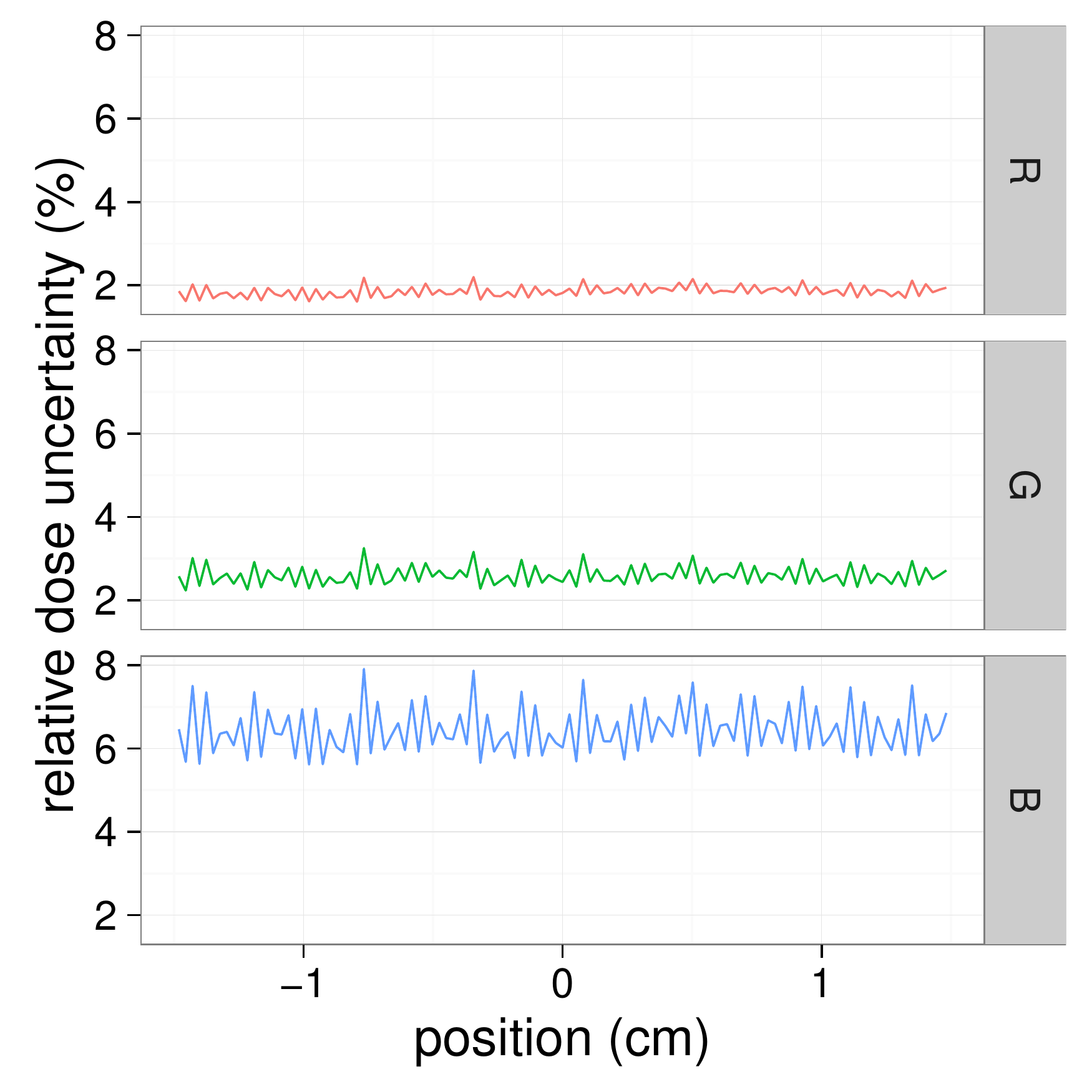}\\
(c1)
\end{minipage}
\hfill
\begin{minipage}[b]{0.24\linewidth}
\centering
\includegraphics[width=\linewidth]{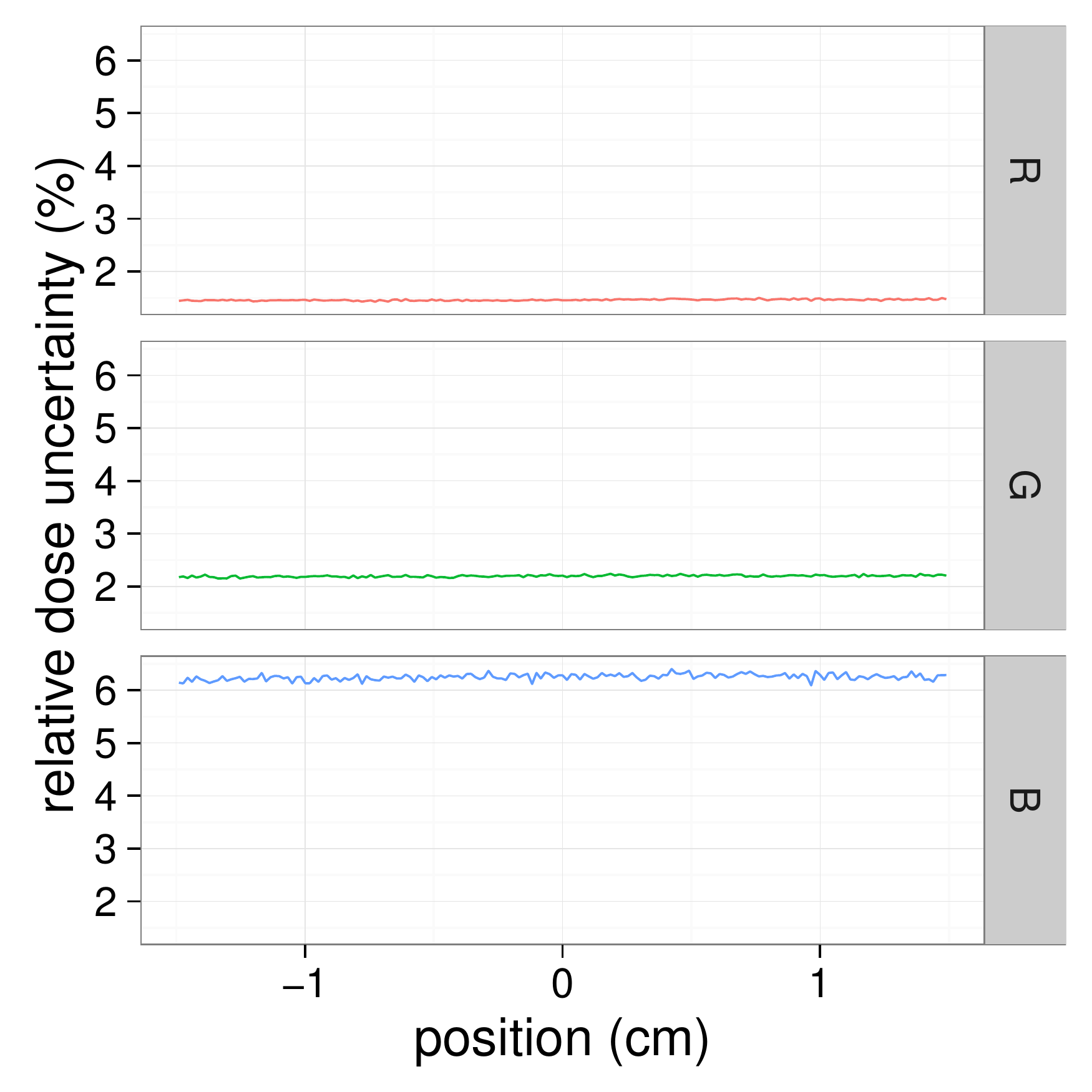}\\
(d1)
\end{minipage}
\begin{minipage}[b]{0.24\linewidth}
\vspace{3ex}
\centering
\includegraphics[width=\linewidth]{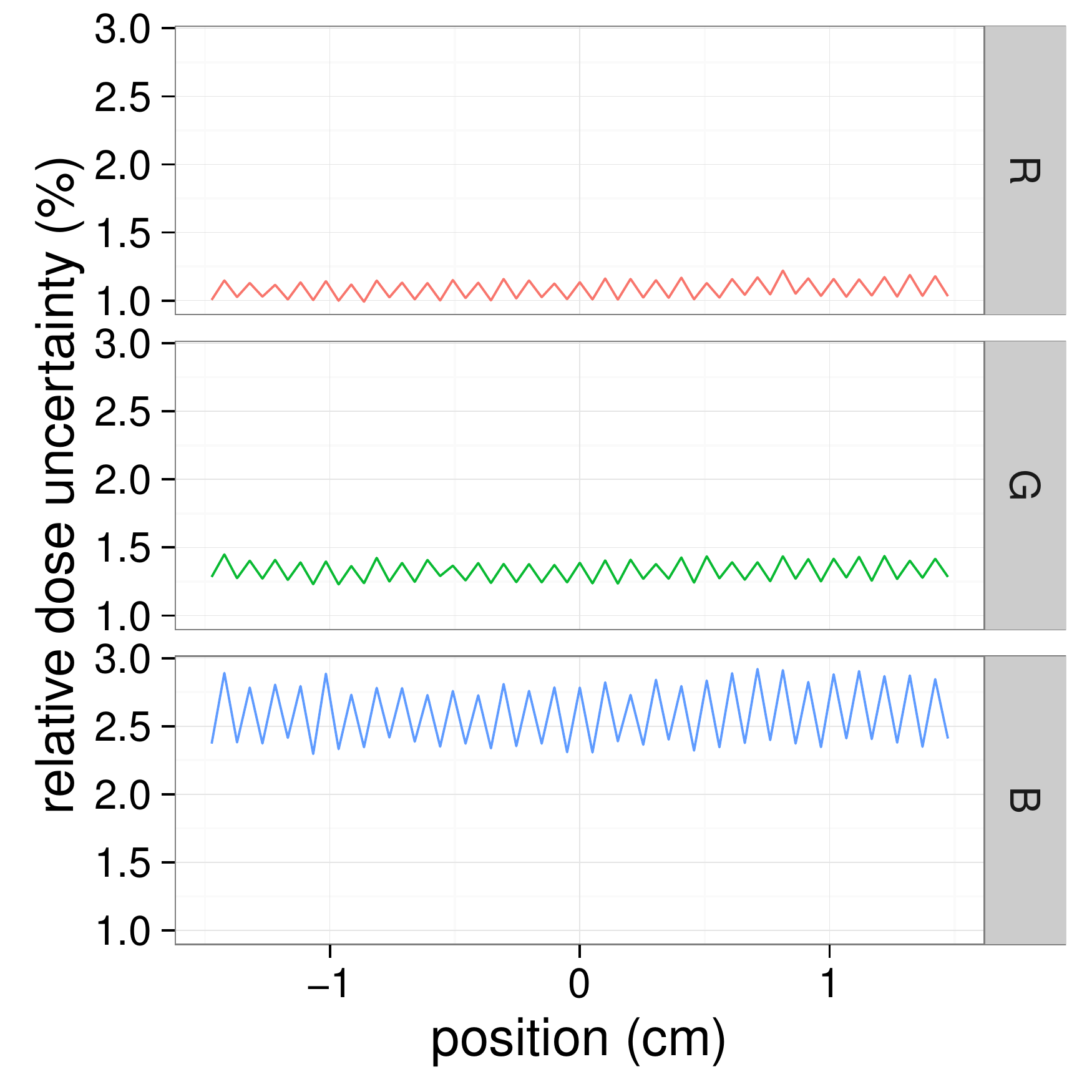}\\
(a2)
\end{minipage}
\hfill
\begin{minipage}[b]{0.24\linewidth}
\centering
\includegraphics[width=\linewidth]{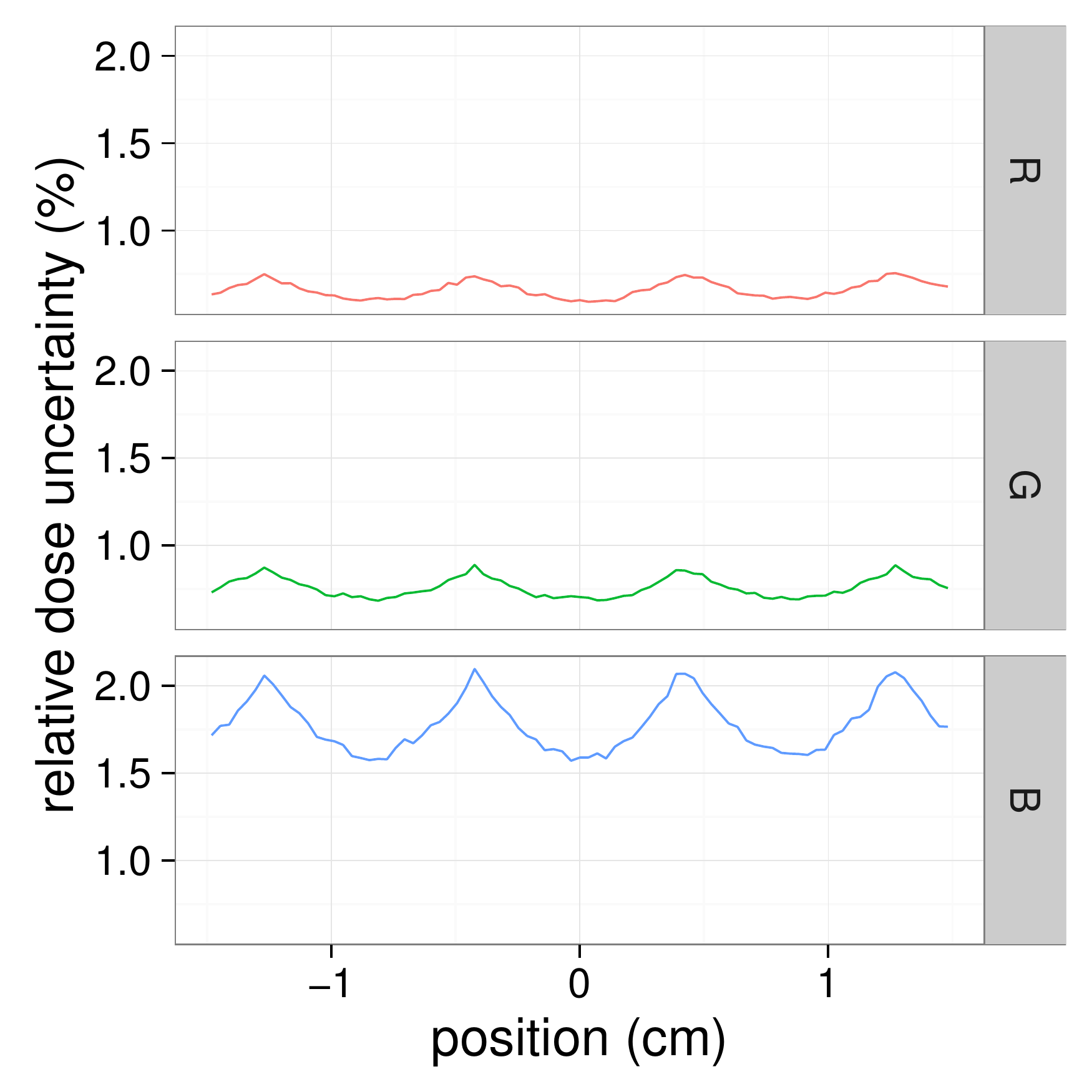}\\
(b2)
\end{minipage}
\hfill
\begin{minipage}[b]{0.24\linewidth}
\centering
\includegraphics[width=\linewidth]{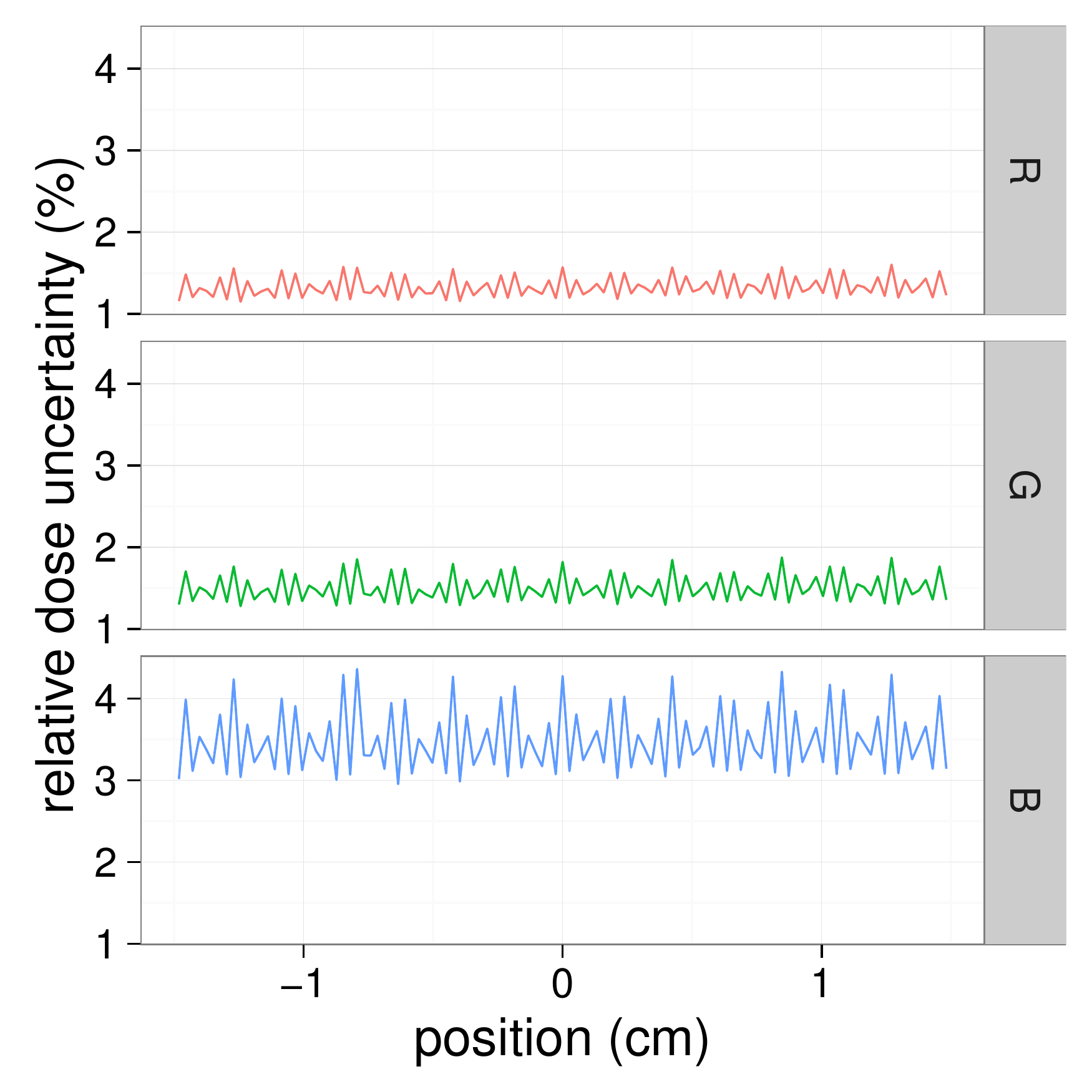}\\
(c2)
\end{minipage}
\hfill
\begin{minipage}[b]{0.24\linewidth}
\centering
\includegraphics[width=\linewidth]{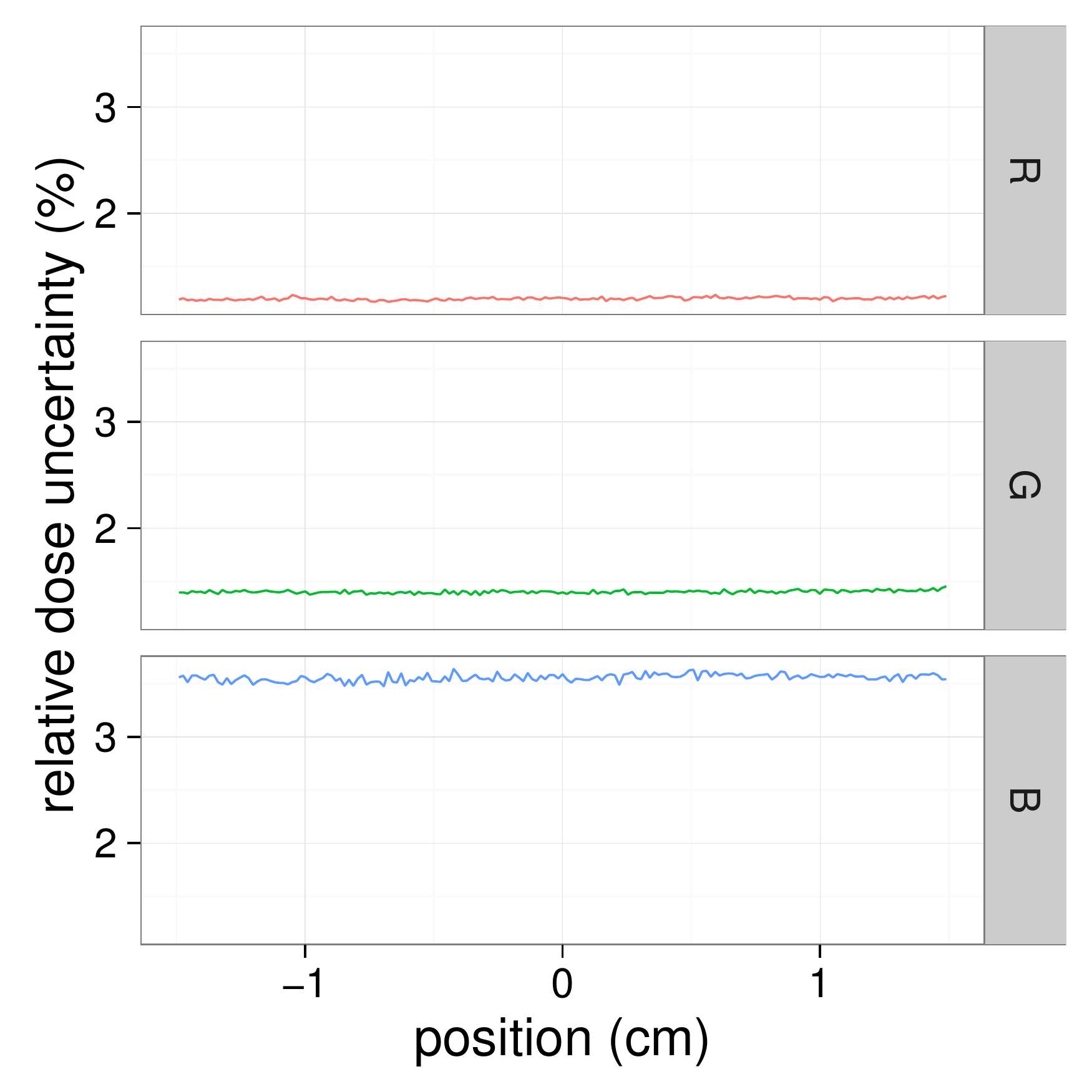}\\
(d2)
\end{minipage}
\begin{minipage}[b]{0.24\linewidth}
\vspace{3ex}
\centering
\includegraphics[width=\linewidth]{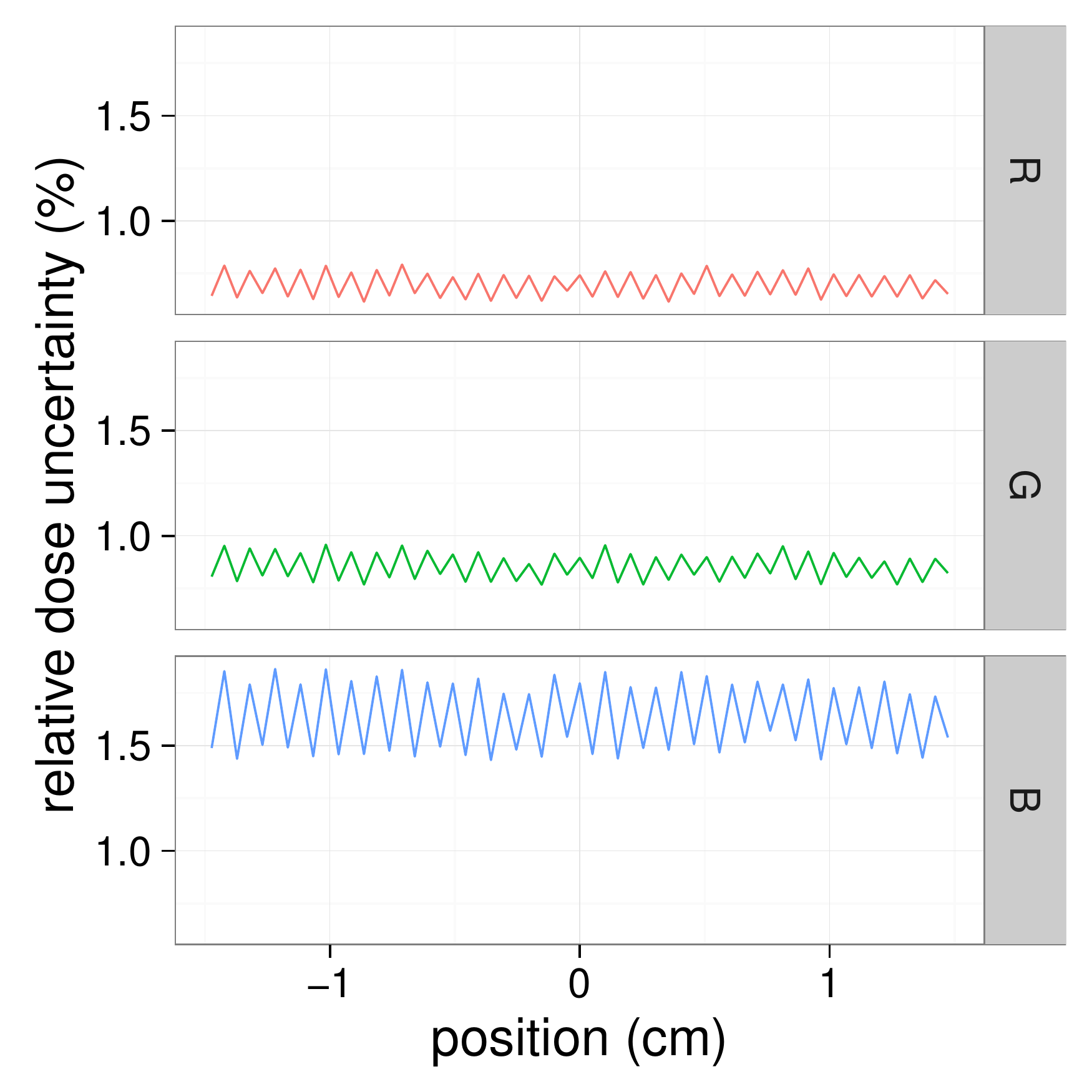}\\
(a4)
\end{minipage}
\hfill
\begin{minipage}[b]{0.24\linewidth}
\centering
\includegraphics[width=\linewidth]{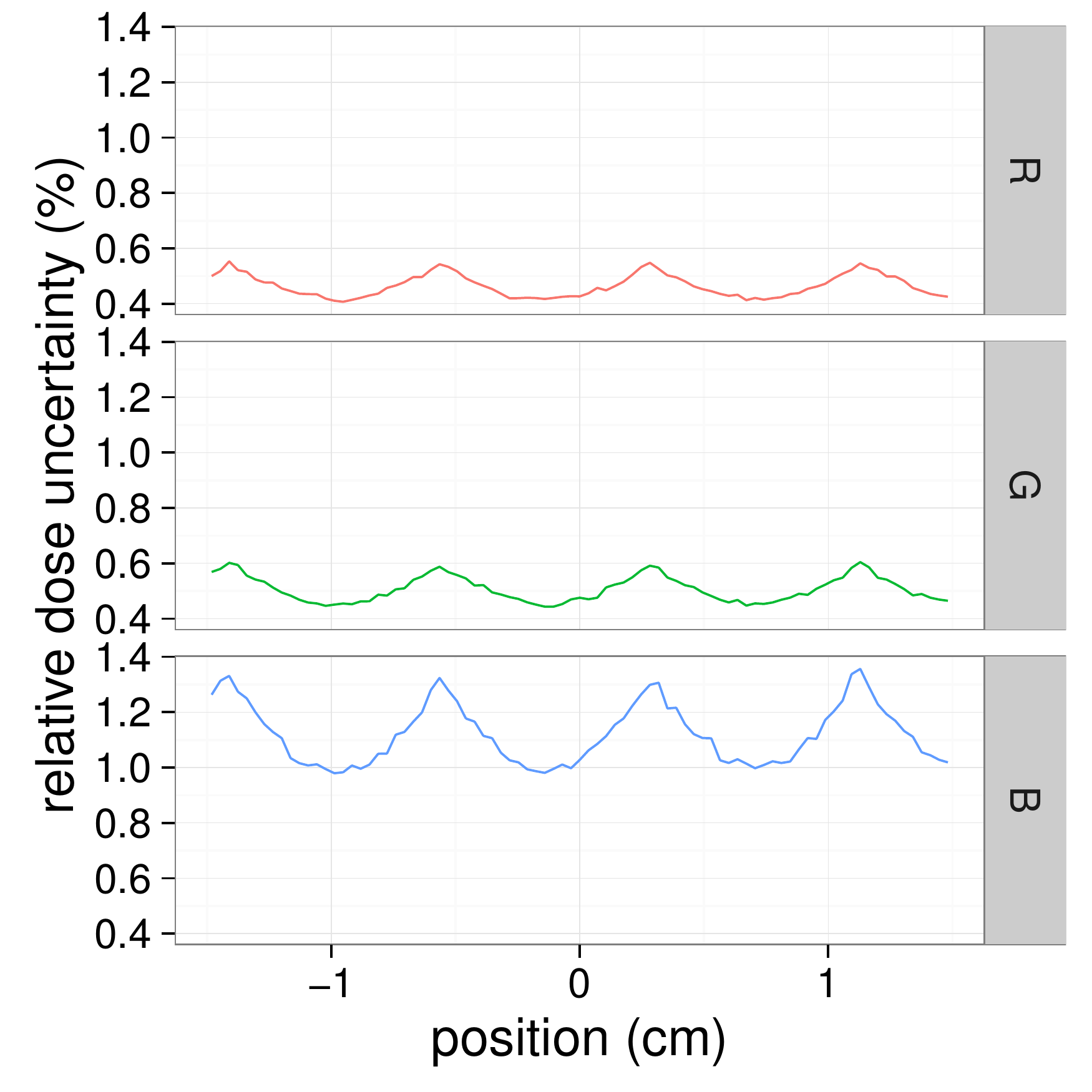}\\
(b4)
\end{minipage}
\hfill
\begin{minipage}[b]{0.24\linewidth}
\centering
\includegraphics[width=\linewidth]{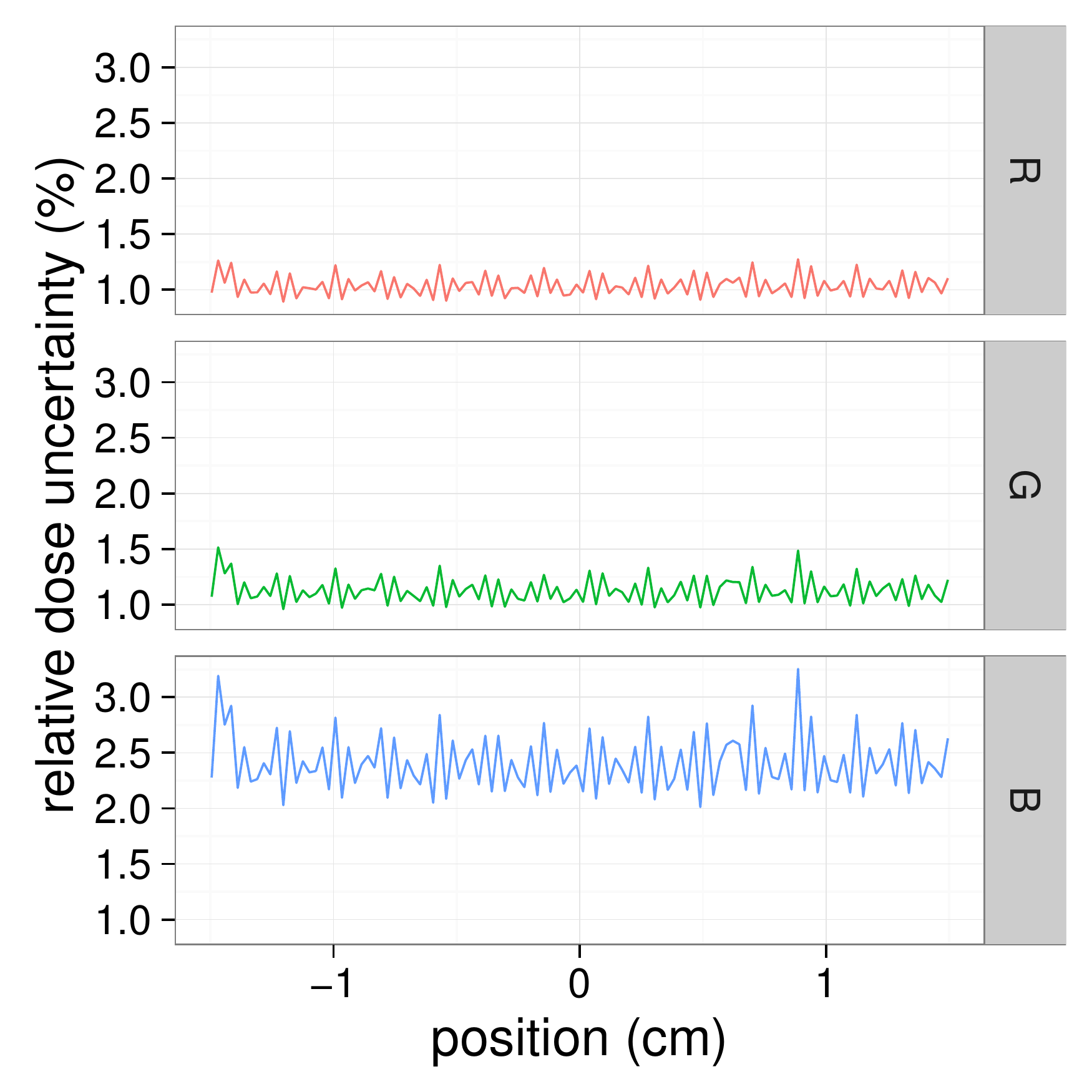}\\
(c4)
\end{minipage}
\hfill
\begin{minipage}[b]{0.24\linewidth}
\centering
\includegraphics[width=\linewidth]{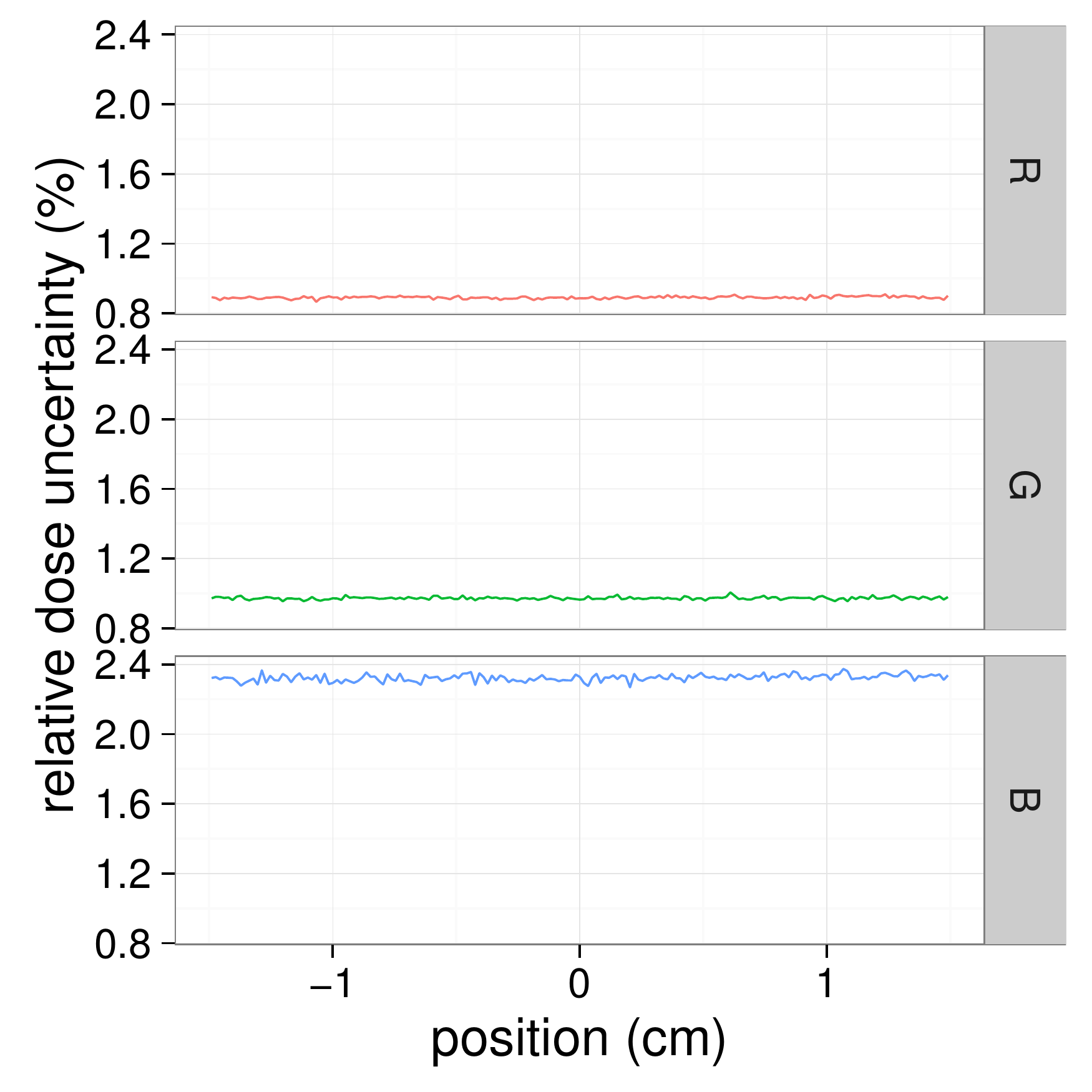}\\
(d4)
\end{minipage}

\caption{\label{fig:res_grid_dose} Relative dose uncertainties as a function of the position for each resolution (50: a, 72: b, 96: c, 150: d), dose (100 cGy: 1, 200 cGy: 2, 400 cGy: 4), and color channel (red: R, green: G, blue: B).}
\end{figure*}

Figure~\ref{fig:res_grid_dose} presents the dosimetric impact of the grid patterns. The dosimetric impact depends on the color channel, the dose, the scanner repeatability and the scanner resolution. For instance: for 100 cGy, 96 dpi and the blue channel, the relative dose uncertainty varied from approximately 6\% to 8\% as a function of the position, while for 400 cGy, 72 dpi and the red channel, it varied from approximately 0.4\% to less than 0.6\%. Even though the dosimetric impact was greater in the blue channel, tipically this channel is not used alone to convert pixel value to dose. Therefore, this impact will be reduced employing multichannel film dosimetry.

\subsection{Spatial inter-scan variability}

\begin{table}
\caption{\label{tab:rel_diffs} Standard deviations of the relative pixel value differences (\%) between the mean scan image and the scans with and without corrections. For ease of analysis, the uncertainties are not included; they were equal to or lower than $2.10^{-4} \%$.}
\begin{ruledtabular}
\begin{tabular}{lcccccccccccc}
Resolution (dpi) & \multicolumn{3}{c}{50} & \multicolumn{3}{c}{72} & \multicolumn{3}{c}{96} & \multicolumn{3}{c}{150}\\
\cline{2-4} \cline{5-7} \cline{8-10} \cline{11-13}
Color channel & R & G & B & R & G & B & R & G & B & R & G & B  \\
\hline 
No correction     & 0.237 & 0.201 & 0.263 & 0.256 & 0.215 & 0.231 & 0.406 & 0.339 & 0.406 & 0.441 & 0.373 & 0.444 \\
Mean correction   & 0.226 & 0.192 & 0.228 & 0.220 & 0.188 & 0.218 & 0.397 & 0.333 & 0.398 & 0.425 & 0.359 & 0.425 \\
Column correction & 0.218 & 0.183 & 0.221 & 0.213 & 0.181 & 0.213 & 0.390 & 0.324 & 0.392 & 0.414 & 0.347 & 0.415 \\
\end{tabular}
\end{ruledtabular}
\end{table}

\begin{table}
\caption{\label{tab:rel_dose_diffs} Standard deviations of the relative dose differences (\%) between the mean scan image and the scans with and without corrections. For ease of analysis, the uncertainties are not included; they were lower than $1.10^{-3} \%$.}
\begin{ruledtabular}
\begin{tabular}{lcccccccccccc}
Resolution (dpi) & \multicolumn{3}{c}{50} & \multicolumn{3}{c}{72} & \multicolumn{3}{c}{96} & \multicolumn{3}{c}{150}\\
\cline{2-4} \cline{5-7} \cline{8-10} \cline{11-13}
Color channel & R & G & B & R & G & B & R & G & B & R & G & B  \\
\hline 
No correction     & 1.0 & 1.9 & 7.8 & 1.1 & 1.9 & 6.9 & 1.6 & 2.8 & 10.5 & 1.8 & 3.1 & 10.8 \\
Mean correction   & 0.9 & 1.6 & 6.2 & 0.9 & 1.5 & 6.2 & 1.6 & 2.7 & 10.2 & 1.7 & 3.0 & 10.2 \\
Column correction & 0.8 & 1.5 & 6.0 & 0.8 & 1.4 & 6.0 & 1.5 & 2.6 & 10.0 & 1.6 & 2.8 & 10.0 \\
\end{tabular}
\end{ruledtabular}
\end{table}

Table~\ref{tab:rel_diffs} and Table~\ref{tab:rel_dose_diffs} contain the standard deviations of the relative differences between the mean scan image and the scan images with and without corrections, for each resolution and each color channel. Table~\ref{tab:rel_diffs} displays pixel value differences and Table~\ref{tab:rel_dose_diffs} dose differences. No image darkening or trend in the inter-scan variability was noticed. 

\begin{figure*}
\begin{minipage}[b]{0.47\linewidth}
\centering
\includegraphics[width=\linewidth]{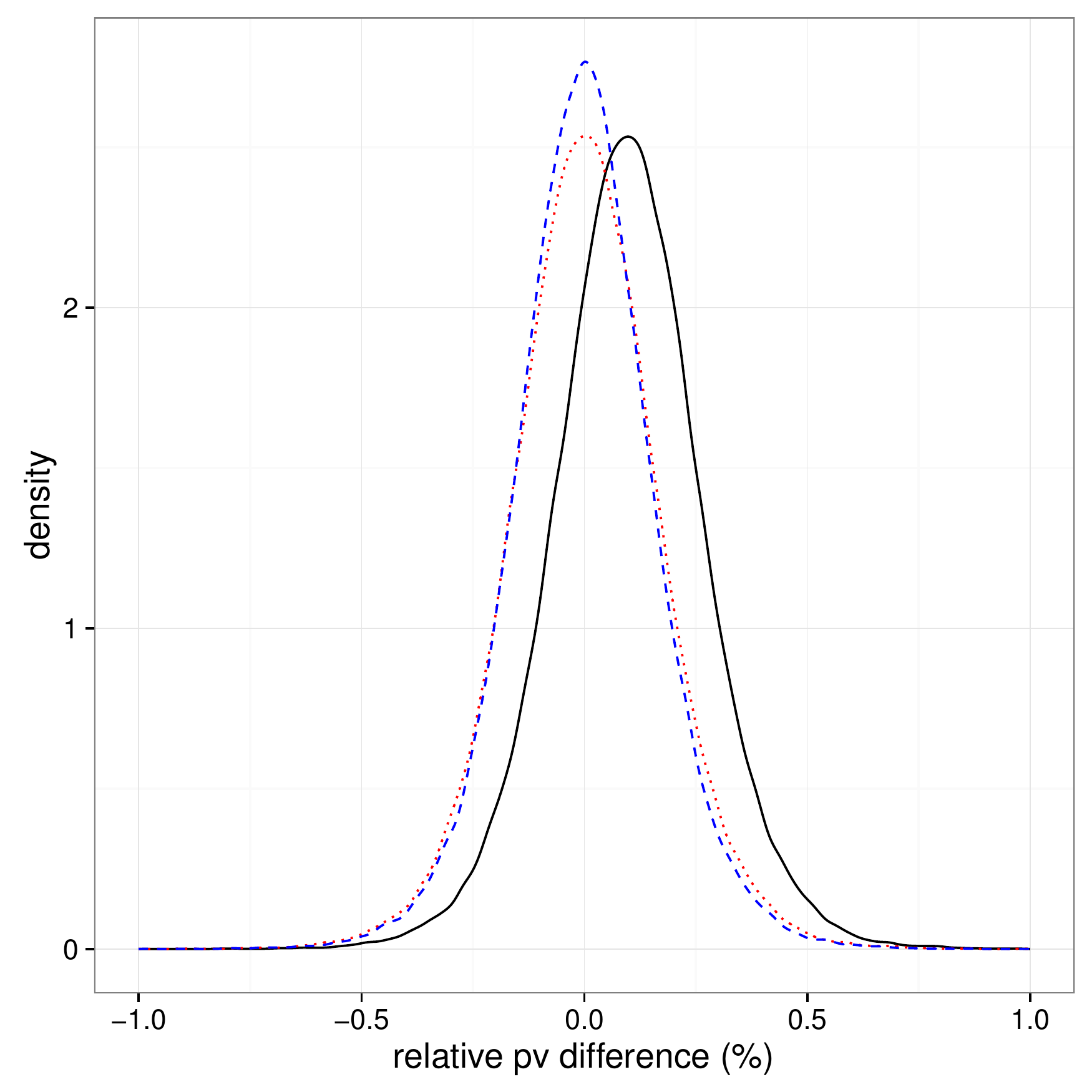}\\
(a)
\end{minipage}
\hfill
\begin{minipage}[b]{0.47\linewidth}
\centering
\includegraphics[width=\linewidth]{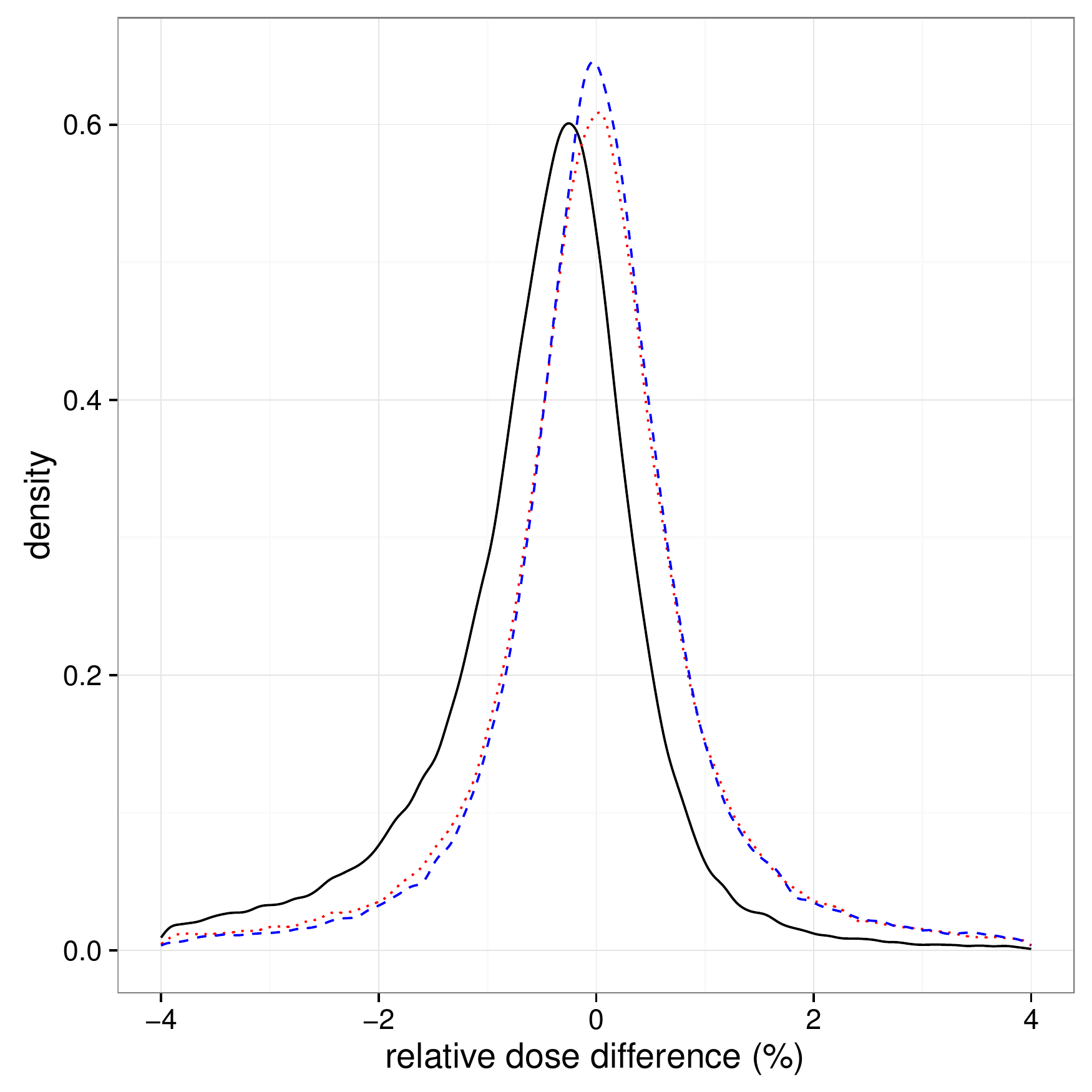}\\
(b)
\end{minipage}
\caption{\label{fig:rel_diffs} Density of the relative differences , a) in pixel value and b) in dose, between the mean image and one of the images, both in the green channel and scanned with a resolution of 72 dpi. The solid line represents the differences without any correction, while the dotted line applies to the mean correction and the dashed line applies to the column correction.}
\end{figure*}

\begin{figure*}
\begin{minipage}[b]{0.32\linewidth}
\centering
\includegraphics[width=\linewidth]{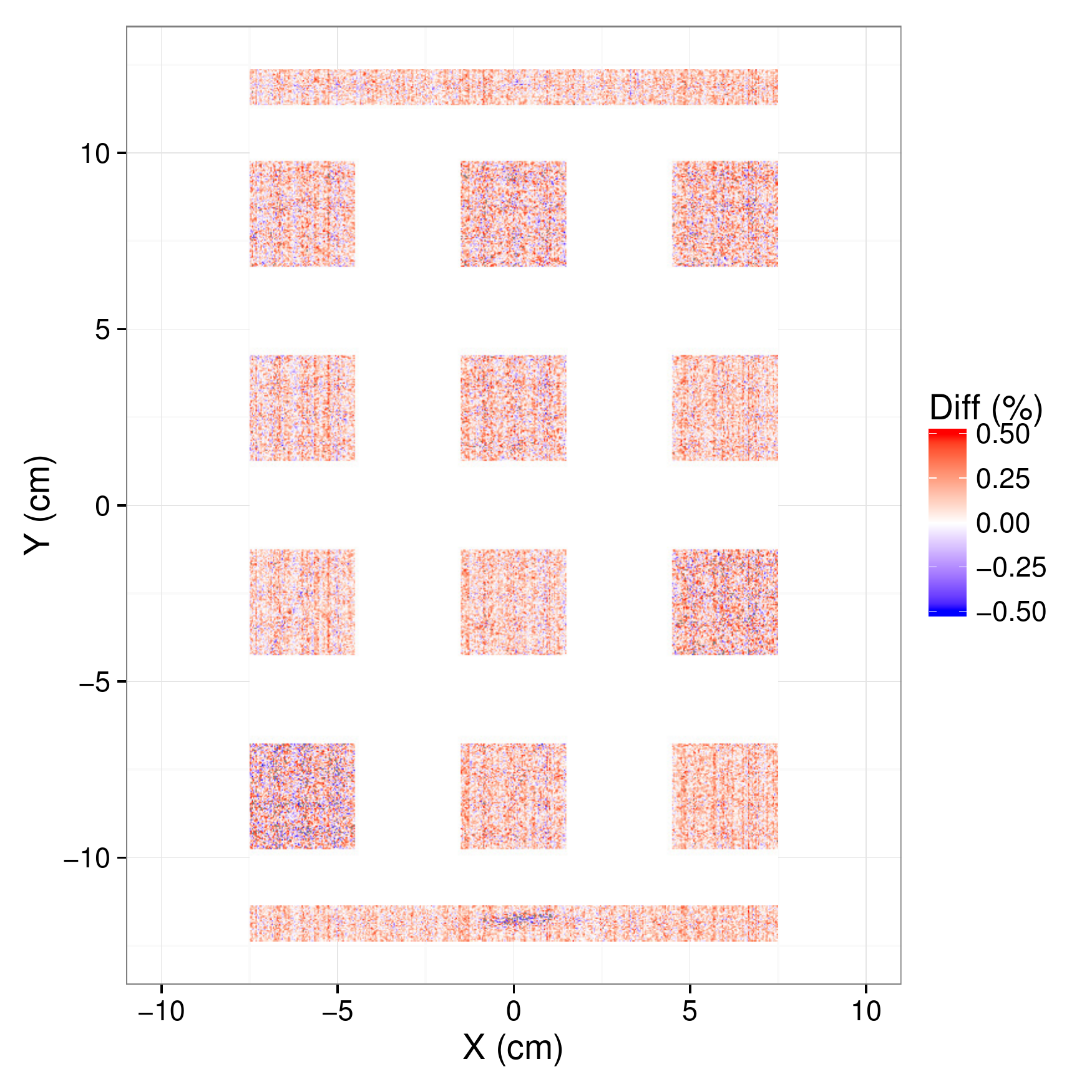}\\
(a)
\end{minipage}
\hfill
\begin{minipage}[b]{0.32\linewidth}
\centering
\includegraphics[width=\linewidth]{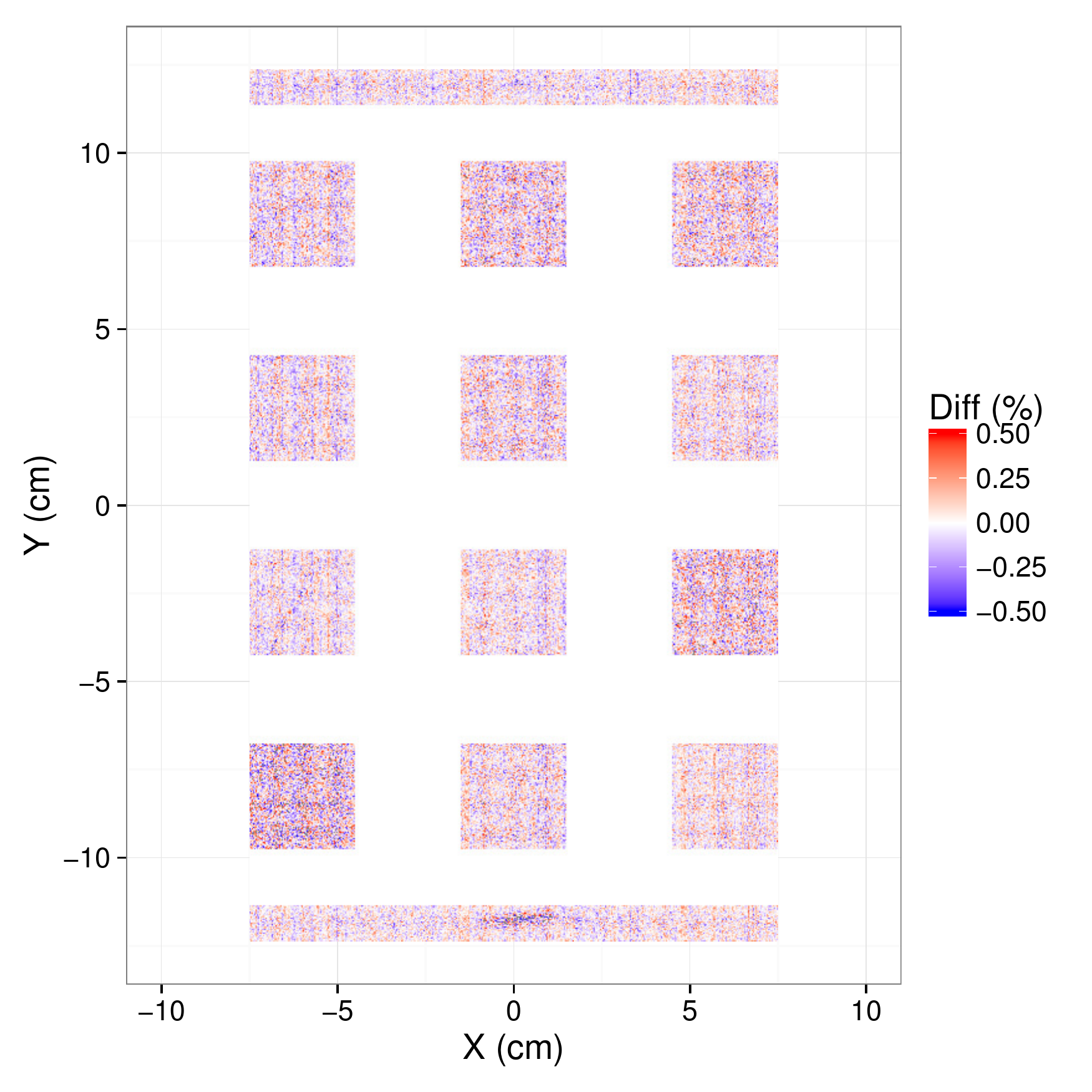}\\
(b)
\end{minipage}
\hfill
\begin{minipage}[b]{0.32\linewidth}
\centering
\includegraphics[width=\linewidth]{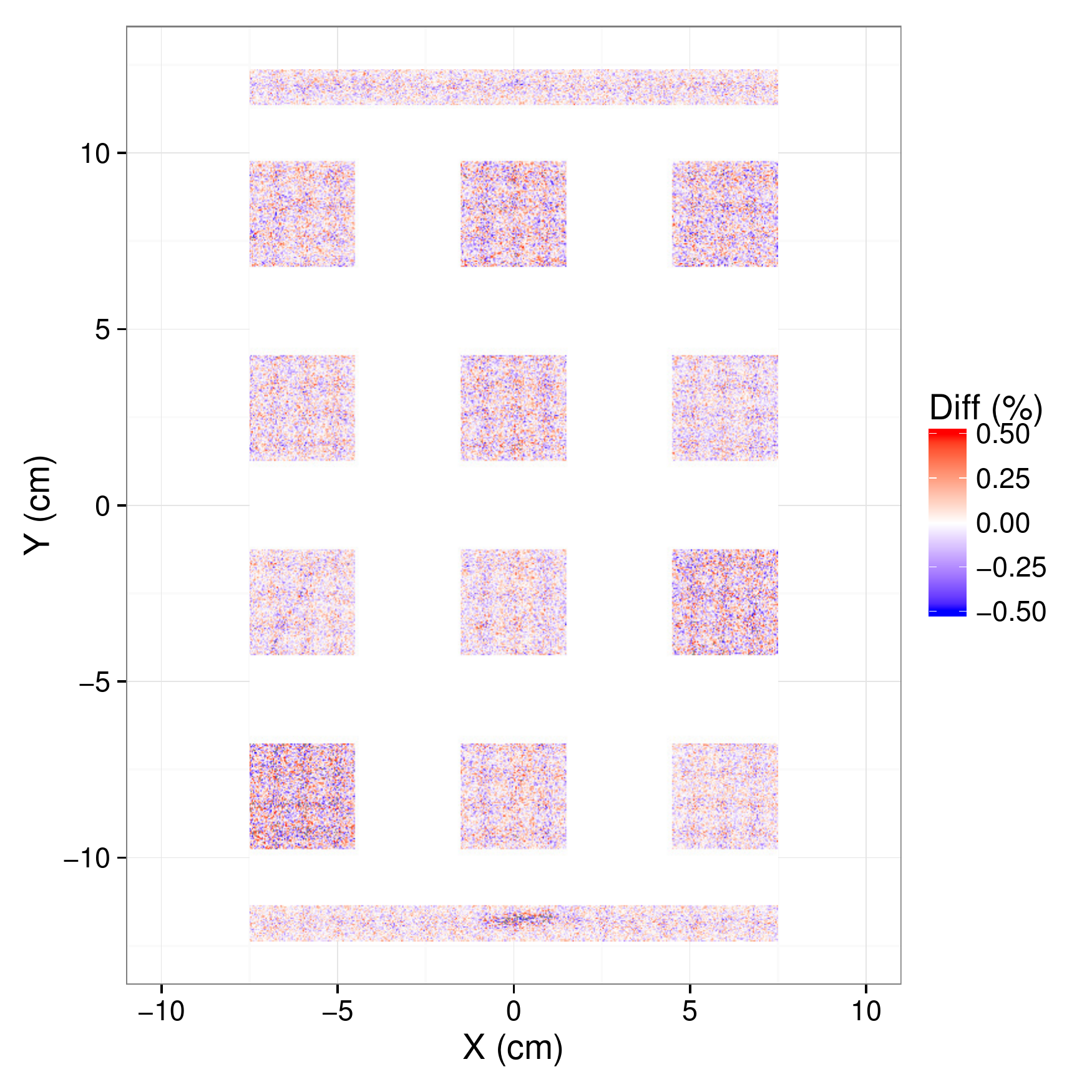}\\
(c)
\end{minipage}
\caption{\label{fig:intrascan_plots} Distribution of the relative differences (\%) between the mean image and one of the images, both in the green channel and scanned with a resolution of 72 dpi: a) without any correction, b) with mean correction, and c) with column correction.}
\end{figure*}

Figure~\ref{fig:rel_diffs} presents the density of the relative differences (in pixel value and dose) between the mean image and one of the images, both in the green channel and scanned with a resolution of 72 dpi. The map of the differences for this same scan is plotted in Figure~\ref{fig:intrascan_plots}. In this case, there was a bias or systematic deviation when no inter-scan correction was applied: the deviation with respect to zero for the mean relative pixel value difference could not be explained by the variance of pixel value differences. Systematic dose deviations were found in many other scans also, independently of the resolution. In 5\% of the red channel images, 9\% of the green and a 51\% of the blue the mean relative dose difference from the reference image was greater than 1\%. No systematic deviation larger than 1\% was found among the corrected images.

\subsection{Scanning reading repeatability}

\begin{figure*}
\begin{minipage}[b]{0.47\linewidth}
\centering
\includegraphics[width=\linewidth]{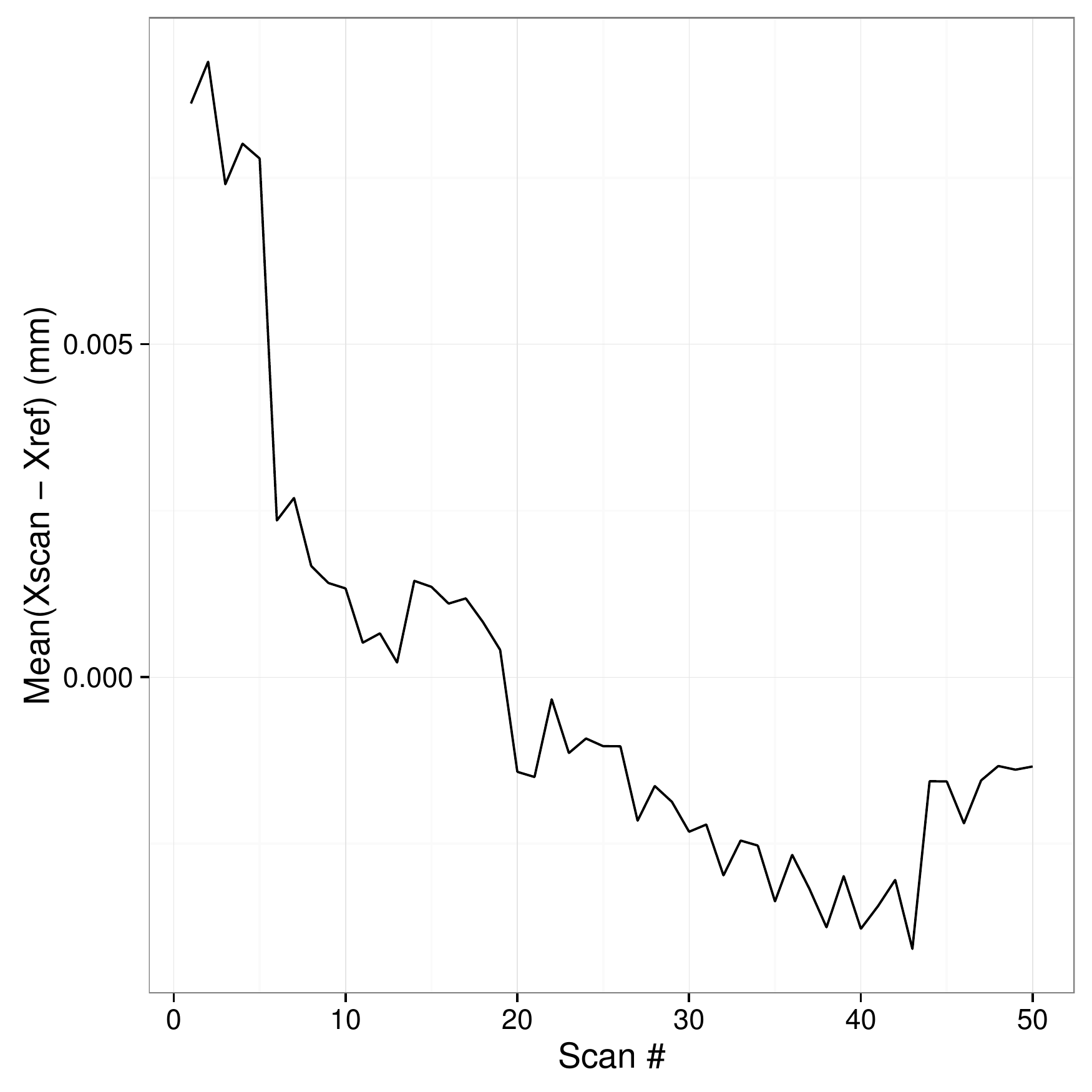}\\
(a)
\end{minipage}
\hfill
\begin{minipage}[b]{0.47\linewidth}
\centering
\includegraphics[width=\linewidth]{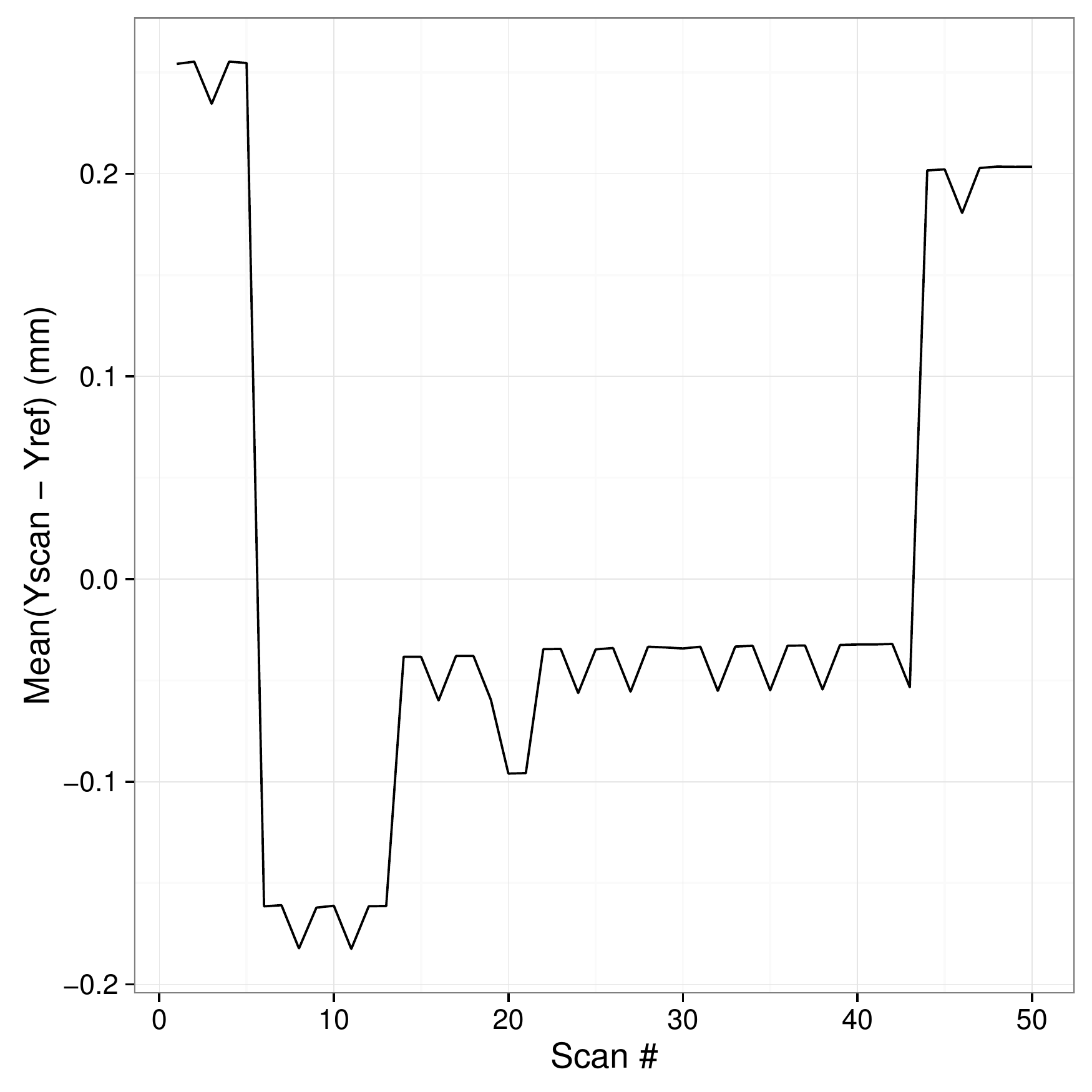}\\
(b)
\end{minipage}

\caption{\label{fig:speed_means} Mean distance between the position of the cross shape in the reference and in each of the scans: a) X axis, b) Y axis.}
\end{figure*}

\begin{figure*}
\begin{minipage}[b]{0.47\linewidth}
\centering
\includegraphics[width=\linewidth]{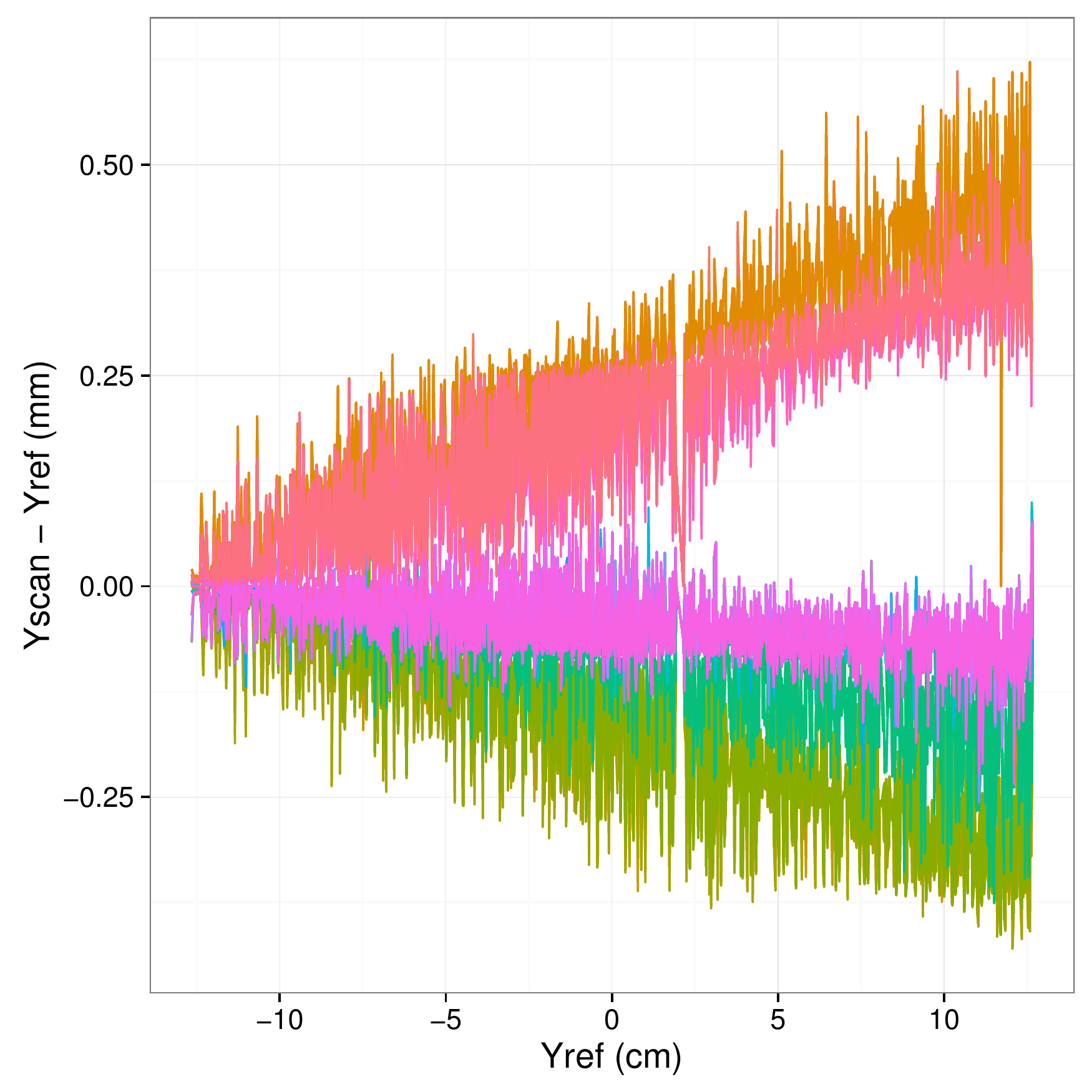}\\
(a)
\end{minipage}
\hfill
\begin{minipage}[b]{0.47\linewidth}
\centering
\includegraphics[width=\linewidth]{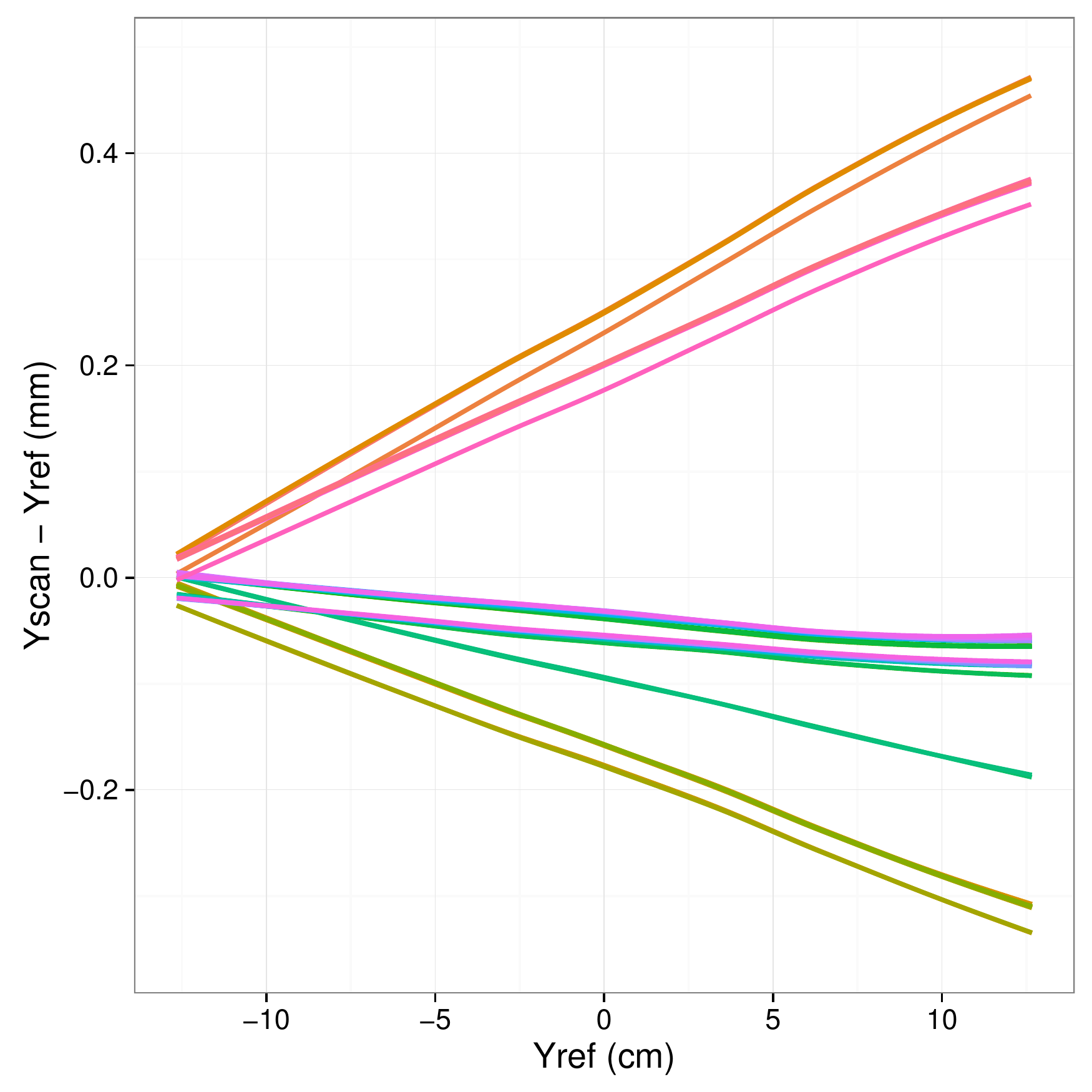}\\
(b)
\end{minipage}

\caption{\label{fig:speed_Y} Distance in the Y axis between pixels of the cross shape in the reference and in each of the scans, as a function of the reference Y position: a) raw differences, b) smoothed differences. Different scans are displayed with different colors.}
\end{figure*}

The mean distance between pixels in the reference cross shape and the same pixels in each of the scans is presented in Figure~\ref{fig:speed_means}. The variations were considered negligible and were, presumably, caused by noise in the X axis. They were not negligible in the Y axis. The distance in the Y axis, as a function of the Y position of the pixel in the reference image, is shown in Figure~\ref{fig:speed_Y}. The signals were noisy, and local polynomial regression fitting was applied to smooth them. Even though 50 scans are represented in this figure, many lines are overlapped. The initial distance in the Y axis between the reference cross and the cross in each scan is neither zero nor unique; it seems to have a set of possible discrete values. Furthermore, this distance does not remain constant, but approximately increases linearly with the lamp movement. Meanwhile, rather than a continuous of possible slopes, a discrete set was found. Both Figure~\ref{fig:speed_means} and Figure~\ref{fig:speed_Y} were obtained from the scans with a resolution of 72 dpi. Still, all the other resolutions produced similar results. 

\section{Discussion}

\subsection{Grid pattern}

Measurements of the scanner are affected by noise. It is well known \cite{ferreira:2009} that the variance of the noise depends on the resolution of the scanner: the larger the resolution, then the larger the variance. However, this variance is not constant throughout the entire scanner bed. For the scanner and scanning software being studied, periodical patterns in both axes have been found using resolutions of 50, 72 and 96 dpi. These patterns are independent of films: they even appear in the absence of transmitted light.

\begin{figure}
\includegraphics[width = 0.47\linewidth]{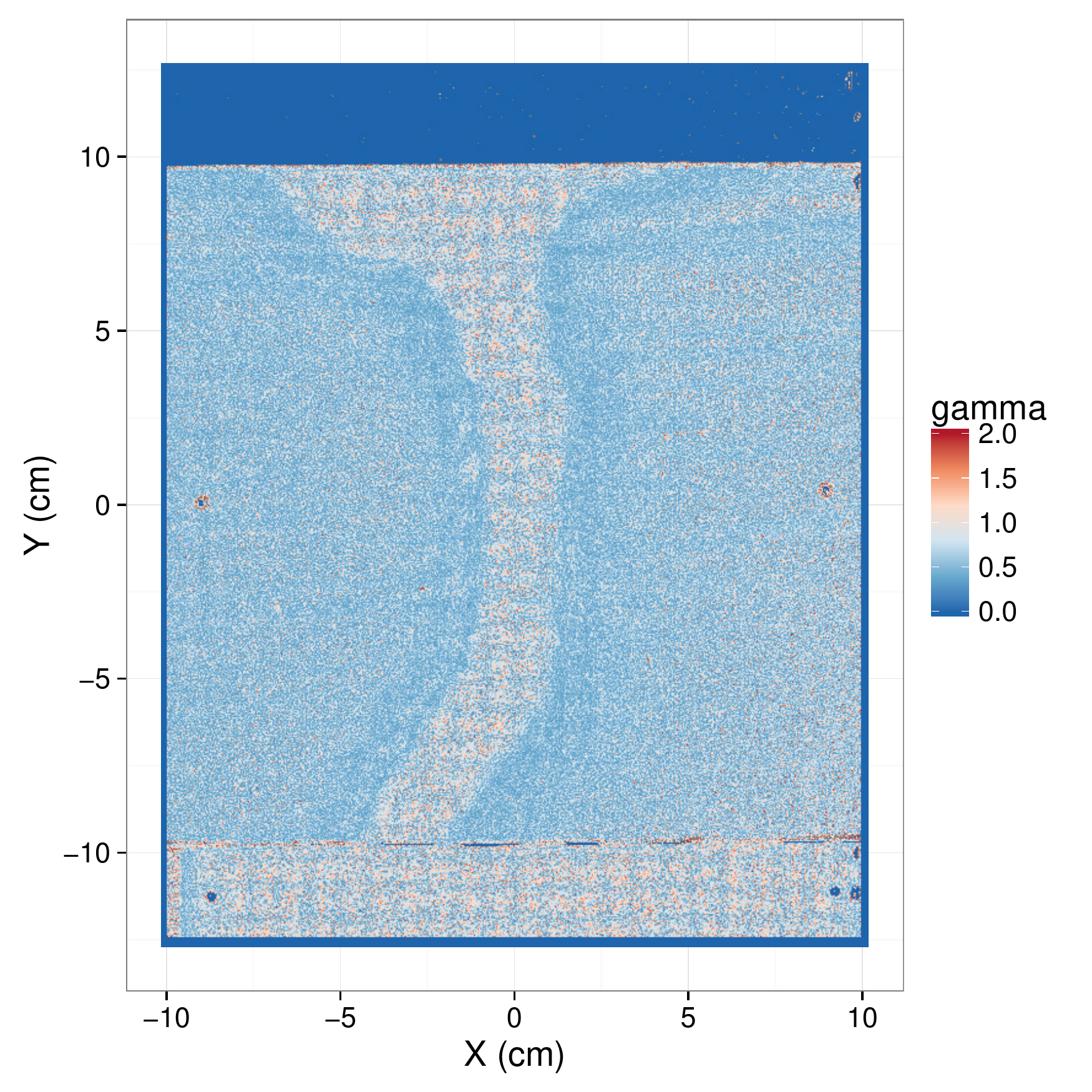}
\caption{\label{fig:disc_grid} Grid pattern in gamma comparison.}
\end{figure}

The dosimetric impact of the grid patterns depends on the slope of the sensitometric curve (and, consequently, on the dose and the color channel), the scanner repeatability and the noise variance, which in turn depends on the scanner resolution. Still, the mean dose uncertainty was at least two times greater than the difference between the maximum and the minimum dose uncertainty, for each case under study. This can explain why grid patterns are rarely detected. Nevertheless, more important than the amplitude of the differences is their periodicity, which can occasionally produce misleading grid artifacts in film dose distributions or gamma index comparisons which could have clinical implications. An example of a gamma analysis, which is affected by the grid pattern, is presented in Figure~\ref{fig:disc_grid}. It is a gamma 1\% 1 ${\rm mm}$ comparing the dose distributions calculated with two different multichannel dosimetry models, the only difference being the shape of the probability density function (pdf) of the perturbation term\cite{mendez:2014}. The film was scanned with 72 dpi. With this resolution, the pattern has a sinusoidal shape with a period of 8.5 ${\rm mm}$, which makes it particularly apparent.

Devic \emph{et al}\cite{devic:2016} proposed scanning at a high resolution ({\it e.g.}, with 150 dpi) and downscale to obtain the resolution of interest. In this way, the standard deviations associated with the average pixel values can be computed. In light of the results of the present research, this approach offers the additional benefit of preventing grid artifacts.  

The spatial variation of the pdf of pixel value differences between repeated scan images determines both grid patterns and spatial inter-scan variability. The pdf variance causes grid patterns, and the pdf mean causes the spatial inter-scan variability.

\subsection{Spatial inter-scan variability}

Applying the mean correction reduced the differences between the scan images and the mean scan image for each color channel and resolution. A larger reduction was achieved by applying the column correction.

The mean correction is equivalent to the correction proposed by Lewis and Devic\cite{lewis:2015}, who recommended the use of an unexposed film piece as reference for the scanner response in each scan image. We support this recommendation, as neglecting this correction can cause systematic errors in the determination of the dose with radiochromic films. Additionally, the response correction can be enhanced including one or several pieces irradiated with known doses to rescale the sensitometric curves ({\it e.g.}, using the efficient protocol for radiochromic film dosimetry proposed by Lewis \emph{et al}\cite{lewis:2012}).

Even though these methods mitigate the inter-scan variability of the scanner, they neglect spatial discrepancies in the repeatability. The column correction method presented in this study mitigates the spatial inter-scan variations caused by deviations in the autocalibration of the individual CCD detectors with respect to their reference state. This method is superior to the mean correction method reducing response inter-scan variations while also removing the systematic errors caused by these variations.

Even though this work employed the mean scan image as reference, as long as the Ref ROI stays in the same position between scans, any other scan or scan average could be used as reference for the correction. If the reference is the average of the scans taken for the calibration, employing the average of repeated scans in subsequent cases should reduce discrepancies in the dose-response relationship. Still, dosimetrically relevant errors caused by scans with large systematic deviations cannot be excluded. Thus, any average of scans should also be corrected using either the mean or the column correction.

\begin{figure}
\includegraphics[width = 0.47\linewidth]{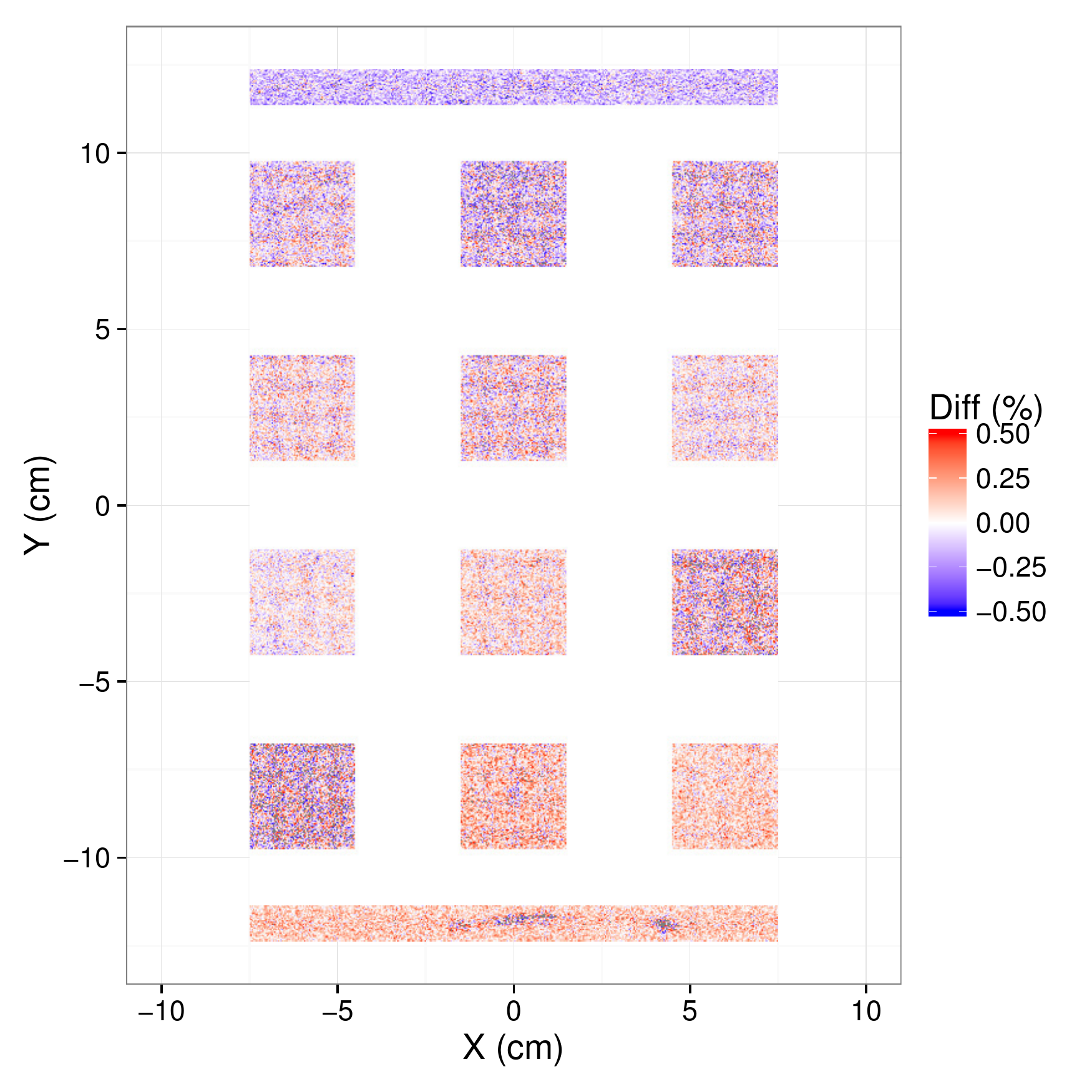}
\caption{\label{fig:warmup} Spatial inter-scan variation in the axis perpendicular to the scanner lamp.}
\end{figure}

Several other correction methods were tested and discarded in the preparation of this work. Some of them were aimed at reducing possible spatial inter-scan variations present in the axis perpendicular to the scanner lamp. None of them improved the results achieved with the column correction. As a consequence, spatial inter-scan variations in this axis were considered to be negligible. Nevertheless, it was observed that they were not negligible in the initial five scans, which were employed for warming up the scan lamp, as can be seen in Figure~\ref{fig:warmup}. This image corresponds to the red channel of one of the warm up scans after applying column corrections; the resolution of the image was 72 dpi. Spatial inter-scan variability in both axes was frequent in the warm up scans, which confirms the importance of warming up the scanner lamp before using it or after long pauses \cite{ferreira:2009}.

\subsection{Scanning reading repeatability}

The initial reading positioning (in the Y axis) and the reading speed of the scanner lamp vary between scans. Differences in the initial positioning were found to be less than 0.1 ${\rm mm}$. However, 20 ${\rm cm}$ away from the initial position, the variations in the speed produced differences of 0.7 ${\rm mm}$. Calculating the average result of several scans reduces the noise; nevertheless, the scanning reading repeatability blurs the resulting image. The blurring increases with the distance from the initial position of the scanning.

The distribution of reading positioning differences can be conservatively estimated as a uniform distribution with a support of length $\Delta_{Y} = 0.1 + 0.003 y$, where $y$ is the distance from the initial reading positioning in ${\rm mm}$. The dosimetric impact of this distribution depends on the dose gradient. For instance, let us consider a film irradiated with a $60^{\circ}$ Enhanced Dynamic Wedge field of dimensions 20$\times$20 ${\rm cm^2}$, which has been scanned several times. Excluding penumbras and out of field areas, the maximum relative dose difference between two scans would be 2.0\%, which corresponds to a point 25 ${\rm cm}$ away from the initial reading positioning in the extreme of the wedge with the lowest dose. The relative dose uncertainty associated with this point according to the uniform distribution would be 0.6\%, and the uncertainty of the reading positioning would be 0.2 ${\rm mm}$. This conservative estimation of the maximum relative dose uncertainty would be substantially reduced simply by placing the lowest dose of the wedge at the beginning of the scanning path. The scanning reading repeatability should also be considered when films are used to measure penumbras. For instance, let us consider a beam profile that is an ideal step function with zero dose out of the field. The penumbra, which is defined as the distance between the points with 20\% and 80\% of the field dose, measures 0 ${\rm mm}$ for the step function. However, if it is calculated employing the mean of several scans of this field, and is situated 25 ${\rm cm}$ away from the initial reading positioning, it will measure 0.5 ${\rm mm}$. While it is true that this value is a conservative estimation of an ideal worst case scenario,  it is a value comparable with the maximum broadening of the penumbra observed by Agostinelli \emph{et al}\cite{agostinelli:2008} for a type 31014 PinPoint ion chamber (PTW, Freiburg, Germany), which was 0.6 ${\rm mm}$. Nevertheless, the broadening of the penumbra when using radiochromic films can be prevented by employing single scans for the measurements.

\section{Conclusions}

For the scanner and scanning software under study, three new sources of uncertainty in radiochromic film dosimetry have been identified and analyzed: the grid pattern, spatial inter-scan variations and scanning reading repeatability. 

The grid patterns appear because the variance of noise is not constant throughout the entire scanner bed: it follows periodical patterns in both axes. These patterns have been identified using resolutions of 50, 72 and 96 dpi. The mean dose uncertainty due to noise and scanner repeatability was found to be at least two times greater than the difference between the maximum and the minimum dose uncertainty caused by the grid patterns, which explains why grid patterns are usually undetected. However, they can produce misleading grid artifacts in film dose distributions or gamma comparisons, with potential clinical implications.

Inter-scan variations produce discrepancies in the dose-response relationship between the calibration and subsequent scans. Response correction methods mitigate these variations and eliminate the systematic errors. In this work, a novel correction method has been proposed to reduce inter-scan variations addressing the deviations in the response of individual CCD detectors with respect to their reference state.

The initial positioning (in the Y axis) and the speed of the scanner lamp vary between
scans. The differences in the initial positioning were found negligible; however, they increase with the distance from the initial position due to the variations in reading speed. As a consequence, average scans are less accurate at the end of the scanning reading than at the beginning. Given the submillimetric scale of the positioning uncertainty, the dosimetric impact is usually negligible. Still, in some measurements this uncertainty can be relevant and actions should be taken to reduce it. 

\begin{acknowledgments}
The authors would like to thank Primo\v{z} Peterlin and Juan Jos\'{e} Rovira for their contributions to this work.

One of the authors (I.M.) is co-founder of Radiochromic.com. 
\end{acknowledgments}

\providecommand{\noopsort}[1]{}\providecommand{\singleletter}[1]{#1}%


\begin{thebibliography}{28}%
\makeatletter
\providecommand \@ifxundefined [1]{%
 \@ifx{#1\undefined}
}%
\providecommand \@ifnum [1]{%
 \ifnum #1\expandafter \@firstoftwo
 \else \expandafter \@secondoftwo
 \fi
}%
\providecommand \@ifx [1]{%
 \ifx #1\expandafter \@firstoftwo
 \else \expandafter \@secondoftwo
 \fi
}%
\providecommand \natexlab [1]{#1}%
\providecommand \enquote  [1]{``#1''}%
\providecommand \bibnamefont  [1]{#1}%
\providecommand \bibfnamefont [1]{#1}%
\providecommand \citenamefont [1]{#1}%
\providecommand \href@noop [0]{\@secondoftwo}%
\providecommand \href [0]{\begingroup \@sanitize@url \@href}%
\providecommand \@href[1]{\@@startlink{#1}\@@href}%
\providecommand \@@href[1]{\endgroup#1\@@endlink}%
\providecommand \@sanitize@url [0]{\catcode `\\12\catcode `\$12\catcode
  `\&12\catcode `\#12\catcode `\^12\catcode `\_12\catcode `\%12\relax}%
\providecommand \@@startlink[1]{}%
\providecommand \@@endlink[0]{}%
\providecommand \url  [0]{\begingroup\@sanitize@url \@url }%
\providecommand \@url [1]{\endgroup\@href {#1}{\urlprefix }}%
\providecommand \urlprefix  [0]{URL }%
\providecommand \Eprint [0]{\href }%
\providecommand \doibase [0]{http://dx.doi.org/}%
\providecommand \selectlanguage [0]{\@gobble}%
\providecommand \bibinfo  [0]{\@secondoftwo}%
\providecommand \bibfield  [0]{\@secondoftwo}%
\providecommand \translation [1]{[#1]}%
\providecommand \BibitemOpen [0]{}%
\providecommand \bibitemStop [0]{}%
\providecommand \bibitemNoStop [0]{.\EOS\space}%
\providecommand \EOS [0]{\spacefactor3000\relax}%
\providecommand \BibitemShut  [1]{\csname bibitem#1\endcsname}%
\let\auto@bib@innerbib\@empty
%</preamble>
\bibitem [{\citenamefont {Devic}(2011)}]{Devic:2011}%
  \BibitemOpen
  \bibfield  {author} {\bibinfo {author} {\bibfnamefont {S.}~\bibnamefont
  {Devic}},\ }\bibfield  {title} {\enquote {\bibinfo {title} {Radiochromic film
  dosimetry: past, present, and future},}\ }\href {\doibase
  10.1016/j.ejmp.2010.10.001} {\bibfield  {journal} {\bibinfo  {journal}
  {Physica medica}\ }\textbf {\bibinfo {volume} {27}},\ \bibinfo {pages}
  {122--134} (\bibinfo {year} {2011})}\BibitemShut {NoStop}%
\bibitem [{\citenamefont {Hartmann}, \citenamefont {Marti\v{s}\'{\i}kov\'{a}},\
  and\ \citenamefont {J\"{a}kel}(2010)}]{hartmann:2010}%
  \BibitemOpen
  \bibfield  {author} {\bibinfo {author} {\bibfnamefont {B.}~\bibnamefont
  {Hartmann}}, \bibinfo {author} {\bibfnamefont {M.}~\bibnamefont
  {Marti\v{s}\'{\i}kov\'{a}}}, \ and\ \bibinfo {author} {\bibfnamefont
  {O.}~\bibnamefont {J\"{a}kel}},\ }\bibfield  {title} {\enquote {\bibinfo
  {title} {Technical {Note}: {Homogeneity} of {Gafchromic EBT2} film},}\ }\href
  {\doibase 10.1118/1.3368601} {\bibfield  {journal} {\bibinfo  {journal}
  {Medical Physics}\ }\textbf {\bibinfo {volume} {37}},\ \bibinfo {pages}
  {1753--1756} (\bibinfo {year} {2010})}\BibitemShut {NoStop}%
\bibitem [{\citenamefont {Andr\'{e}s}\ \emph {et~al.}(2010)\citenamefont
  {Andr\'{e}s}, \citenamefont {del Castillo}, \citenamefont {Tortosa},
  \citenamefont {Alonso},\ and\ \citenamefont {Barquero}}]{andres:2010}%
  \BibitemOpen
  \bibfield  {author} {\bibinfo {author} {\bibfnamefont {C.}~\bibnamefont
  {Andr\'{e}s}}, \bibinfo {author} {\bibfnamefont {A.}~\bibnamefont {del
  Castillo}}, \bibinfo {author} {\bibfnamefont {R.}~\bibnamefont {Tortosa}},
  \bibinfo {author} {\bibfnamefont {D.}~\bibnamefont {Alonso}}, \ and\ \bibinfo
  {author} {\bibfnamefont {R.}~\bibnamefont {Barquero}},\ }\bibfield  {title}
  {\enquote {\bibinfo {title} {A comprehensive study of the {Gafchromic EBT2}
  radiochromic film. {A comparison with EBT}},}\ }\href {\doibase
  10.1118/1.3512792} {\bibfield  {journal} {\bibinfo  {journal} {Medical
  Physics}\ }\textbf {\bibinfo {volume} {37}},\ \bibinfo {pages} {6271--6278}
  (\bibinfo {year} {2010})}\BibitemShut {NoStop}%
\bibitem [{\citenamefont {Rink}\ \emph {et~al.}(2008)\citenamefont {Rink},
  \citenamefont {Lewis}, \citenamefont {Varma}, \citenamefont {Vitkin},\ and\
  \citenamefont {Jaffray}}]{rink:2008}%
  \BibitemOpen
  \bibfield  {author} {\bibinfo {author} {\bibfnamefont {A.}~\bibnamefont
  {Rink}}, \bibinfo {author} {\bibfnamefont {D.~F.}\ \bibnamefont {Lewis}},
  \bibinfo {author} {\bibfnamefont {S.}~\bibnamefont {Varma}}, \bibinfo
  {author} {\bibfnamefont {I.~A.}\ \bibnamefont {Vitkin}}, \ and\ \bibinfo
  {author} {\bibfnamefont {D.~A.}\ \bibnamefont {Jaffray}},\ }\bibfield
  {title} {\enquote {\bibinfo {title} {Temperature and hydration effects on
  absorbance spectra and radiation sensitivity of a radiochromic medium},}\
  }\href@noop {} {\bibfield  {journal} {\bibinfo  {journal} {Medical physics}\
  }\textbf {\bibinfo {volume} {35}},\ \bibinfo {pages} {4545--4555} (\bibinfo
  {year} {2008})}\BibitemShut {NoStop}%
\bibitem [{\citenamefont {Girard}, \citenamefont {Bouchard},\ and\
  \citenamefont {Lacroix}(2012)}]{Girard:2012}%
  \BibitemOpen
  \bibfield  {author} {\bibinfo {author} {\bibfnamefont {F.}~\bibnamefont
  {Girard}}, \bibinfo {author} {\bibfnamefont {H.}~\bibnamefont {Bouchard}}, \
  and\ \bibinfo {author} {\bibfnamefont {F.}~\bibnamefont {Lacroix}},\
  }\bibfield  {title} {\enquote {\bibinfo {title} {Reference dosimetry using
  radiochromic film},}\ }\href
  {http://www.jacmp.org/index.php/jacmp/article/view/3994} {\bibfield
  {journal} {\bibinfo  {journal} {Journal of Applied Clinical Medical Physics}\
  }\textbf {\bibinfo {volume} {13}} (\bibinfo {year} {2012})}\BibitemShut
  {NoStop}%
\bibitem [{\citenamefont {Niroomand-Rad}\ \emph {et~al.}(1998)\citenamefont
  {Niroomand-Rad}, \citenamefont {Blackwell}, \citenamefont {Coursey},
  \citenamefont {Gall}, \citenamefont {Galvin}, \citenamefont {McLaughlin},
  \citenamefont {Meigooni}, \citenamefont {Nath}, \citenamefont {Rodgers},\
  and\ \citenamefont {Soares}}]{aapm:55}%
  \BibitemOpen
  \bibfield  {author} {\bibinfo {author} {\bibfnamefont {A.}~\bibnamefont
  {Niroomand-Rad}}, \bibinfo {author} {\bibfnamefont {C.~R.}\ \bibnamefont
  {Blackwell}}, \bibinfo {author} {\bibfnamefont {B.~M.}\ \bibnamefont
  {Coursey}}, \bibinfo {author} {\bibfnamefont {K.~P.}\ \bibnamefont {Gall}},
  \bibinfo {author} {\bibfnamefont {J.~M.}\ \bibnamefont {Galvin}}, \bibinfo
  {author} {\bibfnamefont {W.~L.}\ \bibnamefont {McLaughlin}}, \bibinfo
  {author} {\bibfnamefont {A.~S.}\ \bibnamefont {Meigooni}}, \bibinfo {author}
  {\bibfnamefont {R.}~\bibnamefont {Nath}}, \bibinfo {author} {\bibfnamefont
  {J.~E.}\ \bibnamefont {Rodgers}}, \ and\ \bibinfo {author} {\bibfnamefont
  {C.~G.}\ \bibnamefont {Soares}},\ }\bibfield  {title} {\enquote {\bibinfo
  {title} {Radiochromic film dosimetry: {Recommendations of AAPM Radiation
  Therapy Committee Task Group} 55},}\ }\href {\doibase 10.1118/1.598407}
  {\bibfield  {journal} {\bibinfo  {journal} {Medical Physics}\ }\textbf
  {\bibinfo {volume} {25}},\ \bibinfo {pages} {2093--2115} (\bibinfo {year}
  {1998})}\BibitemShut {NoStop}%
\bibitem [{\citenamefont {Schoenfeld}\ \emph {et~al.}(2014)\citenamefont
  {Schoenfeld}, \citenamefont {Poppinga}, \citenamefont {Harder}, \citenamefont
  {Doerner},\ and\ \citenamefont {Poppe}}]{Schoenfeld:2014}%
  \BibitemOpen
  \bibfield  {author} {\bibinfo {author} {\bibfnamefont {A.~A.}\ \bibnamefont
  {Schoenfeld}}, \bibinfo {author} {\bibfnamefont {D.}~\bibnamefont
  {Poppinga}}, \bibinfo {author} {\bibfnamefont {D.}~\bibnamefont {Harder}},
  \bibinfo {author} {\bibfnamefont {K.-J.}\ \bibnamefont {Doerner}}, \ and\
  \bibinfo {author} {\bibfnamefont {B.}~\bibnamefont {Poppe}},\ }\bibfield
  {title} {\enquote {\bibinfo {title} {The artefacts of radiochromic film
  dosimetry with flatbed scanners and their causation by light scattering from
  radiation-induced polymers},}\ }\href
  {http://stacks.iop.org/0031-9155/59/i=13/a=3575} {\bibfield  {journal}
  {\bibinfo  {journal} {Physics in Medicine and Biology}\ }\textbf {\bibinfo
  {volume} {59}},\ \bibinfo {pages} {3575--3597} (\bibinfo {year}
  {2014})}\BibitemShut {NoStop}%
\bibitem [{\citenamefont {van Battum}\ \emph {et~al.}(2015)\citenamefont {van
  Battum}, \citenamefont {Huizenga}, \citenamefont {Verdaasdonk},\ and\
  \citenamefont {Heukelom}}]{vanbattum:2015}%
  \BibitemOpen
  \bibfield  {author} {\bibinfo {author} {\bibfnamefont {L.}~\bibnamefont {van
  Battum}}, \bibinfo {author} {\bibfnamefont {H.}~\bibnamefont {Huizenga}},
  \bibinfo {author} {\bibfnamefont {R.}~\bibnamefont {Verdaasdonk}}, \ and\
  \bibinfo {author} {\bibfnamefont {S.}~\bibnamefont {Heukelom}},\ }\bibfield
  {title} {\enquote {\bibinfo {title} {How flatbed scanners upset accurate film
  dosimetry},}\ }\href@noop {} {\bibfield  {journal} {\bibinfo  {journal}
  {Physics in medicine and biology}\ }\textbf {\bibinfo {volume} {61}},\
  \bibinfo {pages} {625} (\bibinfo {year} {2015})}\BibitemShut {NoStop}%
\bibitem [{\citenamefont {Dreindl}, \citenamefont {Georg},\ and\ \citenamefont
  {Stock}(2014)}]{dreindl:2014}%
  \BibitemOpen
  \bibfield  {author} {\bibinfo {author} {\bibfnamefont {R.}~\bibnamefont
  {Dreindl}}, \bibinfo {author} {\bibfnamefont {D.}~\bibnamefont {Georg}}, \
  and\ \bibinfo {author} {\bibfnamefont {M.}~\bibnamefont {Stock}},\ }\bibfield
   {title} {\enquote {\bibinfo {title} {Radiochromic film dosimetry:
  Considerations on precision and accuracy for {EBT2 and EBT3} type films},}\
  }\href@noop {} {\bibfield  {journal} {\bibinfo  {journal} {Zeitschrift
  f{\"u}r Medizinische Physik}\ }\textbf {\bibinfo {volume} {24}},\ \bibinfo
  {pages} {153--163} (\bibinfo {year} {2014})}\BibitemShut {NoStop}%
\bibitem [{\citenamefont {Butson}, \citenamefont {Cheung},\ and\ \citenamefont
  {Yu}(2009)}]{butson:2009}%
  \BibitemOpen
  \bibfield  {author} {\bibinfo {author} {\bibfnamefont {M.~J.}\ \bibnamefont
  {Butson}}, \bibinfo {author} {\bibfnamefont {T.}~\bibnamefont {Cheung}}, \
  and\ \bibinfo {author} {\bibfnamefont {P.}~\bibnamefont {Yu}},\ }\bibfield
  {title} {\enquote {\bibinfo {title} {Evaluation of the magnitude of {EBT
  Gafchromic} film polarization effects},}\ }\href@noop {} {\bibfield
  {journal} {\bibinfo  {journal} {Australasian Physics \& Engineering Sciences
  in Medicine}\ }\textbf {\bibinfo {volume} {32}},\ \bibinfo {pages} {21--25}
  (\bibinfo {year} {2009})}\BibitemShut {NoStop}%
\bibitem [{\citenamefont {Lewis}\ and\ \citenamefont
  {Devic}(2015)}]{lewis:2015}%
  \BibitemOpen
  \bibfield  {author} {\bibinfo {author} {\bibfnamefont {D.}~\bibnamefont
  {Lewis}}\ and\ \bibinfo {author} {\bibfnamefont {S.}~\bibnamefont {Devic}},\
  }\bibfield  {title} {\enquote {\bibinfo {title} {Correcting scan-to-scan
  response variability for a radiochromic film-based reference dosimetry
  system},}\ }\href@noop {} {\bibfield  {journal} {\bibinfo  {journal} {Medical
  physics}\ }\textbf {\bibinfo {volume} {42}},\ \bibinfo {pages} {5692--5701}
  (\bibinfo {year} {2015})}\BibitemShut {NoStop}%
\bibitem [{\citenamefont {Palmer}, \citenamefont {Bradley},\ and\ \citenamefont
  {Nisbet}(2015)}]{palmer:2015}%
  \BibitemOpen
  \bibfield  {author} {\bibinfo {author} {\bibfnamefont {A.~L.}\ \bibnamefont
  {Palmer}}, \bibinfo {author} {\bibfnamefont {D.~A.}\ \bibnamefont {Bradley}},
  \ and\ \bibinfo {author} {\bibfnamefont {A.}~\bibnamefont {Nisbet}},\
  }\bibfield  {title} {\enquote {\bibinfo {title} {Evaluation and mitigation of
  potential errors in radiochromic film dosimetry due to film curvature at
  scanning},}\ }\href@noop {} {\bibfield  {journal} {\bibinfo  {journal}
  {Journal of Applied Clinical Medical Physics}\ }\textbf {\bibinfo {volume}
  {16}} (\bibinfo {year} {2015})}\BibitemShut {NoStop}%
\bibitem [{\citenamefont {Bouchard}\ \emph {et~al.}(2009)\citenamefont
  {Bouchard}, \citenamefont {Lacroix}, \citenamefont {Beaudoin}, \citenamefont
  {Carrier},\ and\ \citenamefont {Kawrakow}}]{bouchard:2009}%
  \BibitemOpen
  \bibfield  {author} {\bibinfo {author} {\bibfnamefont {H.}~\bibnamefont
  {Bouchard}}, \bibinfo {author} {\bibfnamefont {F.}~\bibnamefont {Lacroix}},
  \bibinfo {author} {\bibfnamefont {G.}~\bibnamefont {Beaudoin}}, \bibinfo
  {author} {\bibfnamefont {J.-F.}\ \bibnamefont {Carrier}}, \ and\ \bibinfo
  {author} {\bibfnamefont {I.}~\bibnamefont {Kawrakow}},\ }\bibfield  {title}
  {\enquote {\bibinfo {title} {On the characterization and uncertainty analysis
  of radiochromic film dosimetry},}\ }\href {\doibase 10.1118/1.3121488}
  {\bibfield  {journal} {\bibinfo  {journal} {Medical Physics}\ }\textbf
  {\bibinfo {volume} {36}},\ \bibinfo {pages} {1931--1946} (\bibinfo {year}
  {2009})}\BibitemShut {NoStop}%
\bibitem [{\citenamefont {van Hoof}\ \emph {et~al.}(2012)\citenamefont {van
  Hoof}, \citenamefont {Granton}, \citenamefont {Landry}, \citenamefont
  {Podesta},\ and\ \citenamefont {Verhaegen}}]{vanHoof:2012}%
  \BibitemOpen
  \bibfield  {author} {\bibinfo {author} {\bibfnamefont {S.~J.}\ \bibnamefont
  {van Hoof}}, \bibinfo {author} {\bibfnamefont {P.~V.}\ \bibnamefont
  {Granton}}, \bibinfo {author} {\bibfnamefont {G.}~\bibnamefont {Landry}},
  \bibinfo {author} {\bibfnamefont {M.}~\bibnamefont {Podesta}}, \ and\
  \bibinfo {author} {\bibfnamefont {F.}~\bibnamefont {Verhaegen}},\ }\bibfield
  {title} {\enquote {\bibinfo {title} {Evaluation of a novel triple-channel
  radiochromic film analysis procedure using {EBT2}},}\ }\href
  {http://stacks.iop.org/0031-9155/57/i=13/a=4353} {\bibfield  {journal}
  {\bibinfo  {journal} {Physics in Medicine and Biology}\ }\textbf {\bibinfo
  {volume} {57}},\ \bibinfo {pages} {4353--4368} (\bibinfo {year}
  {2012})}\BibitemShut {NoStop}%
\bibitem [{\citenamefont {Paelinck}, \citenamefont {Neve},\ and\ \citenamefont
  {Wagter}(2007)}]{Paelinck:2007}%
  \BibitemOpen
  \bibfield  {author} {\bibinfo {author} {\bibfnamefont {L.}~\bibnamefont
  {Paelinck}}, \bibinfo {author} {\bibfnamefont {W.~D.}\ \bibnamefont {Neve}},
  \ and\ \bibinfo {author} {\bibfnamefont {C.~D.}\ \bibnamefont {Wagter}},\
  }\bibfield  {title} {\enquote {\bibinfo {title} {Precautions and strategies
  in using a commercial flatbed scanner for radiochromic film dosimetry},}\
  }\href {http://stacks.iop.org/0031-9155/52/i=1/a=015} {\bibfield  {journal}
  {\bibinfo  {journal} {Physics in Medicine and Biology}\ }\textbf {\bibinfo
  {volume} {52}},\ \bibinfo {pages} {231--242} (\bibinfo {year}
  {2007})}\BibitemShut {NoStop}%
\bibitem [{\citenamefont {Ferreira}, \citenamefont {Lopes},\ and\ \citenamefont
  {Capela}(2009)}]{ferreira:2009}%
  \BibitemOpen
  \bibfield  {author} {\bibinfo {author} {\bibfnamefont {B.}~\bibnamefont
  {Ferreira}}, \bibinfo {author} {\bibfnamefont {M.}~\bibnamefont {Lopes}}, \
  and\ \bibinfo {author} {\bibfnamefont {M.}~\bibnamefont {Capela}},\
  }\bibfield  {title} {\enquote {\bibinfo {title} {Evaluation of an {Epson}
  flatbed scanner to read {Gafchromic EBT} films for radiation dosimetry},}\
  }\href@noop {} {\bibfield  {journal} {\bibinfo  {journal} {Physics in
  medicine and biology}\ }\textbf {\bibinfo {volume} {54}},\ \bibinfo {pages}
  {1073} (\bibinfo {year} {2009})}\BibitemShut {NoStop}%
\bibitem [{\citenamefont {Micke}, \citenamefont {Lewis},\ and\ \citenamefont
  {Yu}(2011)}]{AMicke:2011}%
  \BibitemOpen
  \bibfield  {author} {\bibinfo {author} {\bibfnamefont {A.}~\bibnamefont
  {Micke}}, \bibinfo {author} {\bibfnamefont {D.~F.}\ \bibnamefont {Lewis}}, \
  and\ \bibinfo {author} {\bibfnamefont {X.}~\bibnamefont {Yu}},\ }\bibfield
  {title} {\enquote {\bibinfo {title} {Multichannel film dosimetry with
  nonuniformity correction},}\ }\href {\doibase 10.1118/1.3576105} {\bibfield
  {journal} {\bibinfo  {journal} {Medical Physics}\ }\textbf {\bibinfo {volume}
  {38}},\ \bibinfo {pages} {2523--2534} (\bibinfo {year} {2011})}\BibitemShut
  {NoStop}%
\bibitem [{\citenamefont {Mayer}\ \emph {et~al.}(2012)\citenamefont {Mayer},
  \citenamefont {Ma}, \citenamefont {Chen}, \citenamefont {Miller},
  \citenamefont {Belard}, \citenamefont {McDonough},\ and\ \citenamefont
  {O'Connell}}]{mayer:2012}%
  \BibitemOpen
  \bibfield  {author} {\bibinfo {author} {\bibfnamefont {R.~R.}\ \bibnamefont
  {Mayer}}, \bibinfo {author} {\bibfnamefont {F.}~\bibnamefont {Ma}}, \bibinfo
  {author} {\bibfnamefont {Y.}~\bibnamefont {Chen}}, \bibinfo {author}
  {\bibfnamefont {R.~I.}\ \bibnamefont {Miller}}, \bibinfo {author}
  {\bibfnamefont {A.}~\bibnamefont {Belard}}, \bibinfo {author} {\bibfnamefont
  {J.}~\bibnamefont {McDonough}}, \ and\ \bibinfo {author} {\bibfnamefont
  {J.~J.}\ \bibnamefont {O'Connell}},\ }\bibfield  {title} {\enquote {\bibinfo
  {title} {Enhanced dosimetry procedures and assessment for {EBT2} radiochromic
  film},}\ }\href {\doibase 10.1118/1.3694100} {\bibfield  {journal} {\bibinfo
  {journal} {Medical Physics}\ }\textbf {\bibinfo {volume} {39}},\ \bibinfo
  {pages} {2147--2155} (\bibinfo {year} {2012})}\BibitemShut {NoStop}%
\bibitem [{\citenamefont {M{\'e}ndez}\ \emph {et~al.}(2014)\citenamefont
  {M{\'e}ndez}, \citenamefont {Peterlin}, \citenamefont {Hudej}, \citenamefont
  {Strojnik},\ and\ \citenamefont {Casar}}]{mendez:2014}%
  \BibitemOpen
  \bibfield  {author} {\bibinfo {author} {\bibfnamefont {I.}~\bibnamefont
  {M{\'e}ndez}}, \bibinfo {author} {\bibfnamefont {P.}~\bibnamefont
  {Peterlin}}, \bibinfo {author} {\bibfnamefont {R.}~\bibnamefont {Hudej}},
  \bibinfo {author} {\bibfnamefont {A.}~\bibnamefont {Strojnik}}, \ and\
  \bibinfo {author} {\bibfnamefont {B.}~\bibnamefont {Casar}},\ }\bibfield
  {title} {\enquote {\bibinfo {title} {On multichannel film dosimetry with
  channel-independent perturbations},}\ }\href {\doibase
  http://dx.doi.org/10.1118/1.4845095} {\bibfield  {journal} {\bibinfo
  {journal} {Medical Physics}\ }\textbf {\bibinfo {volume} {41}},\ \bibinfo
  {pages} {011705 (10pp.)} (\bibinfo {year} {2014})}\BibitemShut {NoStop}%
\bibitem [{\citenamefont {Azor{\'\i}n}, \citenamefont {Garc{\'\i}a},\ and\
  \citenamefont {Mart{\'\i}-Climent}(2014)}]{perez:2014}%
  \BibitemOpen
  \bibfield  {author} {\bibinfo {author} {\bibfnamefont {J.~F.~P.}\
  \bibnamefont {Azor{\'\i}n}}, \bibinfo {author} {\bibfnamefont {L.~I.~R.}\
  \bibnamefont {Garc{\'\i}a}}, \ and\ \bibinfo {author} {\bibfnamefont {J.~M.}\
  \bibnamefont {Mart{\'\i}-Climent}},\ }\bibfield  {title} {\enquote {\bibinfo
  {title} {A method for multichannel dosimetry with {EBT3} radiochromic
  films},}\ }\href {\doibase http://dx.doi.org/10.1118/1.4871622} {\bibfield
  {journal} {\bibinfo  {journal} {Medical Physics}\ }\textbf {\bibinfo {volume}
  {41}},\ \bibinfo {pages} {062101 (10pp.)} (\bibinfo {year}
  {2014})}\BibitemShut {NoStop}%
\bibitem [{\citenamefont {Lewis}\ \emph {et~al.}(2012)\citenamefont {Lewis},
  \citenamefont {Micke}, \citenamefont {Yu},\ and\ \citenamefont
  {Chan}}]{lewis:2012}%
  \BibitemOpen
  \bibfield  {author} {\bibinfo {author} {\bibfnamefont {D.}~\bibnamefont
  {Lewis}}, \bibinfo {author} {\bibfnamefont {A.}~\bibnamefont {Micke}},
  \bibinfo {author} {\bibfnamefont {X.}~\bibnamefont {Yu}}, \ and\ \bibinfo
  {author} {\bibfnamefont {M.~F.}\ \bibnamefont {Chan}},\ }\bibfield  {title}
  {\enquote {\bibinfo {title} {An efficient protocol for radiochromic film
  dosimetry combining calibration and measurement in a single scan},}\ }\href
  {\doibase 10.1118/1.4754797} {\bibfield  {journal} {\bibinfo  {journal}
  {Medical Physics}\ }\textbf {\bibinfo {volume} {39}},\ \bibinfo {pages}
  {6339--6350} (\bibinfo {year} {2012})}\BibitemShut {NoStop}%
\bibitem [{\citenamefont {Palmer}, \citenamefont {Nisbet},\ and\ \citenamefont
  {Bradley}(2013)}]{palmer:2013}%
  \BibitemOpen
  \bibfield  {author} {\bibinfo {author} {\bibfnamefont {A.~L.}\ \bibnamefont
  {Palmer}}, \bibinfo {author} {\bibfnamefont {A.}~\bibnamefont {Nisbet}}, \
  and\ \bibinfo {author} {\bibfnamefont {D.}~\bibnamefont {Bradley}},\
  }\bibfield  {title} {\enquote {\bibinfo {title} {Verification of high dose
  rate brachytherapy dose distributions with {EBT3 Gafchromic} film quality
  control techniques},}\ }\href@noop {} {\bibfield  {journal} {\bibinfo
  {journal} {Physics in medicine and biology}\ }\textbf {\bibinfo {volume}
  {58}},\ \bibinfo {pages} {497} (\bibinfo {year} {2013})}\BibitemShut
  {NoStop}%
\bibitem [{\citenamefont {M{\'e}ndez}(2015)}]{mendez:2015}%
  \BibitemOpen
  \bibfield  {author} {\bibinfo {author} {\bibfnamefont {I.}~\bibnamefont
  {M{\'e}ndez}},\ }\bibfield  {title} {\enquote {\bibinfo {title} {Model
  selection for radiochromic film dosimetry},}\ }\href@noop {} {\bibfield
  {journal} {\bibinfo  {journal} {Physics in medicine and biology}\ }\textbf
  {\bibinfo {volume} {60}},\ \bibinfo {pages} {4089} (\bibinfo {year}
  {2015})}\BibitemShut {NoStop}%
\bibitem [{\citenamefont {Lewis}\ and\ \citenamefont
  {Chan}(2015)}]{lewis:2015b}%
  \BibitemOpen
  \bibfield  {author} {\bibinfo {author} {\bibfnamefont {D.}~\bibnamefont
  {Lewis}}\ and\ \bibinfo {author} {\bibfnamefont {M.~F.}\ \bibnamefont
  {Chan}},\ }\bibfield  {title} {\enquote {\bibinfo {title} {Correcting lateral
  response artifacts from flatbed scanners for radiochromic film dosimetry},}\
  }\href@noop {} {\bibfield  {journal} {\bibinfo  {journal} {Medical physics}\
  }\textbf {\bibinfo {volume} {42}},\ \bibinfo {pages} {416--429} (\bibinfo
  {year} {2015})}\BibitemShut {NoStop}%
\bibitem [{\citenamefont {Marti\v{s}\'{\i}kov\'{a}}, \citenamefont
  {Ackermann},\ and\ \citenamefont {J\"{a}kel}(2008)}]{Martisikova:2008}%
  \BibitemOpen
  \bibfield  {author} {\bibinfo {author} {\bibfnamefont {M.}~\bibnamefont
  {Marti\v{s}\'{\i}kov\'{a}}}, \bibinfo {author} {\bibfnamefont
  {B.}~\bibnamefont {Ackermann}}, \ and\ \bibinfo {author} {\bibfnamefont
  {O.}~\bibnamefont {J\"{a}kel}},\ }\bibfield  {title} {\enquote {\bibinfo
  {title} {Analysis of uncertainties in {Gafchromic EBT} film dosimetry of
  photon beams},}\ }\href {http://stacks.iop.org/0031-9155/53/i=24/a=001}
  {\bibfield  {journal} {\bibinfo  {journal} {Physics in Medicine and Biology}\
  }\textbf {\bibinfo {volume} {53}},\ \bibinfo {pages} {7013--7027} (\bibinfo
  {year} {2008})}\BibitemShut {NoStop}%
\bibitem [{\citenamefont {{R Core Team}}(2012)}]{R:software}%
  \BibitemOpen
  \bibfield  {author} {\bibinfo {author} {\bibnamefont {{R Core Team}}},\
  }\href {http://www.R-project.org/} {\emph {\bibinfo {title} {R: A Language
  and Environment for Statistical Computing}}},\ \bibinfo {organization} {R
  Foundation for Statistical Computing},\ \bibinfo {address} {Vienna, Austria}
  (\bibinfo {year} {2012}),\ \bibinfo {note} {{ISBN} 3-900051-07-0}\BibitemShut
  {NoStop}%
\bibitem [{\citenamefont {Devic}, \citenamefont {Tomic},\ and\ \citenamefont
  {Lewis}(2016)}]{devic:2016}%
  \BibitemOpen
  \bibfield  {author} {\bibinfo {author} {\bibfnamefont {S.}~\bibnamefont
  {Devic}}, \bibinfo {author} {\bibfnamefont {N.}~\bibnamefont {Tomic}}, \ and\
  \bibinfo {author} {\bibfnamefont {D.}~\bibnamefont {Lewis}},\ }\bibfield
  {title} {\enquote {\bibinfo {title} {Reference radiochromic film dosimetry:
  Review of technical aspects},}\ }\href@noop {} {\bibfield  {journal}
  {\bibinfo  {journal} {Physica Medica}\ }\textbf {\bibinfo {volume} {32}},\
  \bibinfo {pages} {541--556} (\bibinfo {year} {2016})}\BibitemShut {NoStop}%
\bibitem [{\citenamefont {Agostinelli}\ \emph {et~al.}(2008)\citenamefont
  {Agostinelli}, \citenamefont {Garelli}, \citenamefont {Piergentili},\ and\
  \citenamefont {Foppiano}}]{agostinelli:2008}%
  \BibitemOpen
  \bibfield  {author} {\bibinfo {author} {\bibfnamefont {S.}~\bibnamefont
  {Agostinelli}}, \bibinfo {author} {\bibfnamefont {S.}~\bibnamefont
  {Garelli}}, \bibinfo {author} {\bibfnamefont {M.}~\bibnamefont
  {Piergentili}}, \ and\ \bibinfo {author} {\bibfnamefont {F.}~\bibnamefont
  {Foppiano}},\ }\bibfield  {title} {\enquote {\bibinfo {title} {Response to
  high-energy photons of {PTW31014 PinPoint} ion chamber with a central
  aluminum electrode},}\ }\href@noop {} {\bibfield  {journal} {\bibinfo
  {journal} {Medical physics}\ }\textbf {\bibinfo {volume} {35}},\ \bibinfo
  {pages} {3293--3301} (\bibinfo {year} {2008})}\BibitemShut {NoStop}%
\end{thebibliography}
\end{document}